\documentclass[twocolumn,english,twocolumn,prl]{revtex4-1}
\usepackage[T1]{fontenc}
\usepackage[latin9]{inputenc}
\usepackage{geometry}
\usepackage{verbatim}
\usepackage{mathrsfs}
\usepackage{mathtools}
\usepackage{amsmath}
\usepackage{amssymb}
\usepackage{graphicx}

\makeatletter
\usepackage{amsmath}

\usepackage{graphicx, mathtools}
\usepackage[normalem]{ulem}
\usepackage[bottom]{footmisc}
\usepackage[colorlinks=true,linkcolor=MidnightBlue,urlcolor=MidnightBlue,citecolor=MidnightBlue,anchorcolor=MidnightBlue]{hyperref}
\usepackage[dvipsnames]{xcolor}

\usepackage{chngcntr}
\usepackage{listings}
\usepackage{xcolor, tikz-cd}

\usepackage{tikz-cd} 
\usepackage{amsthm}
\usepackage{bbm}
\usepackage{mathrsfs}
\usepackage{verbatim}
\usepackage{alphalph}
\usepackage{yhmath}
\usepackage{listings}
\usepackage{courier} 

\definecolor{codegreen}{rgb}{0,0.6,0}
\definecolor{codegray}{rgb}{0.5,0.5,0.5}
\definecolor{codepurple}{rgb}{0.58,0,0.82}
\definecolor{backcolour}{rgb}{0.98,0.98,0.98}

\lstdefinestyle{mystyle}{
	backgroundcolor=\color{backcolour},   
	commentstyle=\color{codegreen},
	keywordstyle=\color{magenta},
	numberstyle=\tiny\color{codegray},
	stringstyle=\color{codepurple},
	basicstyle=\linespread{0.8}\footnotesize\ttfamily
	breakatwhitespace=false,         
	breaklines=true,                 
	captionpos=b,                    
	keepspaces=true,                 
	numbersep=5pt,                  
	showspaces=false,                
	showstringspaces=false,
	showtabs=false,                  
	tabsize=2
}

\makeatother

\usepackage{babel}
\begin{document}
	\title{Cartan media: geometric continuum mechanics in homogeneous spaces}
	\author{Lukas Kikuchi}
	\email{ltk26@cam.ac.uk}
	
	\affiliation{Department of Applied Mathematics and Theoretical Physics, Centre
		for Mathematical Sciences, University of Cambridge, Wilberforce Road,
		Cambridge CB3 0WA, United Kingdom}
	\author{Ronojoy Adhikari}
	\affiliation{Department of Applied Mathematics and Theoretical Physics, Centre
		for Mathematical Sciences, University of Cambridge, Wilberforce Road,
		Cambridge CB3 0WA, United Kingdom}
	\begin{abstract}
		We present a geometric formulation of the mechanics of a field that
		takes values in a homogeneous space $\mathbb{X}$ on which a Lie group
		$G$ acts transitively. This generalises the mechanics of Cosserat
		media where $\mathbb{X}$ is the frame bundle of Euclidean space and
		$G$ is the special Euclidean group. Kinematics is described by a
		map from a space-time manifold to the homogeneous space. This map
		is characterised locally by generalised strains (representing spatial
		deformations) and generalised velocities (representing temporal motions).
		These are, respectively, the spatial and temporal components of the
		Maurer-Cartan one-form in the Lie algebra of $G$. Cartan\textquoteright s
		equation of structure provides the fundamental kinematic relationship
		between generalised strains and velocities. Dynamics is derived from
		a Lagrange-d\textquoteright Alembert principle in which generalised
		stresses and momenta, taking values in the dual Lie algebra of $G$,
		are paired, respectively, with generalised strains and velocities.
		For conservative systems, the dynamics can be expressed completely
		through a generalised Euler-Poincare action principle. The geometric
		formulation leads to accurate and efficient structure-preserving integrators
		for numerical simulations. We provide an unified description of the
		mechanics of Cosserat solids, surfaces and rods using our formulation.
		We further show that, with suitable choices of $\mathbb{X}$ and $G$,
		a variety of systems in soft condensed matter physics and beyond can
		be understood as instances of a class of materials we provisionally
		call Cartan media.
	\end{abstract}
	\maketitle

	\section{Introduction}
	
	Cosserat media are models of continuua in which the elementary constituent
	is notionally a rigid body. The configuration of a Cosserat medium
	is determined by the position and orientation of each constituent
	in Euclidean space. Pairs of configurations are considered equivalent
	if a single rigid transformation, consisting of a translation and
	a rotation, takes every constituent in the first configuration to
	the corresponding constituent in the second configuration. Pairs of
	configurations are deformed with respect to each other if different
	translations and rotations are needed to bring the constituents into
	correspondence. A measure of deformation can then be constructed by
	examining how the translations and rotations differ from a single
	uniform rigid transformation. The kinematic theory of strain in Cosserat
	media can be constructed entirely on this basis. A dynamic theory
	of stress can be constructed in duality with the kinematic theory
	of strain, in which stress and strain are paired, in a precise mathematical
	sense, to yield a virtual work. The mechanical equations of motions
	can then be obtained by imposing suitable conditions on the virtual
	work.
	
	At the heart of this elegant formulation of the mechanics of Cosserat
	media \citep{schaeferAnalysisMotorfelderIm1967,schaeferBasicAffineConnection1967}
	is a symmetry, namely that of rigid transformations, or isometries
	of Euclidean space. Mathematically, the configuration of the Cosserat
	medium can be mapped to a space $\mathbb{X}=\mathcal{F}(\mathbb{E}^{3})$,
	the frame bundle of Euclidean space consisting of the union of pairs
	of points and rigid frames. A continuous symmetry group $G=SE(3)$,
	the Euclidean group, acts on the frame bundle, such that any two elements
	of the bundle are related by a rigid transformation. Like Euclidean
	space itself, its frame bundle is a homogeneous space, in which, informally
	speaking, there is no ``distinguished point''. The homogeneity of
	the frame bundle and the existence of a group of symmetries in terms
	of which ``displacements'' in the configuration space can be characterised
	is the basis of the geometric structure of Cosserat media.
	
	The question we ask and address here is the following: what is the
	general geometric structure of the mechanics of a continuum whose
	configuration takes values in a homogeneous space $\mathbb{X}$ on
	which a Lie group $G$ acts transitively? As we shall show below,
	this seemingly abstract question arises repeatedly in concrete form
	in the modelling of many continuum systems but seems not to have been
	consciously recognised. The Cosserat medium provokes merely one instance
	of this question, namely when $\mathbb{X}$ is the frame bundle of
	Euclidean space and $G$ is the special Euclidean group. In the remainder
	of this paper, we show that the mechanics of such continuua can be
	constructed by analogy with, and as a generalisation of, the geometric
	mechanics of Cosserat media. The configuration in the homogeneous
	space $\mathbb{X}$ can be ``lifted'' to the group $G$ and then
	the infinitesimal structure of $G,$ encoded in its Lie algebra $\mathfrak{g}$
	and its Lie-algebra valued Maurer-Cartan one-form $\xi$, can be used
	to study infinitesimal deformations of the medium. The dynamical theory
	can be constructed analogously by pairing the infinitesimal deformations
	with corresponding generalised momenta and generalised stresses to
	obtain the virtual work from which the dynamical equations follow.
	In this manner, we recognise a single geometric framework within which
	to study a multitude of continuum systems that share the properties
	of homogeneity and symmetry.
	
	The study of the geometry of space curves through the use of infinitesimal
	rigid motions of orthogonal frames in Euclidean space was initiated
	by Bartels and developed by Frenet and Serret \citep{clellandFrenetCartanMethod2017}.
	Darboux extended it to the study of surfaces embedded in Euclidean
	space \citep{darbouxLeconsTheorieGenerale1887}. Cartan recognised
	the essential underlying principle as the action of a Lie group on
	a homogeneous space and, in a remarkable generalisation, developed
	his method of the \emph{repère mobile} to study the geometry of submanifolds
	of homogeneous spaces \citep{cartanMethodeRepereMobile1935}. Cartan
	made \citep{cosseratTheorieCorpsDeformables2009} a significant reference
	to the symmetries and mechanics of Cosserat media when presenting
	his method. Given this historical background, it does not seem inappropriate
	to recognise the class of systems studied in this paper as ``Cartan
	media''.
	
	The remainder of the paper is organised as follows. In Section II
	we introduce the mathematical notation used in the paper. In Section
	III we construct the kinematics of Cartan media and derive the kinematic
	equations of motion using the local properties of the Lie group $G$.
	The resulting equations of motion are defined in terms of intrinsic
	quantities that are invariant to global group action. In Section IV
	we derive the geometrised dynamics of Cartan media, in both the conservative
	and non-conservative settings. Sections V applies the formulation
	to unify the presentation of Cosserat solids, surfaces and rods. Section
	VI shows how surfaces and filaments are also instances of Cartan media,
	and how to derive their geometrised mechanics in terms of their intrinsic
	and extrinsic curvatures. Section VII provides further examples in
	soft matter physics and beyond. Finally, in Section VII we summarise
	and conclude our work.
	
	\section{Notation}
	
	The set of positive real numbers is denoted $\mathbb{R}_{+}$, and
	the set of $m\times m$ matrices as $\mathbb{R}^{m\times m}$, and
	positive definite matrices as $\mathbb{R}_{+}^{m\times m}$. Euclidean
	space $\mathbb{E}^{3}$ is the vector space $\mathbb{R}^{3}$ equipped
	with the standard Euclidean inner product. Elements of $\mathbb{E}^{3}$
	are denoted as column vectors $\mathbf{a}=(a_{1}\ a_{2}\ a_{3})^{T}\in\mathbb{E}^{3},\ a_{i}\in\mathbb{R}$,
	such that the inner product is $\mathbf{a}\cdot\mathbf{b}=\mathbf{a}^{T}\mathbf{b}$
	for any $\mathbf{a},\mathbf{b}\in\mathbb{E}^{3}$, and using the isomorphism
	$\mathbb{E}^{3}\cong T\mathbb{E}^{3}$ the inner product extends to
	tangent vectors as well. A vector in a \emph{spatial frame of reference}
	is denoted with a superscript as $\mathbf{v}^{s}=(v_{1}^{s}\ v_{2}^{s}\ v_{3}^{s})^{T}\in T\mathbb{E}^{3}$.
	That is, $\mathbf{v}^{s}=v_{i}^{s}\mathbf{d}_{i}$ where $\mathbf{d}_{i}=(\delta_{i1}\ \delta_{i2}\ \delta_{i3})^{T},\ i=1,2,3$
	is a fixed basis for $T\mathbb{E}^{3}$. This is in contrast to the
	corresponding vector in the \emph{body frame of reference}, a moving
	frame $E=(\mathbf{e}_{1}\ \mathbf{e}_{2}\ \mathbf{e}_{3})$ coincident
	with the material points of a continuum body, which is written without
	the superscript $\mathbf{v}\in T\mathbb{E}^{3}$. This notational
	rule is applied to all tangent vectors, with the exception of basis
	vectors such as $\mathbf{e}_{i}$ and $\mathbf{d}_{i}$, and derivatives
	of $\mathbb{E}^{3}$-valued fields. Corresponding vectors in the two
	frames are related by a rotation $R\in SO(3)$ as $\mathbf{v}^{s}=R\mathbf{v}$,
	where the Lie group $SO(3)$ is the special orthogonal group. We will
	often make the identification $R=E$, such that $\mathbf{v}^{s}=v_{i}^{s}\mathbf{d}_{i}=v_{i}\mathbf{e}_{i}$,
	where repeated indices indicates a summation.. The distinction between
	the spatial and body frames will reoccur throughout this text in relation
	to Euclidean, but also more general, settings. We will often utilise
	the isomorphism between $3$-vectors in $T\mathbb{E}^{3}$and $\mathfrak{so}(3)$,
	where the latter is the Lie algebra of $SO(3)$. For $\mathbf{v},\mathbf{w}\in T\mathbb{E}^{3}$,
	we can write $\mathbf{v}\times\mathbf{w}=\hat{v}\mathbf{w}$, where
	$\hat{v}\in\mathfrak{so}(3)$ denotes the \emph{hat map}, given in
	Eq.~\ref{eq:hat map}. See the appendix for further details on vector
	and matrix operations. We write general Lie groups as $G$, and their
	corresponding Lie algebra as $\mathfrak{g}$. A \emph{homogeneous
		space} is a space $\mathbb{X}$ that admits a transitive action of
	a Lie group $G$. That is, given any two points $q_{1},q_{2}\in\mathbb{X}$
	there exists a Lie group element $g\in G$ such that $g\cdot q_{1}=q_{2}$,
	where the latter denotes the action of $G$ on $\mathbb{X}$. An example
	of a homogeneous space is the $2$-sphere, on which $SO(3)$ acts
	transitively.
	
	\section{Kinematics \label{sec:geometric kinematics}}
	
	We begin with a brief overview of kinematics in classical continuum
	mechanics in the Lagrangian description \citep{batraElementsContinuumMechanics2006},
	which will serve as a reference point for the generalised kinematics
	we develop in this section. In classical continuum mechanics the kinematics
	of a continuum body can be described by a map $\mathbf{x}:[0,T]\times M\to\mathbb{E}^{3}$,
	where $\mathbf{u}\in M$ is a material coordinate of a given reference
	configuration $M$, and $\mathbf{x}(t,\mathbf{u})\in\mathbb{E}^{3}$
	is the current vector of displacement at time $t\in[0,T]$. The reference
	configuration $M$ shares the same topology and dimensionality as
	that of the continuum body itself, but can otherwise differ in shape.
	At time $t$, each material coordinate $\mathbf{u}\in M$ is thus
	assigned a value in $\mathbb{E}^{3}$ via the map $\mathbf{x}(t,\cdot):M\to\mathbb{E}^{3}$,
	such that the current configuration is given by the image $\mathbf{x}(t,M)$.
	Consequently, $M$ can be seen as an index set over the $\mathbb{E}^{3}$-valued
	point-continua of the continuum body, and the material coordinate
	$\mathbf{u}$ a continuous multi-dimensional index over the current
	configuration $\mathbf{x}$. We therefore call $M$ the \emph{material
		base space} and $\mathbb{E}^{3}$ the \emph{configuration space} of
	the system. In what follows we will generalise this to consider continuum
	bodies with alternate configuration spaces, material dimensionalities
	and topologies.
	
	Cosserat point-continua have configuration spaces that include an
	internal micropolarity \citep{cosseratTheoryDeformableBodies1909,eremeyevFoundationsMicropolarMechanics2012,altenbachCosseratMedia2013,rubinCosseratTheoriesShells2000,NonlinearProblemsElasticity2005},
	as well as their external configuration in $\mathbb{E}^{3}$. For
	instance, in addition to the displacement, we may associate a unit
	vector $\mathbf{p}^{s}\in T\mathbb{E}^{3}$, satisfying $\mathbf{p}^{s}\cdot\mathbf{p}^{s}=1$,
	to the point-continua, representing its micropolarity. The configuration
	space of such a system is the product $\mathbb{E}^{3}\times S^{2}$,
	where $S^{2}$ is the 2-sphere, such that the configurations are $q=(\mathbf{x},\mathbf{p}^{s})\in\mathbb{E}^{3}\times S^{2}$.
	If the orientation around the polar vector is a degree of freedom,
	the internal configuration can be represented with an orthonormal
	triad of vectors $E=(\mathbf{e}_{1}\ \mathbf{e}_{2}\ \mathbf{e}_{3})$.
	The configuration space for such a system can be identified with the
	\emph{frame bundle} $\mathcal{F}(\mathbb{E}^{3})$ of $\mathbb{E}^{3}$,
	where elements $q=(\mathbf{x},E)\in\mathcal{F}(\mathbb{E}^{3})$ are
	called \emph{trihedrons}. The above are all examples of homogeneous
	spaces, on which $SE(3)$, the special Euclidean group, acts transitively.
	We therefore call $SE(3)$ the \emph{symmetry group} of $\mathbb{E}^{3}$,
	$\mathbb{E}^{3}\times S^{2}$ and $\mathcal{F}(\mathbb{E}^{3})$.
	
	There may in general be multiple Lie groups that act transitively
	on a given homogeneous space. The choice of symmetry group $G$ determines
	how we treat the symmetries of the configuration space. For example,
	we could equivalently choose the product $G=T(3)\times SO(3)$ as
	the symmetry group of $\mathbb{X}=\mathbb{E}^{3}\times S^{2}$, where
	$T(3)$ is the group of translations on $\mathbb{E}^{3}$. The two
	terms of $G$ then act individually on the two respective terms of
	$\mathbb{X}$, thus treating $\mathbb{E}^{3}$ and $S^{2}$ as two
	unrelated spaces. This separation does not reflect most applications
	of Cosserat media, where the micropolarity is often considered, either
	implicitly or explicitly, as tangent vectors on the ambient space.
	That is, $S^{2}$ is often considered the subset of unit vectors in
	$T\mathbb{E}^{3}$. The implication of this is that transformations
	on $\mathbb{E}^{3}$ and $T\mathbb{E}^{3}$ should be done consistently;
	this reflects the fact that the point-continua are \emph{rigid}. This
	is accomodated by choosing the symmetry group to be a semidirect product
	$G=T(3)\rtimes SO(3)=SE(3)$, which in this case acts on $\mathbb{E}^{3}\times T\mathbb{E}^{3}\supset\mathbb{E}^{3}\times S^{2}$,
	as will be shown below. This example highlights why we bundle the
	external and internal (micropolar) configuration spaces of a continuum
	system as one collective homogeneous space $\mathbb{X}$, and the
	utility, and importance, of this will be borne out through the results
	of this text.
	
	Filaments \citep{nordgrenComputationMotionElastic1974,goldsteinNonlinearDynamicsStiff1995,sodaDynamicsStiffChains1973}
	and surfaces \citep{powersDynamicsFilamentsMembranes2010} are examples
	of continuum bodies that are embedded in three-dimensional Euclidean
	space but with a lower intrinsic dimensionality. That is, the configuration
	space of filaments and surfaces is $\mathbb{E}^{3}$, but their material
	base spaces are one and two-dimensional respectively. Similarly, \emph{Cosserat
		rods} and \emph{surfaces} \citep{rubinCosseratTheoriesShells2000,altenbachCosseratMedia2013}
	are continua with micropolar configuration spaces, but with one- and
	two-dimensional material base spaces respectively. For all continua,
	the topology of the continuum body must be reflected in the topology
	of the material base space. For example, a closed surface will be
	topologically equivalent to the unit sphere $S^{2}$, and we may thus
	let $M=S^{2}$. Similarly, an adequate material base space for a closed
	rod is the the periodic unit interval.
	
	We will now generalise the kinematics of classical continuum mechanics,
	by considering systems of general material base spaces and homogeneous
	configuration spaces. Let the material base space $M$ be a topological
	manifold of dimension $d$, and let the configuration space $\mathbb{X}$
	be an $n$-dimensional homogeneous space. At time $t\in[0,T]$, each
	\emph{material point} $p\in M$ is mapped to a point in $\mathbb{X}$
	via the \emph{spatio-temporal configuration} $q:W\to\mathbb{X}$,
	where we have defined the \emph{kinematic base space} $W=[0,T]\times M$.
	The configuration of the system at time $t$ is thus given by the
	image $q(t,M)$. Let $G$ be a \emph{symmetry group }of $\mathbb{X}$\emph{,
	}an $r$-dimensional Lie group that acts transitively on $\mathbb{X}$
	under a given group action $g\cdot q\in\mathbb{X}$, for $g\in G$
	and $q\in\mathbb{X}$. In general the choice of symmetry group is
	not unique, and may also be of larger dimensionality than the configuration
	space $r\geq n$. In the special case when $r=n$, the symmetry group
	is diffeomorphic to the configuration space $\mathbb{X}$. \footnote{Note that a Lie group $G$ may be considered a homogeneous space,
		as it acts on itself transitively.} For any Lie group $G$ there is an associated Lie algebra $\mathfrak{g}$,
	which is an $r$-dimensional vector space equipped with a \emph{Lie
		bracket }$[\cdot,\cdot]:\mathfrak{g}\times\mathfrak{g}\to\mathfrak{g}$.
	
	\begin{figure*}
		\begin{minipage}[t]{0.48\linewidth}%
			\includegraphics[width=0.9\linewidth]{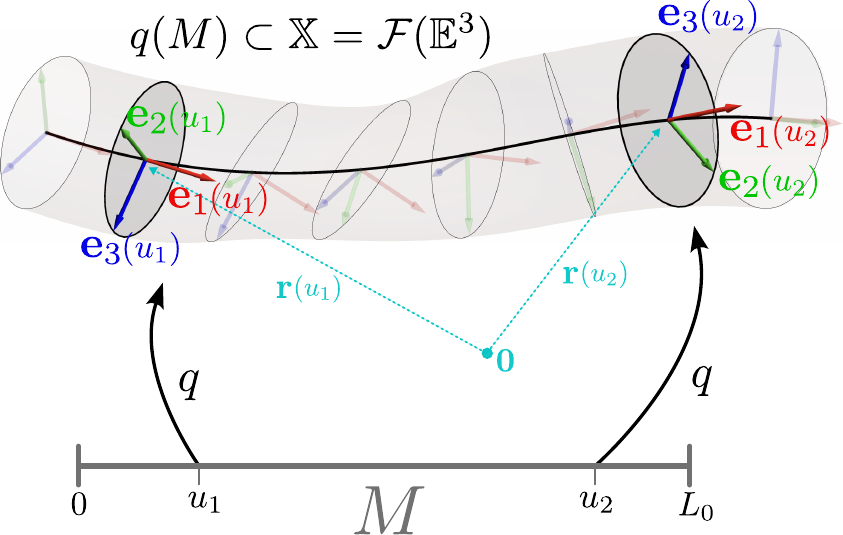}
			\caption{{\small{}The figure depicts an open Cosserat rod, mathematically described
					as a mapping between a material base space $M=[0,L_{0}]$ and the
					Lie group configuration space $\mathbb{X}=\mathcal{F}(\mathbb{E}^{3})$,
					where $\mathcal{F}(\mathbb{E}^{3})$ is the frame bundle of }$\mathbb{E}^{3}$,{\small{}
					the set of all orthonormal frames of Euclidean space. At any time
					$t\in[0,T]$, the Cosserat rod is the image $q(t,M)\subset\mathcal{F}(\mathbb{E}^{3})$.
					In the figure, two material points $u_{1},u_{2}\in M$ are shown mapped
					to configurations $(\mathbf{r}(t,u_{1}),E(t,u_{1})),(\mathbf{r}(t,u_{2}),E(t,u_{2}))\in\mathbb{X}$.
					The temporal argument $t$ is suppressed in the figure.}}
			\label{fig:M to Cosserat rod}%
		\end{minipage}\hfill{}%
		\begin{minipage}[t]{0.48\linewidth}%
			\includegraphics[width=0.9\linewidth]{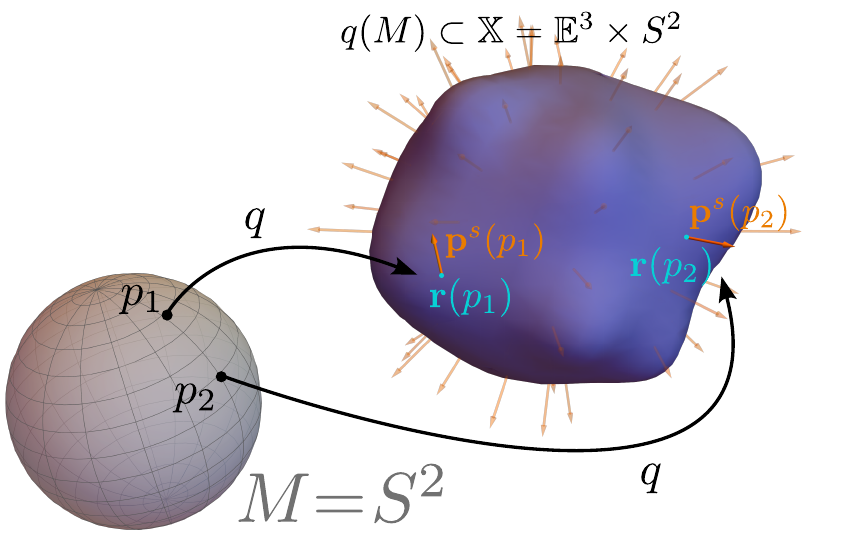}
			\caption{{\small{}The figure depicts a Cosserat surface, with material base
					space $M=S^{2}$ and configuration space $\mathbb{X}=\mathbb{E}^{3}\times S^{2}$.
					The material base space is the $2$-sphere $S^{2}$, which ensuresthat
					the surface is closed. At any time $t\in[0,T]$, the Cosserat surface
					is the image $q(t,M)\subset\mathbb{E}^{3}\times S^{2}$. In the figure,
					two material points $p_{1},p_{2}\in M$ are shown mapped to $(\mathbf{r}(p_{1}),\mathbf{p}^{s}(p_{2})),(\mathbf{r}(p_{2}),\mathbf{p}^{s}(p_{2}))\in\mathbb{X}$.
					The temporal argument $t$ is suppressed in the figure.}}
			\label{fig:M to Cosserat surface}%
		\end{minipage}
	\end{figure*}
	
	We will refer to systems that can be kinematically described using
	the formalism outlined above as \emph{Cartan media}. Classical three-dimensional
	continuum bodies, surfaces and rods, as well as their corresponding
	Cosserat variants, all conform to this class of systems. See Fig.~\ref{fig:M to Cosserat rod}
	and Fig.~\ref{fig:M to Cosserat surface} for illustrations of an
	open Cosserat rods and a closed Cosserat surface respectively. In
	Sec.~\ref{sec:Further-examples-of-cartan-media} we will give further
	examples of Cartan media, including relativistic rods and field theories.
	
	Let $W=[0,T]\times M$ be the \emph{kinematic base space}. The transitive
	action of $G$ on $\mathbb{X}$, guarantees the existence of a map
	$\Phi:W\to G$, which we call a \emph{structure field}, that satisfies
	\begin{equation}
		q(t,p)=\Phi(t,p)\cdot q_{r}\label{eq:x from phi}
	\end{equation}
	for all $p\in M$ and $t\in[0,T]$, where $q_{r}\in\mathbb{X}$ is
	a fixed reference configuration. Equation \ref{eq:x from phi} shows
	that the spatio-temporal configuration of the system is encapsulated
	in its entirety by the structure field. The reference configuration
	$q_{r}$ will often carry the interpretation of being the configuration
	of each point continua $p\in M$ as observed \emph{within} its own
	body frame of reference.
	
	As an example to illustrate Eq.~\ref{eq:x from phi}, consider a
	Cosserat surface with configuration space $\mathbb{X}=\mathbb{E}^{3}\times S^{2}$
	and material base space $M$. The spatial configuration at time $t$
	will be a map $q(t,\cdot):M\to\mathbb{E}^{3}\times S^{2}$, which
	we write as $q=(\mathbf{r},\mathbf{p}^{s})$. Let $q_{r}=(\mathbf{0},\mathbf{p})$,
	where $\mathbf{p}\in S^{2}$ is a fixed unit vector. Then the configuration
	$q(t,\cdot)$ at each material point $p\in M$ can be constructed
	by rotation and translation of $q_{r}$, which is precisely a group
	action of $SE(3)$ on $\mathbb{E}^{3}\times S^{2}$. Now, let $R:W\to SO(3)$
	be $R=(\mathbf{e}_{1}\ \mathbf{e}_{2}\ \mathbf{e}_{3})$, which is
	an orthonormal frame we define to satisfy $\mathbf{p}^{s}=R\mathbf{p}=p_{i}\mathbf{e}_{i}$.
	Now, consider an observer located at $\mathbf{r}(t,p)$ and with a
	frame of reference $R(t,p)$. Then, the material point $p\in M$ will
	be located at $\mathbf{0}$, and oriented as $\mathbf{p}$, relative
	to the observer. We thus see that $q_{r}$ can be interpreted as the
	observed configuration of each material point $p$ relative to an
	observer that is coincident and co-moving with $p$.
	
	It should be noted that if $r>n$ there is no unique choice of structure
	field. The gauge freedom in the choice of structure field will be
	discusssed in Sec.~\ref{subsec:Adapted-frames}. A brief mathematical
	interlude: we note that the spatio-temporal configuration of Cartan
	media can be be considered sections on the trivial fibre bundle $M\times\mathbb{X}$.
	Similarly, $\Phi$ is a section on the trivial principal bundle $M\times G$.
	
	There is flexibility in how we represent elements of the configuration
	space. For instance, in the above we may have alternatively parameterised
	configurations as $(\mathbf{r},\theta,\phi)$, where $\theta$ and
	$\phi$ are angular coordinates on the sphere. Similarly, when $\mathbb{X}=\mathcal{F}(\mathbb{E}^{3})$
	we wrote configurations as pairs $(\mathbf{r},E)\in\mathbb{X}$, though
	we could have also parameterised the orthonormal triad in terms of
	Euler angles $E=E(\alpha,\beta,\gamma)$, such that configurations
	are specified by six scalars $(\mathbf{r},\alpha,\beta,\gamma)$.
	Strictly, these should be understood as coordinates on $\mathbb{X}$,
	rather than elements of the configuration space itself. However, we
	will not stress this distinction, and we will furthermore make use
	of abuses of notations such as $(\mathbf{r},\alpha,\beta,\gamma)\in\mathbb{X}$,
	identifying coordinates with configurations. In full generality, we
	may express configurations as a vector of real numbers $q\in\mathbb{R}^{m}$,
	where $m\geq n$. This in turn induces a representation $\Pi:G\to GL(\mathbb{R}^{m})$
	of the group action on the configuration space, such that we can write
	$x(t,p)=\Pi(\Phi(t,p))x_{0}$. For example, consider again a Cosserat
	surface $\mathbb{X}=\mathbb{E}^{3}\times S^{2}$, with symmetry group
	$G=SE(3)$. We write group elements as pairs $g=\left(\mathbf{a};A\right)\in G$,
	where $\mathbf{a}\in\mathbb{E}^{3}$ is a translation and $A\in SO(3)$
	a rotation, and let $y=(\mathbf{s},\mathbf{q}^{s})\in\mathbb{X}$.
	A reasonable choice of group action would then be $g\cdot y=\Pi(g)y=(A\mathbf{s}+\mathbf{a},A\mathbf{q}^{s})$.
	We can thus construct the structure field as $\Phi(t,p)=\left(\mathbf{r};R\right)$
	which satisfies Eq.~\ref{eq:x from phi}, where $R\in SO(3)$ satisfies
	$R\mathbf{p}=\mathbf{p}^{s}$. Different choices of how the configuration
	space is expressed, whether as coordinates or explicit elements, will
	lead to different representations of the Lie group. The framework
	we present in this paper is agnostic with respect to this choice,
	as long as the Lie group representation is constructed consistently.
	
	We have shown that the spatio-temporal configuration of a continuum
	body in a homogeneous space, which we refer to as Cartan media, can
	be parameterised in terms of a Lie group-valued structure field. In
	the following subsection we will make use of the Lie group-Lie algebra
	correspondence, to furthermore express $\Phi$ in terms of Lie algebra-valued
	fields. The latter are generalised strain and velocity fields, which
	are analogous to those of classical continuum mechanics. We call this
	process \emph{geometrisation}. In Sec.~\ref{subsec:Generalised-strain-and-velocity-fields}
	we derive kinematic equations of motion for Cartan media, in terms
	of generalised strain and velocity fields. The resulting kinematics
	is geometric in the sense that the equations of motion are expressed
	in terms of the differential geometry of the continuum body, and are
	constructed to respect the symmetries of the configuration space.
	In Sec.~\ref{subsec:Adapted-frames} we consider the case of $\text{dim}(G)>\text{dim}(\mathbb{X})$,
	when there is a gauge freedom in the choice of $\Phi$, and how the
	redundancy in the parameterisation can be removed by \emph{kinematic
		adaptation}.
	
	\subsection{Generalised strain and velocity fields \label{subsec:Generalised-strain-and-velocity-fields}}
	
	Let $u^{\alpha}:M\to\mathbb{R},\ \alpha=1,\dots,d$, be coordinates
	on $M$. For the sake of simplicity, we have assumed that the coordinates
	cover $M$. This is not true in general, as for example $M=S^{2}$
	would require at least two overlapping coordinate charts. We will
	consider the case of multiple charts in Sec.~\ref{subsec:Local-charts}.
	
	The exterior derivative of the structure field is $d\Phi=\dot{\Phi}dt+\partial_{\alpha}\Phi du^{\alpha}$,
	where $\dot{\Phi}=\partial_{t}\Phi$ and $\partial_{\alpha}=\frac{\partial}{\partial u^{\alpha}}$,
	and where repeated indices denote an Einstein summation. Here $\dot{\Phi}:M\to TG$
	and $\partial_{\alpha}\Phi:M\to TG$ are the infinitesimal generators
	of $\Phi$, and should be seen as vectors fields on the symmetry group
	$G$, where $TG$ is the tangent bundle of $G$.
	
	A unique property of Lie groups is the ability to relate vector fields
	to corresponding Lie algebra-valued fields. That is, the tangent space
	$T_{g}G$ at $g\in G$ can be related to the tangent space at the
	identity $\mathfrak{g}=T_{e}G$. In other words, we have the diffeomorphism
	$TG\cong G\times\mathfrak{g}$, such that sections on $TG$ can be
	related to sections on $G\times\mathfrak{g}$. Via left-translation
	to the identity $T_{e}G\cong\mathfrak{g}$, we define the $\mathfrak{g}$-fields
	\begin{subequations}
		\label{eq:generalised strain and velocity fields}
		\begin{align}
			X_{\alpha} & =\Phi^{-1}\partial_{\alpha}\Phi, & \alpha=1,\dots,d\label{eq:generalised strain fields}\\
			N & =\Phi^{-1}\dot{\Phi},\label{eq:generalised velocity fields}
		\end{align}
	\end{subequations}
	which we call respectively the \emph{generalised strain }and \emph{velocity
		fields}. Respectively, $X_{\alpha}$ and $N$ encode the spatial structure
	of $\Phi$ and its temporal evolution. The generalised strain and
	velocity fields together form a $\mathfrak{g}$-valued $1$-form
	\begin{equation}
		\begin{aligned}\xi & =\Phi^{-1}d\Phi\\
			& =Ndt+X_{\alpha}du^{\alpha},
		\end{aligned}
		\label{eq:frame generator}
	\end{equation}
	which we call the \emph{structure generator}. Equation \ref{eq:frame generator}
	completes the process of geometrisation, having gone from the spatio-temporal
	configuration, to the structure field $\Phi$, and finally to $X_{\alpha}$
	and $N$.
	
	It is notable that $\xi$ is left-invariant under any global transformation
	in $G$. That is if $\Phi'=g\Phi$, for some $g\in G$, then $\xi'=\Phi'^{-1}d\Phi'=\xi$.
	Conversely, if the respective generators $\xi_{1}$ and $\xi_{2}$
	of two structure fields $\Phi_{1}$ and $\Phi_{2}$ satisfy $\xi_{1}=\xi_{2}$,
	then there exists a $g\in G$ such that $\Phi_{1}=g\Phi_{2}$. The
	implication of this fact is that $\xi$ does not contain global information
	about the configuration of the system, and it further indicates that
	the components of $\xi$ are \emph{differential invariants} \citep{clellandFrenetCartanMethod2017},
	and is therefore the appropriate mathematical object to define the
	kinematics of the system in terms of its intrinsic, and extrinsic,
	geometry \footnote{To provide further mathematical context, we also note that within
		differential geometry $\Phi$ is called an \emph{immersion} of $G$
		into $M$, and we have $\xi=\Phi^{*}\omega$, where $\Phi^{*}$ denotes
		the pull-back, and where $\omega=g^{-1}dg,\ g\in G$ is known as the
		\emph{Maurer-Cartan form}. See \citep{clellandFrenetCartanMethod2017}
		for a detailed proof, and for further mathematical exposition.}. The structure field can be reconstructed from the structure generator,
	up to global transformation in $G$ by solving Eq.~\ref{eq:frame generator}
	for $\Phi$. The numerical algorithm for the reconstruction is described
	in detail in Sec.~\ref{subsec:Spatio-temporal reconstruction}.
	
	We should emphasise that, at each time $t$, the spatial component
	of $\xi$ is a $\mathfrak{g}$-valued $1$-form on $M$ that generates
	the spatial structure of $\Phi$. Let $d_{M}$ denote the exterior
	derivative on $M$, then for a given time $t$, the spatial structure
	generator is $\xi_{M}=\Phi(t,\cdot)^{-1}d_{M}\Phi(t,\cdot)$, which
	in local coordinates is $X_{\alpha}(t,\cdot)du^{\alpha}$. As $\xi_{M}$
	is a coordinate-independent object, this emphasises that only the
	topological properties of $M$ are of importance for the kinematics,
	whilst the the particular choice of coordinates are not. The structure
	generator can thus be decomposed as $\xi=Ndt+\xi_{M}$, which is made
	possible by the fact that the kinematic base space has a product structure
	$W=[0,T]\times M$.
	
	The generalised strain and velocity fields should be seen as rates-of-deformation
	defined in the body frame of reference of the system. In contrast,
	the corresponding fields in the spatial frame of reference are $N^{s}=\dot{\Phi}\Phi^{-1}$
	and $X_{\alpha}^{s}=(\partial_{\alpha}\Phi)\Phi^{-1}$which are found
	by right-translation to the identity. To see this, first let $\rho:\mathfrak{g}\to\mathfrak{gl}(\mathbb{R}^{m})$
	be the representation of the action of the Lie algebra on the configuration
	$\mathbb{X},$ induced by $\Pi$ as $\rho(N^{s})=\partial_{t}\Pi(\Phi)\Pi(\Phi)^{-1}$
	and $\rho(X_{\alpha}^{s})=\partial_{\alpha}\Pi(\Phi)\Pi(\Phi)^{-1}$.
	Then, from Eq.~\ref{eq:x from phi}, we then find that $\dot{q}=\rho(N^{s})x$
	and $\partial_{\alpha}q=\rho(X_{\alpha}^{s})x$. We can relate $N^{s}$
	and $X_{\alpha}^{s}$ to their corresponding kinematic fields in the
	body frame as $N^{s}=\Phi N\Phi^{-1}=\text{Ad}_{\Phi}N$ and $X_{\alpha}^{s}=\text{Ad}_{\Phi}X_{\alpha}$,
	where $\text{Ad}:G\times\mathfrak{g}\to\mathfrak{g}$ is the adjoint
	action of $G$ on $\mathfrak{g}$. Therefore, we will make us of the
	composition $\rho\circ\text{Ad}_{\Phi}$, which is a representation
	of the action of the body frame fields on the configuration. We have
	that $\dot{q}=\rho(\text{Ad}_{\Phi}N)q$ and $\partial_{\alpha}q=\rho(\text{Ad}_{\Phi}X_{\alpha})q$
	or, more compactly
	\begin{equation}
		\begin{aligned}dq & =\rho(\text{Ad}_{\Phi}N)qdt+\rho(\text{Ad}_{\Phi}X_{\alpha})qdu^{\alpha}\\
			& =\rho(\text{Ad}_{\Phi}\xi)q
		\end{aligned}
		\label{eq:dx}
	\end{equation}
	which expresses the derivatives of the spatio-temporal configuration
	in terms of the generalised strain and velocity fields. Though Eq.~\ref{eq:dx}
	are kinematic equations of motion for the system, they will not be
	used as such, as a geometric formulation of the kinematics will be
	developed in the following subsection. Equation \ref{eq:dx} is however
	useful in two main regards: 1) It relates the often more conceptually
	intuitive expressions for $\dot{q}$ and $\partial_{\alpha}q$, to
	the Lie algebraic $N$ and $X_{\alpha}$, so as to lend this intuition
	in understanding the latter. 2) If the spatio-temporal configuration
	and its derivatives are assigned physical units, we can infer the
	units of the components of $N$ and $X_{\alpha}$ by dimensional analysis.
	An example of the former use-case will be shown at the end of this
	subsection, and further examples of both use-cases will be shown in
	Sec.~\ref{sec:cosserat-as-cartan-media}-\ref{sec:Further-examples-of-cartan-media}.
	
	It is worth reiterating the requisite steps required to reach Eq.~\ref{eq:dx}.
	After having identified the configuration space $\mathbb{X}$ of a
	given system, and an appropriate symmetry group $G$. We must then
	specify the action of $G$ on configurations $\mathbb{X}\subseteq\mathbb{R}^{m}$.
	That is, we must find an appropriate representation $\Pi:G\to GL(\mathbb{R}^{m}$)
	that acts on $\mathbb{X}$. Note that in practice, we often identify
	the group elements with their matrix representations. Once $\Pi$
	has been specified, we must identify the corresponding matrix representation
	of the Lie algebra $\rho$. Finally, we can relate the differential
	of the configuration $dx$ with the action of the kinematic fields
	as Eq.~\ref{eq:dx}. The reader should note that the choice of group
	action, though in principle arbitrary as long as it satisfies Eq.~\ref{eq:x from phi},it
	is often desirable, in practice, to choose it such that the generalised
	strain and velocity have intuitive physical intepretations.
	
	To illustrate the above, we briefly provide an example (see Sec.~\ref{sec:cosserat-as-cartan-media}-\ref{sec:Further-examples-of-cartan-media}
	for further examples. In particular Sec\@.~\ref{sec:cosserat 3d bodies}
	is a good example of the procedure). Consider a filament lying on
	the surface of the unit-sphere. Its spatio-temporal configuration
	is $\mathbf{p}^{s}:W\to\mathbb{X}$, where $\mathbb{X}=S^{2}$ and
	$M=[0,L_{0}]$, and $L_{0}\in\mathbb{R}^{+}$. For fixed $t$, the
	image $\mathbf{p}^{s}(t,[0,L_{0}])$ is then a curve on the sphere.
	The spatio-temporal configuration can be written in terms of a structure
	field $R:W\to SO(3)$, such that $\mathbf{p}^{s}=R\mathbf{p}$, where
	$\mathbf{p}\in S^{2}$. We write the generalised strain and velocity
	fields as $\hat{\Omega}=R^{-1}\dot{R}$ and $\hat{\pi}=R^{-1}\partial_{u}R$,
	where the hat denotes the hat map, defined in Eq.~\ref{eq:hat map},
	such that $\boldsymbol{\Omega},\boldsymbol{\pi}:W\to\mathbb{R}^{3}$
	are angular velocity vectors along the time and material direction
	respectively. We then have that $\dot{\mathbf{p}}^{s}=\dot{R}R^{-1}\mathbf{p}^{s}=\hat{\Omega}^{s}\mathbf{p}^{s}=\boldsymbol{\Omega}^{s}\times\mathbf{p}^{s}$
	and $\partial_{u}\mathbf{p}^{s}=\boldsymbol{\pi}^{s}\times\mathbf{p}^{s}$,
	where $\boldsymbol{\Omega}^{s}=R\boldsymbol{\Omega}$ and $\boldsymbol{\pi}^{s}=R\boldsymbol{\pi}$.
	Alternatively, we can write these in terms of the body frame velocity
	and strain as $\dot{\mathbf{p}}^{s}=\text{Ad}_{R}\hat{\Omega}\mathbf{p}^{s}$
	and $\partial_{u}\mathbf{p}^{s}=\text{Ad}_{R}\hat{\pi}\mathbf{p}^{s}$.
	In the framework we present here we will prioritise the body frame.
	Physically, this is often the appropriate choice as, for instance,
	the moment of inertia of a rotating rigid body is a constant matrix
	in its body frame. Therefore, the relation Eq.\ref{eq:dx} will be
	useful in Sec.~\ref{sec:geometric dynamics}, where we derive the
	dynamical force balance equations for Cartan media.
	
	In the two-step procedure we call \emph{geometrisation}, we have gone
	from a kinematic description in terms of a spatio-temporal configuration,
	and replaced it with a structure field satisfying Eq.~\ref{eq:x from phi},
	and finally a structure generator Eq.~\ref{eq:frame generator}.
	As will be demonstrated through multiple examples, the generalised
	strain provides a description of the system in terms of its intrinsic
	geometry. Where the geometrisation does not lead to a fully intrinsic
	description, this can be ameliorated through \emph{kinematic adaptation},
	discussed in Sec.~\ref{subsec:Adapted-frames}. The geometric formulation
	lends itself to the study of the constitutive mechanics of continuum
	systems, which is fundamentally the dynamics of the differential properties
	of manifolds, where stress is the dynamical response to strain. In
	the setting of Cartan media, which we have defined as sub-manifolds
	of homogeneous spaces, we are thus developing a generalised notion
	of strains (and their conjugate stresses, in Sec.~\ref{sec:geometric dynamics}).
	In the following subsection, we show how the geometrised description,
	encoded in $X_{\alpha}$, evolves over time in response to the generalised
	velocity $N$.
	
	\subsection{Kinematic equations of motion\label{subsec:The-kinematic-equations}}
	
	Thus far we have explicitly considered the structure field as defined
	globally over the material base space $M$ and time $[0,T]$. However,
	in applications, we often want the temporal evolution of the frame
	given an initial configuration at $t=0$. That is, we are interested
	in how the spatial configuration deforms under an applied velocity
	field. Here we derive the kinematic equations of motion of Cartan
	media in a geometric form, the solution of which yields the structure
	generator $\xi$, from which $\Phi$ can be reconstructed using Eq.~\ref{eq:frame generator}.
	
	As it turns out, the equations of motion can be found by deriving
	the conditions under which a $\mathfrak{g}$-valued $1$-form is a
	structure generator. That is, the condition under which a $\mathfrak{g}$-valued
	$1$-form $\xi$ on $M$ is related to a $G$-valued function on $M$
	via Eq.~\ref{eq:frame generator}. To find this condition, we take
	the exterior derivative of both sides of Eq.~\ref{eq:frame generator},
	to find
	\begin{equation}
		\begin{aligned}d\xi & =d(\Phi^{-1}d\Phi)\\
			& =-(\Phi^{-1}d\Phi\Phi^{-1})\wedge d\Phi+\Phi^{-1}d(d\Phi)\\
			& =-\xi\Phi^{-1}\wedge d\Phi
		\end{aligned}
	\end{equation}
	where $\wedge$ is the wedge product and where we have used that $d\Phi$
	must be an exact differential $d(d\Phi)=0$. The matrix wedge product
	is defined element-wise using the standard definition of the wedge
	product. For any two matrices of forms $A$ and $B$, their wedge
	product $C=A\wedge B$ is given element-wise as $C_{ij}=A_{ij}\wedge B_{ij}$.
	Using the fact that the wedge product commutes with matrix multiplication,
	we arrive at
	\begin{equation}
		d\xi+\xi\wedge\xi=0,\label{eq:xi integrability condition}
	\end{equation}
	which is an integrability condition on $\xi$. It should be seen as
	analogous to the requirement of the equality of mixed partials of
	an exact differential in multi-variate calculus. By substituting Eq.~\ref{eq:frame generator}
	into Eq.~\ref{eq:xi integrability condition}, we find
	\begin{subequations}
		\label{eq:kinematic equations of motion and integrability conditions}
		\begin{align}
			\dot{X}_{\alpha} & =\mathcal{D}_{\alpha}N, & \alpha=1,\dots,d\label{eq:kinematic equations of motion}\\
			\partial_{\beta}X_{\alpha} & =\mathcal{D}_{\alpha}X_{\beta}, & \alpha=1,\dots,d-1,\label{eq:spatial integrability conditions}\\
			&  & \beta=\alpha+1,\dots,d.\nonumber 
		\end{align}
	\end{subequations}
	where $\mathcal{D}_{\alpha}=\partial_{\alpha}+\text{ad}_{X_{\alpha}}$
	is a covariant derivative with respect to $G$ along the $\alpha$th
	material direction, and $\text{ad}_{Z}:\mathfrak{g}\to\mathfrak{g}$
	is the adjoint action of any $Z\in\mathfrak{g}$, given by $\text{ad}_{Z}Y=[Z,Y],\ Y\in\mathfrak{g}$.
	Equation \ref{eq:kinematic equations of motion} are the geometrised
	kinematic equations of motion of Cartan continua, describing the temporal
	evolution of the differential geometry of the system, as encoded in
	the generalised strain fields $X_{\alpha}$, as a response to the
	generalised velocity field $N$. Equation \ref{eq:spatial integrability conditions}
	are not equations of motion, but are rather spatial integrability
	conditions. See the subsequent subsection for more details. The kinematics
	and the spatial integrability conditions can also be expressed in
	a non-coordinate and invariant manner. Inserting the decomposition
	$\xi=Ndt+\xi_{M}$ into Eq.~\ref{eq:xi integrability condition},
	we find
	\begin{subequations}
		\label{eq:kinematic equations of motion and integrability conditions, invariant}
		\begin{align}
			\dot{\xi}_{M}= & \ \mathcal{D}_{M}N,\label{eq:invariant kinematic eom}\\
			d_{M}\xi_{M} & +\xi_{M}\wedge\xi_{M}=0,\label{eq:invariant spatial integrability}
		\end{align}
	\end{subequations}
	which are respectively the kinematic equation of motion of the system,
	and the spatial integrability conditions on the generalised strain,
	and where $\mathcal{D}_{M}=d_{M}+\text{ad}_{\xi_{M}}$. In local coordinates,
	the time derivative of the material structure generator is to be understood
	as $\dot{\xi}_{M}=\dot{X}_{\alpha}du^{\alpha}$. We see that the spatial
	integrability condition Eq.~\ref{eq:invariant spatial integrability}
	mirrors Eq.~\ref{eq:xi integrability condition}. This is to be expected
	as integrability needs to be obeyed irrespective of the base, whether
	$M$ or $W$. We can thus also interpret Eq.~\ref{eq:invariant kinematic eom}
	as an integrability condition in the temporal direction.
	
	In simulations the coordinatised formulation of the kinematics is
	simpler to implement. However, Eq.~\ref{eq:invariant kinematic eom}
	may be particularly useful for more sophisiticated discretisation
	techniques, that preserve the geometric properties of the exterior
	product \citep{desbrunDiscreteExteriorCalculus2005,hiraniDiscreteExteriorCalculusa}.
	
	Equation \ref{eq:kinematic equations of motion}, or Eq.~\ref{eq:invariant kinematic eom},
	can be used to derive the equations of motion of any Cartan media.
	Consider again the example of the filament on the unit-sphere. From
	Eq.~\ref{eq:kinematic equations of motion} we have that $\dot{\boldsymbol{\pi}}=\partial_{u}\boldsymbol{\Omega}+\boldsymbol{\pi}\times\boldsymbol{\Omega}$,
	where we have used that $\text{ad}_{\hat{a}}\hat{b}=[\hat{a},\hat{b}]=\widehat{\mathbf{a}\times\mathbf{b}}$,
	for any $\hat{a},\hat{b}\in\mathfrak{so}(3)$. Further examples are
	presented in Sec.~\ref{sec:cosserat-as-cartan-media}-\ref{sec:Further-examples-of-cartan-media}.
	
	From an initial boundary value of the structure field $\Phi_{0}:M\to G$,
	the initial conditions of the kinematic equations of motion can be
	computed using $X_{\alpha}|_{t=0}=\Phi_{0}^{-1}\partial_{\alpha}\Phi_{0}$.
	Then, given a generalised velocity field $N:W\to\mathfrak{g}$, Eq.~\ref{eq:kinematic equations of motion}
	can be solved to find the structure generator $\xi$. It is important
	to recall that $\xi$ only captures the kinematics of the differential
	geometry of the system. That is, $\xi$ only specifies $\Phi$ at
	any given time $t$ up to a global transformation $g\in G$. Therefore
	a separate equation must be solved to track the global movement of
	the system. Let $p_{r}\in M$ be a given reference material point,
	with material coordinate $\mathbf{u}_{r}$, and let $\Phi_{r}(t)=\Phi(t,p_{r})$
	be the structure field evaluated at $p_{r}$. The equation of motion
	of $\Phi_{r}$ is then given by
	\begin{equation}
		\dot{\Phi}_{r}(t)=\Phi_{r}(t)N(t,p_{r})\label{eq:(app) global frame equation of motion-1}
	\end{equation}
	which is a matrix ODE, that can be solved using standard methods \citep{haleOrdinaryDifferentialEquations2009}.
	The full global spatio-temporal configuration can thus be constructed
	by solving Eq.~\ref{eq:frame generator} with boundary conditions
	$\Phi(t,p_{r})=\Phi_{r}(t),\ t\in[0,T]$. The reconstruction algorithm
	is described in detail in Sec.~\ref{subsec:Spatio-temporal reconstruction}.
	
	\subsection{Spatial integrability \label{subsec:Spatial-integrability}}
	
	Equation \ref{eq:spatial integrability conditions} are a set of $(d+1)d/2$
	spatial integrability conditions that must be simultaneously satisfied
	at all times $t\in[0,T]$. Note that though there are $d$ strain
	fields $X_{\alpha}$, adding up to a seeming total of $rd$ degrees
	of freedom, Eq.~\ref{eq:spatial integrability conditions} shows
	that these are not independent. As expected, only $r$ independent
	degrees of freedom remain to determine the structure generator once
	the spatial integrability conditions have been imposed.
	
	At first glance it may seem that Eq.~\ref{eq:kinematic equations of motion and integrability conditions}
	overdetermines the system, but Eq. ~\ref{eq:kinematic equations of motion}
	and Eq.~\ref{eq:spatial integrability conditions} are compatible.
	To see this, we compute the time-derivative of the latter to get 
	\begin{equation}
		\begin{aligned} & \partial_{t}\left(\partial_{\beta}X_{\alpha}-(\partial_{\alpha}+\text{ad}_{X^{\alpha}})X_{\beta}\right)\\
			& =\partial_{\beta}\dot{X}_{\alpha}-\partial_{\alpha}\dot{X}_{\beta}-[\dot{X}_{\alpha},X_{\beta}]-[X_{\alpha},\dot{X}_{\beta}]\\
			& =(\partial_{\beta}+\text{ad}_{X_{\beta}})\dot{X}_{\alpha}-(\partial_{\alpha}+\text{ad}_{X_{\alpha}})\dot{X}_{\beta}\\
			& =\partial_{\beta}([X_{\alpha},N])+[X_{\beta},\partial_{\alpha}N]+[X_{\beta},[X_{\alpha},N]]\\
			& -\partial_{\alpha}([X_{\beta},N])-[X_{\alpha},\partial_{\beta}N]-[X_{\alpha},[X_{\beta},N]]\\
			& =-[\partial_{\alpha}X_{\beta},N]+[X_{\beta},[X_{\alpha},N]]\\
			& +[\partial_{\beta}X_{\alpha},N]-[X_{\alpha},[X_{\beta},N]]\\
			& =[\partial_{\beta}X_{\alpha},N]-[\partial_{\alpha}X_{\beta},N]+[[X_{\beta},X_{\alpha}],N]
		\end{aligned}
	\end{equation}
	where we used the Jacobi identity $[A,[B,C]]=-[C,[A,B]]-[B,[C,A]]$.
	Finally, we get 
	\begin{equation}
		\partial_{t}\Delta_{\alpha\beta}^{\text{int}}=[\Delta_{\alpha\beta}^{\text{int}},N].\label{eq:time derivative of spatial integrability conditions-1}
	\end{equation}
	where 
	\begin{equation}
		\Delta_{\alpha\beta}^{\text{int}}=\partial_{\beta}X_{\alpha}-\mathcal{D}_{\alpha}X_{\beta}
	\end{equation}
	is the residual error in spatial integrability $\Delta_{\alpha\beta}^{\text{int}}:W\to\mathbb{R}$.
	If $X_{\alpha}$ satisfies Eq.~\ref{eq:spatial integrability conditions}
	at time $t=0$, then the right-hand side of Eq.~\ref{eq:time derivative of spatial integrability conditions-1}
	vanishes. Therefore we see that the kinematic equations of motion
	Eq.~\ref{eq:kinematic equations of motion} preserves the spatial
	integrability conditions Eq.~\ref{eq:spatial integrability conditions}
	at all future times. Equation \ref{eq:time derivative of spatial integrability conditions-1}
	also serves as a scaling law of the amplitude of the residual error
	in spatial integrability, which can be studied by considering the
	eigenvalues of the adjoint operator $-\text{ad}_{N}$.
	
	\subsection{Local charts on the material base space \label{subsec:Local-charts}}
	
	In general, the material base space does not admit a global chart.
	Let $\mathcal{A}_{M}$ be an atlas over $M$, with elements $\mathcal{A}_{M}=\{(U_{a},\mathbf{u}_{a})\ |\ a\in I\}$,
	where $U_{a}\subset M$ and $\mathbf{u}_{a}:M\to\mathbb{R}^{d}$ and
	$I$ is an index set. Due to the product structure of the kinematic
	base space $W$, $\mathcal{A}_{M}$ extends trivially to an atlas
	$\mathcal{A}_{W}=\{([0,T]\times U_{a},\mathbf{w}_{a})\ |\ a\in I\}$
	over $W$, where $\mathbf{w}_{a}=(t,u_{a}^{1},\dots,u_{a}^{d})$.
	
	Consider two local charts $(U,\mathbf{u}),(U',\mathbf{u}')\in\mathcal{A}_{M}$
	with intersecting domains $U'\cap U\neq\emptyset$, such that the
	spatial structure generator can be written as $\xi_{M}=X_{\alpha}du^{\alpha}=X'_{\alpha}du'^{\alpha}$
	on $U'\cap U$. Then we must have
	
	\begin{equation}
		X'_{\beta}=X_{\alpha}\frac{\partial u^{\alpha}}{\partial u'^{\beta}}\label{eq:X_alpha transformation law}
	\end{equation}
	on $U'\cap U$. If the material base space $M$ does not admit global
	coordinates, then it is necessary to construct a patchwork of generalised
	strain fields over coordinate charts that cover $M$, satisfying Eq.~\ref{eq:X_alpha transformation law}.
	
	\subsection{Adapted structure fields \label{subsec:Adapted-frames}}
	
	If $r>n$, that is if the dimensionality of the symmetry group is
	larger than that of the configuration space, then there is no unique
	structure field that satisfies Eq.~\ref{eq:x from phi} for a given
	spatio-temporal configuration. This can be shown as follows. For any
	homogeneous space $\mathbb{X}$ with symmetry group $G$, there exists
	a $(r-n)$-dimensional subgroup $H\subseteq G$ such that $G/H\cong\mathbb{X}$,
	where $G/H=\left\{ gH:g\in G\right\} $ is the left coset space. The
	subgroup $H$ is called a \emph{stabiliser}, and for any $q\in\mathbb{X}$
	we have that $h\cdot q=q$, for all $h\in H$. Therefore if $\Phi(t,p)$
	satisfies Eq.~\ref{eq:x from phi} at a point $(t,p)\in W$, then
	so does $\Phi(t,p)h$ for all $h\in H$.
	
	There are therefore $r-n$ redundant degrees of freedom in the kinematic
	configuration, once the system configuration has been expressed in
	terms of the generalised strain fields. This is in principle not of
	consequence, as the redundancies will have no effect on the reconstruction
	of the true spatio-temporal configuration, via Eq.~\ref{eq:x from phi}.
	However, it is possible to eliminate these redundant degrees of freedom
	by choosing a 'gauge' in $H$. Formally, a gauge is a prescription
	that selects a unique structure field $\Phi$ that solves Eq.~\ref{eq:x from phi},
	given a spatio-temporal configuration $q:W\to\mathbb{X}$. We call
	this a \emph{kinematic adaption}, which is a concept closely related
	to the notion of an \emph{adapted frame }of space curves and surfaces
	\citep{frenetCourbesDoubleCourbure1852,darbouxLeconsTheorieGenerale1887}
	and the \emph{theory of moving frames }\citep{clellandFrenetCartanMethod2017,cartanTheorieGroupesFinis1951,felsMovingCoframesPractical1998,felsMovingCoframesII1999,olverModernDevelopmentsTheory,olverSurveyMovingFrames2005}.
	For space curves and surfaces, the adaptation leads to the desirable
	result of a parameterisation in terms of the intrinsic and extrinsic
	curvatures of the systems. In the following we outline the required
	steps to construct a kinematic adaptation, which we will then briefly
	illustrate using an example. The reader may also find the detailed
	derivations of some examples systems in Sec.~\ref{sec:Examples-of-Cartan-media-with-adapted-frames}
	helpful to understand the procedure.
	
	The construction of an adapted structure field is equivalent to choosing
	an element $h\in H$ at each time and material point $(t,p)\in W$,
	which is formally a function $h=h(q,\partial_{t}q,\partial_{\alpha}q,\dots)$,
	leading to a one-to-one ``map-between-maps'' from spatio-temporal
	configurations $q:W\to\mathbb{X}$ to structure fields $\Phi:W\to\mathbb{X}$.
	The reduction of the degrees of freedom of the structure field results
	in the elimination of $n-r$ constraints on the components of the
	generalised strain and velocity fields each, which are to be inferred
	from Eq.~\ref{eq:generalised strain and velocity fields} and Eq.~\ref{eq:kinematic equations of motion and integrability conditions}.
	
	In most applications, as in the example we will give below, the adapted
	structure field is adapted to the \emph{spatial} kinematics of the
	system. That is, the gauge is chosen as $h=h(q,\partial_{\alpha}q,\dots)$,
	without $\partial_{t}x$. However, this is not always the case, see
	Sec.~\ref{sec:Relativistic Cosserat rods} for an example. To simplify
	the discussion, we will presume we are working with a system in $d=1$
	material dimensions, with a spatially adapted structure field (see
	Sec.~\ref{sec:Examples-of-Cartan-media-with-adapted-frames} for
	higher-dimensional examples). In this case, from Eq.~\ref{eq:generalised strain and velocity fields}
	we find constraints on the strain, such that it takes value in a vector
	subset $X\in V\subseteq\mathfrak{g}$, where $\text{dim}(V)=n$. Let
	$V^{\perp}$ be the orthogonal complement of $V$ respectively, where
	$\text{dim}(V^{\perp})=r-n$, and let $b_{i}\in V,\ i=1,\dots,n,$
	and $b_{j}^{\perp}\in V,\ j=1,\dots,r-n$ be basis vectors for $V$
	and $V^{\perp}$ respectively. A given Lie algebra element $C\in\mathfrak{g}$
	can be expanded as $C=C_{i}b_{i}+C_{j}^{\perp}b_{j}^{\perp}$, where
	$C_{i}$ and $C_{i}^{\perp}$ are the components of $C$ in the respective
	bases. Then Eq.~\ref{eq:kinematic equations of motion and integrability conditions}
	decomposes into
	\begin{subequations}
		\begin{align}
			\dot{X}_{i} & =(\mathcal{D}_{\alpha}N)_{i}, & i=1,\dots,n,\label{eq:kinematic equations of motion under adapted frames}\\
			0 & =(\mathcal{D}_{\alpha}N)_{j}^{\perp}, & j=1,\dots,r-n,\label{eq:adapted frames constraints}
		\end{align}
		\label{eq:eoms and constraints due to adapted frames}
	\end{subequations}
	which is the kinematic equation of motion for the strain, and a set
	of constraints on $N$, respectively. In effect, Eq.~\ref{eq:adapted frames constraints}
	constrains $N$ onto the subsurface of $\mathfrak{g}$ that ensures
	that the structure field remains adapted in time. Note that the kinematic
	equations of motions in its original unconstrained form, Eq.~\ref{eq:kinematic equations of motion},
	hold as before, under the condition that Eq.~\ref{eq:adapted frames constraints}
	is satisfied. The remaining components of the generalised strain will
	in general comprise the extrinsic, in addition to the intrinsic, geometry
	of the system.
	
	To illustrate the above, let us again consider a filament on the unit
	sphere. The configuration space of the filament is $\mathbb{X}=S^{2}$,
	and its symmetry group $G=SO(3)$; that is, $n=2$ and $r=3$. This
	will lead to an $SO(2)$ gauge freedom in the kinematic description.
	To see this, we write the structure field as $R:W\to SO(3)$ as an
	orthonormal frame $R=(\mathbf{e}_{1}\ \mathbf{e}_{2}\ \mathbf{e}_{3})$
	that satisfies $R\mathbf{p}=\mathbf{p}^{s}$, where $\mathbf{p}\in S^{2}$
	and $\mathbf{p}^{s}:W\to S^{2}$. We furthermore let $\mathbf{p}=(0\ 0\ 1)^{T}$,
	such that $\mathbf{p}^{s}=R\mathbf{p}=\mathbf{e}_{3}$. The intrinsic
	geometry of the filament corresponds to the longitudinal deformations
	along its length, and the extrinsic geometry its curvature. We see
	Eq.~\ref{eq:x from phi} remains invariant under rotations of $R$
	around $\mathbf{e}_{3}$, at any point $u\in M$. The corresponding
	generalised strain $\hat{\pi}:W\to\mathfrak{so}(3)$ will thus have
	a superfluous degree of freedom. To rid the geometric kinematic description
	of the system of this gauge freedom, we can construct a kinematic
	adaptation by constraining $\mathbf{e}_{1}$ to be tangent to the
	filament at all times; that is, we set $\mathbf{e}_{1}=h^{-1}\partial_{u}\mathbf{p}^{s}$,
	where $h=|\partial_{u}\mathbf{p}^{s}|$. This effectively chooses
	a unique element $R=R(\mathbf{p}^{s},\partial_{u}\mathbf{p}^{s})$
	for a given filament $\mathbf{p}^{s}(t,\cdot)$. From Eq.~\ref{eq:generalised strain and velocity fields}
	we then find that $\pi_{1}=0$, and we write $\boldsymbol{\pi}=(0\ h\ \kappa)^{T}$,
	where $\kappa$ is the scalar curvature of the filament, and $h^{2}$
	the metric along the filament induced from ambient Euclidean space.
	The components $\kappa$ and $h$ correspond to the extrinsic and
	intrinsic geometry of the filament respectively. To see the latter,
	note that the length of the filament is given by $L=\int_{0}^{L_{0}}|\partial_{u}\mathbf{p}^{s}|du=\int_{0}^{L_{0}}|\boldsymbol{\theta}|du=\int_{0}^{L_{0}}hdu$.
	Now, let us assume that the structure field is adapted at $t=0$,
	such that $\mathbf{e}_{1}(0,u)=(h^{-1}\partial_{u}\mathbf{e}_{3})\big|_{t=0}$
	and $\boldsymbol{\pi}(0,u)=(R^{-1}\partial_{u}R)_{t=0}=(0\ h(0,u)\ \kappa(0,u))^{T}$.
	The generalised velocity must be constrained so as the maintain the
	adaption of the system in time, which we can find from Eq.~\ref{eq:kinematic equations of motion and integrability conditions},
	leading to $\partial_{u}\Omega_{1}-\Omega_{2}\kappa+\Omega_{3}h=0$.
	Solving the constraint for $\Omega_{2},$we have that $\boldsymbol{\Omega}=\boldsymbol{\Omega}(h,\kappa,\Omega_{3})$;
	that is, the generalised velocity is a function the generalised strain
	and $\Omega_{3}$.
	
	The above example shows that we can see kinematic adaptation as a
	way to ``emulate'' $\mathbb{X}$-configured Cartan media, with symmetry
	group $G$, using a kinematically constrained Cartan system in a larger
	configuration space $\tilde{\mathbb{X}}\supset\mathbb{X}$ with the
	same symmetry group, for which $\text{dim}(\tilde{\mathbb{X}})=\text{dim}(G)$.
	The homogeneous space $\tilde{\mathbb{X}}$ is essentially $G$ itself,
	and can formally be written as a coset-space $\tilde{\mathbb{X}}=G/\{e\}$,
	where $e\in G$ is the identity element. In the above example, we
	had that $\mathbb{X}=S^{2}$ and $\tilde{\mathbb{X}}=SO(3)$, where
	the latter can be seen as the configuration space of a ``Cosserat
	rod'' on the sphere (which is the subject of Sec.~\ref{sec:Cosserat rods on 2-spheres}).
	From this point-of-view, the adapted structure field amounts to a
	``kinematically constrained'' Cosserat rod on the sphere, where
	it is constrained such that $\mathbf{e}_{1}$ is tangent to the center-line.
	
	\section{Dynamics \label{sec:geometric dynamics}}
	
	In the previous section we found kinematic equations of motion of
	Cartan media in terms of their generalised strain and velocity fields.
	Here we derive geometrised second-order equations of motion, in terms
	of \emph{generalised momentum }and\emph{ stress fields} and \emph{generalised
		body force densities}. Consequently, we obtain momentum balance equations
	analogous to those of classical continuum mechanics. As will be seen
	here, and with more detail in Sec.~\ref{sec:cosserat-as-cartan-media}-\ref{sec:Further-examples-of-cartan-media},
	the form of the balance equations reflect the symmetry group of the
	configuration space. The Cauchy momentum equation of classical continuum
	mechanics, which is the balance of linear momentum, derives from the
	translational symmetry of Euclidean space. Analogously, if $\mathbb{X}=S^{2}$,
	for which the appropriate symmetry group is $G=SO(3)$, then the rotational
	symmetry of the configuration space will lead to angular momentum
	balance equations.
	
	The derivation will begin programmatically by way of Hamilton's principle.
	That is, given a configuration-dependent Lagrangian density, most
	commonly constructed from kinetic and potential energy densities,
	the dynamical trajectory of the system must then be a stationary point
	of the corresponding action functional. The equations of motion of
	such trajectories can then be found by variation of the action in
	the configurational degrees of freedom, leading to the Euler-Lagrange
	equations. However, this will not naturally lead to mechanics expressed
	in terms of the differential geometry of the system, in the spirit
	of the geometrised kinematics developed in the previous section. Instead,
	we relate the variation to the infinitesimal action of the generalised
	strain and velocity fields. In so doing we formulate the mechanics
	in terms of \emph{dual Lie algebra}-valued fields, that are conjugate
	to the Lie algebraic kinematic variables: generalised momentum and
	stress fields, and the generalised body force density.
	
	This approach was first pioneered in \citep{poincareFormeNouvelleEquations1901}
	for point particles, from which a large literature on geometric particle
	mechanics \citep{marleHenriPoincareNote2013,marsdenIntroductionMechanicsSymmetry2013,holmGeometricMechanicsRotating2008}
	has sprung. In those contexts, the resulting equation of motion is
	known as the \emph{Euler-Poincaré} \emph{equation}. The derivation
	we present here can thus be seen as a generalisation of this procedure
	to continuum systems in homogeneous spaces.
	
	The conservative dynamics of a system is fully specified given a Lagrangian
	$L(q,dq)$, which, given a spatio-temporal configuration $q:W\to\mathbb{X}$,
	is a volume form over the material base space. In a local chart $\mathbf{u}:U\to\mathbb{R}^{d},\ U\subseteq M$,
	the Lagrangian can be written as $L(q,dq)=\mathcal{L}(q,\dot{q},\partial_{\alpha}q)\ dt\wedge dV$,
	where we have defined the volume element $dV=du^{1}\wedge\dots\wedge du^{d}$,
	and where $\mathcal{L}$ is the Lagrangian density, which is a scalar
	density $\mathcal{L}(q(\cdot,\cdot),\dot{q}(\cdot,\cdot),\partial_{\alpha}q(\cdot,\cdot)):W\to\mathbb{R}$
	given a spatio-temporal configuration. Under a change-of-coordinates
	we have that volume element transforms as a tensor density of weight
	$-1$,
	
	$dV'=du'^{1}\wedge\dots\wedge du'^{d}=|J|^{-1}dV$, and the Lagrangian
	density is a scalar density of weight $1$,
	
	\begin{equation}
		\mathcal{L}'(q,\dot{q},\partial'_{\alpha}q)=|J|\mathcal{L}(q,\dot{q},\partial_{\alpha}q).\label{eq:L transformation}
	\end{equation}
	where $\partial'_{\alpha}=J_{\alpha}^{\beta}\partial_{\beta}$, $J_{\alpha}^{\beta}=\frac{\partial u^{\beta}}{\partial u'^{\alpha}}$
	is the Jacobian matrix of the coordinate transformation and $|J|=\text{det}\left[J\right]$.
	Therefore we have $\mathcal{L}(q,\dot{q},\partial_{\alpha}q)\ dt\wedge dV=\mathcal{L}'(q,\dot{q},\partial'_{\alpha}q)\ dt\wedge dV'$
	.

	The \emph{action} is a functional of the spatio-temporal configuration,
	formed by integrating $L$ over the kinematic base space $W=[0,T]\times M$.
	A spatio-temporal configuration is called \emph{physical} if it is
	a stationary trajectory of the action functional, that is
	\begin{equation}
		\delta\int_{W}L(q,dq)=\delta\int_{W}\mathcal{L}(q,\dot{q},\partial_{\alpha}q)\ dt\wedge dV=0\label{eq:configurational variation}
	\end{equation}
	under variations $q\to q+\delta q$, where $\delta q:W\to T\mathbb{X}$
	is a variational test function, which must vanish at the temporal
	boundaries $\delta q(0,p)=\delta q(T,p)=0$ for all $p\in M$. Note
	that Eq.~\ref{eq:configurational variation} is an abuse of notation,
	as the Lagrangian density is defined with respect to a chart, which
	may only cover a subset of the material base space $U\subseteq M$.
	Strictly, the Lagrangian must be varied over a set of charts $\{(U_{a},\mathbf{u}_{a})\ |\ a\in K\}=\mathcal{B}\subset\mathcal{A}_{M}$,
	where $K\subset I$ is an index set, such that the domains $U_{\alpha}$
	form a disjoint partition of $M$. That is, $U_{a}\cap U_{b}=\emptyset$
	for $a\neq b$ and $\bigcup_{a\in K}\bar{U}_{a}=M$, where $\bar{U}_{k}$
	denotes the closure of the open set $U_{k}$. Hamilton's principle
	is then $\delta\sum_{a\in J}\int_{[0,T]\times U_{a}}\mathcal{L}^{(a)}(q,\dot{q},\partial_{\alpha}^{(a)}q)\ dt\wedge dV_{(a)}=0$.
	In the following, we will continue to use the notation in Eq.~\ref{eq:configurational variation},
	and address chart-related issues as they arise.
	
	In most physical applications, the dynamics is defined via a kinetic
	$\mathcal{K}(\dot{q})$ and potential $\mathcal{U}(q,\partial_{\alpha}q$)
	energy density over the system, such that the Lagrangian is constructed
	as $\mathcal{L}(q,\dot{q},\partial_{\alpha}q)=\mathcal{K}(\dot{q})-\mathcal{U}(q,\partial_{\alpha}q)$.
	Equation \ref{eq:configurational variation} can be recognised as
	identical in form to the variational principles found in classical
	field theory, where the first term would correspond to local field
	effects, and the latter two correspond to temporal and spatial deformations
	of the field. For example, in scalar field theory a typical Lagrangian
	may be of the form $\mathcal{L}=\frac{1}{2}\dot{\phi}^{2}+\frac{1}{2}|\partial_{\alpha}\phi|^{2}-\frac{1}{2}m^{2}\phi^{2}-\frac{\lambda}{4!}\phi^{4}$,
	where $\phi\in\mathbb{X}=\mathbb{R}$. In Sec.~\ref{sec:The O(3) non-linear sigma model}
	we apply the geometrised mechanical framework presented here to non-linear
	$\sigma$ field theories.
	
	Recall Eq.~\ref{eq:x from phi} and Eq.~\ref{eq:dx}, which gives
	us the spatio-temporal configuration in terms of the structure field
	$q=\Phi\cdot q_{r}$, and its derivatives in terms of the generalised
	strain and velocity $\dot{q}=\rho(\text{Ad}_{\Phi}N)q$ and $\partial_{\alpha}q=\rho(\text{Ad}_{\Phi}X_{\alpha})q$.
	We can rewrite these as $\dot{x}=\Pi(\Phi)\rho(N)q_{0}$ and $\partial_{\alpha}q=\Pi(\Phi)\rho(X_{\alpha})q_{r}$,
	or combined as $dq=\Pi(\Phi)\rho(\xi)q_{r}$. We can therefore define
	the \emph{left-trivialised Lagrangian} \citep{marleHenriPoincareNote2013,marsdenIntroductionMechanicsSymmetry2013,holmGeometricMechanicsRotating2008}
	\begin{equation}
		\check{L}(\Phi,\xi)=L(\Pi(\Phi)q_{r},\Pi(\Phi)\rho(\xi)q_{r}),\label{eq:left-trivialised Lagrangian}
	\end{equation}
	which is, in local coordinates, $\check{\mathcal{L}}(\Phi,N,X_{\alpha})=\mathcal{L}(\Pi(\Phi)q_{0},\Pi(\Phi)\rho(N)q_{r},\Pi(\Phi)\rho(X_{\alpha})q_{r})$,
	such that $\check{L}(\Phi,\xi)=\check{\mathcal{L}}(\Phi,N,X_{\alpha})\ dt\wedge dV$.
	The Lagrangian Eq.~\ref{eq:left-trivialised Lagrangian} is now expressed
	entirely in terms of Lie group and Lie algebraic fields. The left-trivialised
	Lagrangian transforms as $\check{\mathcal{L}}'(\Phi,N,X'_{\alpha})=|J|\check{\mathcal{L}}(\Phi,N,X_{\alpha})$
	under a change-of-coordinates, where $X'_{\alpha}$ is given by Eq.~\ref{eq:X_alpha transformation law}.
	Note that the generalised strain and velocity fields are expressed
	in the body frame in $\check{\mathcal{L}}'(\Phi,N,X'_{\alpha})$,
	therefore the left-trivialised Lagrangian neatly separates non-constitutive
	and constitutive parts of the dynamics, corresponding to the first
	and the latter two arguments respectively.
	
	Variations of Eq.~\ref{eq:left-trivialised Lagrangian} must preserve
	both the spatial and temporal integrability of the system, Eq.~\ref{eq:kinematic equations of motion and integrability conditions}.
	We must therefore derive the space of permissible variations $\delta N$
	and $\delta X_{\alpha}$. We begin by considering the variation of
	the configurational Lagrangian, Eq.~\ref{eq:configurational variation}.
	Applying the variational operator $\delta$ on both sides of Eq.~\ref{eq:x from phi}
	we find that $\delta q=\delta\Pi(\Phi)q_{0}=\delta\Pi(\Phi)\Pi(\Phi^{-1})q=\rho(\text{Ad}_{\Phi}\eta)q$,
	where $\eta=\Phi^{-1}\delta\Phi$, which must satisfy $\eta(0,p)=\eta(T,p)=0$
	for all $p\in M$. We see that the configurational variation can be
	fully expressed in terms of the Lie algebraic variational test function
	$\eta$. We now vary the structure generator $\xi$ to get
	\begin{equation}
		\begin{aligned}\delta\xi & =-\Phi^{-1}\delta\Phi\Phi^{-1}d\Phi+\Phi^{-1}\delta(d\Phi)\\
			& =d\eta-\eta\xi+\xi\eta\\
			& =d\eta+\text{ad}_{\xi}\eta
		\end{aligned}
	\end{equation}
	where we used $d\eta=-\xi\eta+\Phi^{-1}d(\delta\Phi)$ and $d\delta=\delta d$.
	In local coordiantes, we thus find that the variations of the generalised
	strain and velocity fields are
	\begin{subequations}
		\label{eq:variations of strain and velocity}
		\begin{align}
			\delta N & =\mathcal{D}_{t}=\dot{\eta}+\text{ad}_{N}\eta,\label{eq:velocity variation}\\
			\delta X_{\alpha} & =\mathcal{D}_{\alpha}=\partial_{\alpha}\eta+\text{ad}_{X}\eta\label{eq:strain variation}
		\end{align}
	\end{subequations}
	for $\alpha=1,\dots,d$, and where we used Eq.~\ref{eq:frame generator}.
	In the subsequent subsections, we will perform this Lie algebraic
	variation to derive the geometrised dynamical equations of motion
	of Cartan media.
	
	In Sec.~\ref{subsec:Generalised-stress} and Sec.~\ref{subsec:Generalised-body-force}
	we will derive generalised momentum balance equations for Cartan media,
	using the left-trivialised Lagrangian and Hamilton's principle. Section
	\ref{subsec:Generalised-stress} treats the purely constitutive case,
	where the generalised stress fields are derived from strain-dependent
	potential energy densities. Section \ref{subsec:Generalised-body-force}
	introduces generalised body force densities, derived from configuration-dependent
	potential energy densities. In Sec.~\ref{subsec:Non-conservative-dynamics}
	we relax the assumption of conservative dynamics, formulating a generalised
	Lagrange-D'Alembert principle for Cartan media, from which general
	non-conservative dynamical equations of motion can be derived. Finally,
	in Sec.~\ref{subsec:Adapted-frames} we continue the discussion of
	Sec.~\ref{subsec:Adapted-frames} and consider dynamics under kinematic
	adaptation.
	
	\subsection{Generalised stresses \label{subsec:Generalised-stress}}
	
	We will now apply Hamilton's principle assuming purely constitutive
	dynamics. That is, we make the following two assumptions: Firstly,
	the configurational Lagrangian is not dependent on its first argument
	$L(q,dq)=L(q_{0},dq)$. Secondly, the latter argument satisfies \emph{left-invariance}
	$L(q,\Pi(g)\rho(\xi)q_{r})=L(q,\rho(\xi)q_{r})$ for any $g\in G$,
	which in local coordinates corresponds to $\mathcal{L}(q,\Pi(g)\rho(N)q_{r},\Pi(g)\rho(X_{\alpha})q_{r})=\mathcal{L}(q,\rho(N)q_{r},\rho(X_{\alpha})q_{r})$.
	Both of these are equivalent to the condition that the left-trivialised
	Lagrangian satisfies $\check{L}(\Phi,\xi)=\check{L}(e,\xi)$, where
	$e\in G$ is the identity element. If these hold, we can construct
	a \emph{reduced Lagrangian }\citep{marleHenriPoincareNote2013,marsdenIntroductionMechanicsSymmetry2013,holmGeometricMechanicsRotating2008}
	\begin{equation}
		l(\xi)=L(q_{r},\rho(\xi)q_{r})\label{eq:reduced lagrangian}
	\end{equation}
	and $\ell(N,X_{\alpha})=\mathcal{L}(q_{r},\rho(N)q_{r},\rho(X_{\alpha})q_{r})$
	in local coordinates, such that $l(\xi)=\ell(N,X_{\alpha})\ dt\wedge dV$,
	with which we can derive the dynamics using Hamilton's principle.
	The resulting second-order equations of motion will be purely constitutive,
	expressed in terms of generalised momentum and stress fields.
	
	An example of a reducible Lagrangian for the filament on the unit
	sphere is $\mathcal{L}=\frac{m}{2}|\dot{\mathbf{p}}^{s}|^{2}-\frac{k}{2}|\partial_{u}\mathbf{p}^{s}|^{2}$,
	which is only dependent on the derivatives of the configuration. This
	Lagrangian is reducible, as $|\dot{\mathbf{p}}^{s}|^{2}=|\dot{R}\mathbf{p}|^{2}=|R\hat{\Omega}\mathbf{p}|^{2}=|\hat{\Omega}\mathbf{p}|^{2}=|\boldsymbol{\Omega}\times\mathbf{p}|^{2}=\boldsymbol{\Omega}^{T}P_{0}\boldsymbol{\Omega}$
	and $|\partial_{u}\mathbf{p}^{s}|^{2}=|R\hat{\pi}\mathbf{p}|^{2}=\boldsymbol{\pi}^{T}P_{0}\boldsymbol{\pi}$,
	where $\hat{\pi}=\Phi^{-1}\partial_{u}\Phi$ is the spatial angular
	rate-of-change along the material coordinate $u$ and $P_{0}=\hat{p}^{T}\hat{p}$.
	We can write the reduced Lagrangian as $\ell=\frac{m}{2}\boldsymbol{\Omega}^{T}P_{0}\boldsymbol{\Omega}-\frac{k}{2}\boldsymbol{\pi}^{T}P_{0}\boldsymbol{\pi}$.
	Note that $P_{0}$ is a rank-$2$ matrix, reflecting the fact that
	$\text{dim}(S^{2})=n=2$, whilst $\text{dim}(SO(3))=r=3$. Such cases
	will not affect the discussion that follows, but we will consider
	a worked out example in Sec.~\ref{sec:cosserat surfaces}. The Lagrangian
	would be irreducible if it contains a term like $\mathbf{p}^{T}\mathbf{p}^{s}$,
	which breaks the first of the two assumptions outlined above. More
	esoterically, the second assumption does not hold if the strain couples
	with quantities in the spatial frame directly, such as if there is
	a term like $\mathbf{p}\cdot\partial_{u}\mathbf{p}^{s}$. Finally,
	a term like $\mathbf{p}^{s}\cdot\partial_{u}\mathbf{p}^{s}$ would
	violate both assumptions simulatenously.
	
	We now use Eq.~\ref{eq:variations of strain and velocity} to vary
	the reduced Lagrangian. Let $b_{i}\in\mathfrak{g},\ i=1,\dots,r$,
	be a basis for the Lie algebra, such that we can write $C=C_{i}b_{i}$
	for any $C\in\mathfrak{g}$. The variation of the Lagrangian is then
	$\delta\ell=\frac{\partial\ell}{\partial N_{i}}\delta N_{i}+\frac{\partial\ell}{\partial(X_{\alpha})_{i}}\delta(X_{\alpha})_{i}$,
	where we can identify $\frac{\partial\ell}{\partial N_{i}}$ and $\frac{\partial\ell}{\partial(X_{\alpha})_{i}}$
	as components of elements of the dual Lie algebra $\mathfrak{g}^{*}$,
	which are contracting with the variations $\delta N$ and $\delta X_{\alpha}$
	respectively. We rewrite the contractions as an inner product $\langle\cdot,\cdot\rangle:\mathfrak{g}\times\mathfrak{g}^{*}\to\mathbb{R}$,
	which we define as
	\[
	\langle C,Y\rangle=C_{i}Y_{i}
	\]
	for any $C\in\mathfrak{g}$ and $Y=Y_{i}B_{i}\in\mathfrak{g}^{*}$,
	where $B_{i}\in\mathfrak{g}^{*},\ i=1,\dots,r$ is a basis for the
	dual Lie algebra satisfying $\langle b_{i},B_{j}\rangle=\delta_{ij}$.
	We can thus rewrite the variation as $\delta\ell=\left\langle \delta N,S\right\rangle -\left\langle \delta X_{\alpha},Q^{\alpha}\right\rangle $,
	where we have defined the \emph{generalised momentum} $S=\frac{\partial\ell}{\partial N}=\frac{\partial\mathcal{K}}{\partial N}$
	and \emph{generalised stress} $Q^{\alpha}=-\frac{\partial\ell}{\partial X_{\alpha}}=\frac{\partial\mathcal{U}}{\partial X_{\alpha}}$
	\emph{fields}.
	\begin{subequations}
		\label{eq:generalised momentum and stress fields}
		\begin{align}
			S & =\frac{\partial\ell}{\partial N}=\frac{\partial\mathcal{K}}{\partial N}\in\mathfrak{g}^{*},\label{eq:generalised momentum field}\\
			Q^{\alpha} & =-\frac{\partial\ell}{\partial X_{\alpha}}=\frac{\partial\mathcal{U}}{\partial X_{\alpha}}\in\mathfrak{g}^{*}.\label{eq:generalised stress fields}
		\end{align}
	\end{subequations}
	Note that these derivatives are a short-hand for $\frac{\partial\ell}{\partial N}=\frac{\partial\ell}{\partial N_{i}}B_{i}$.
	However, if we consider Lie algebra elements in a matrix representation,
	$\frac{\partial\ell}{\partial N}$ can be interpreted as a matrix
	derivative. See App.~\ref{app:Vector and Matrix operations} for
	more details. The inner product also allows us to dualise linear operators
	on the Lie algebra. Given a linear operator $\mathrm{A}:\mathfrak{g}\to\mathfrak{g}$,
	we define its \emph{dual} as the operator $\mathrm{A}^{*}:\mathfrak{g}^{*}\to\mathfrak{g}^{*}$
	defined by the relation $\left\langle \mathrm{A}C,Y\right\rangle =-\left\langle C,\mathrm{A}^{*}Y\right\rangle $,
	for any $C\in\mathfrak{g}$ and $Y\in\mathfrak{g}^{*}$.
	
	The generalised momentum field transforms as a scalar density of weight
	$1$ under change of charts, and the generalised stress field as a
	vector density of weight $1$. We have
	\begin{subequations}
		\label{eq:generalised momentum and stress fields transformation}
		\begin{align}
			S' & =|J|S,\label{eq:S transformations}\\
			Q'^{\alpha} & =|J|\frac{\partial u'^{\alpha}}{\partial u^{\beta}}Q{}^{\beta}\label{eq:Q^alpha transformation}
		\end{align}
	\end{subequations}
	where we used Eq.~\ref{eq:configurational variation} and Eq.~\ref{eq:X_alpha transformation law}.
	
	We can now proceed to vary the action. We have
	
	\begin{equation}
		\begin{aligned} & \delta\int_{W}l(\xi)=\delta\int_{W}\ell(N,X_{\alpha})\ dt\wedge dV\\
			& =\int_{W}\left\{ \left\langle \delta N,S\right\rangle -\left\langle \delta X_{\alpha},Q^{\alpha}\right\rangle \right\} dt\wedge dV\\
			& =\int_{W}\left\{ \left\langle \mathcal{D}_{t}\eta,S\right\rangle -\left\langle \mathcal{D}_{\alpha}\eta,Q^{\alpha}\right\rangle \right\} dt\wedge dV\\
			& =\int_{W}\left\{ -\left\langle \eta,\mathcal{D}_{t}^{*}S\right\rangle +\left\langle \eta,\mathcal{D}_{\alpha}^{*}Q^{\alpha}\right\rangle \right\} dt\wedge dV\\
			& \ +\int_{W}\left\{ \partial_{t}\langle\eta,S\rangle-\partial_{\alpha}\langle\eta,Q^{\alpha}\rangle\right\} dt\wedge dV
		\end{aligned}
		\label{eq:variation of ell 2}
	\end{equation}
	where we used integration-by-parts, and where the dualised covariant
	derivatives are $\mathcal{D}_{t}^{*}=\partial_{t}+\text{ad}_{N}^{*}$
	and $\mathcal{D}_{\alpha}^{*}=\partial_{t}+\text{ad}_{X_{\alpha}}^{*}$,
	and where $\text{ad}_{C}^{*}:\mathfrak{g}^{*}\to\mathfrak{g}^{*},\ C\in\mathfrak{g}$
	is the dual of the adjoint action. See App.~\ref{app:Vector and Matrix operations}
	for further details on the dual adjoint, and an example of its matrix
	representation for $\mathfrak{se}(3)$. We note here again that we
	are abusing the notation, as the integrands are expressed in local
	charts. This is only of consequence for the final line of Eq.~\ref{eq:variation of ell 2},
	which will lead to boundary terms, and will be massaged into an invariant
	form below.
	
	By imposing that the integral of the penultimate line of Eq.~\ref{eq:variation of ell 2}
	must vanish for all $\eta$, we find the momentum balance equations
	for constitutive dynamics in local coordinates
	\begin{equation}
		\mathcal{D}_{t}^{*}S=\mathcal{D}_{\alpha}^{*}Q^{\alpha},\label{eq:constitutive momentum balance equations}
	\end{equation}
	which, together with the kinematics equations of motion Eq.~\ref{eq:kinematic equations of motion},
	defines the constitutive mechanics of Cartan media. It is of notethat
	$\Phi$ does not appear explicitly in the constitutive momentum balance
	equations derived here, as for the geometrised kinematic equations
	of motion Eq.~\ref{eq:kinematic equations of motion}. Therefore,
	if the dynamics is constitutive, the entire set of kinematic and dynamic
	equations of motion can be simulated purely in terms of the generalised
	strain and velocity fields $X_{\alpha}$and $N$. The balance equation
	can be expressed as equations of motion expressed in terms of $N$
	using Eq.~\ref{eq:generalised momentum and stress fields}. Combined,
	the kinematic and dynamic equations of motion form a second-order
	system for constitutive Cartan media mechanics.
	
	Eq.~\ref{eq:constitutive momentum balance equations} can now be
	used to derive the constitutive momentum balance equations for any
	Cartan media. Let us again consider the filament on the sphere. Let
	the reduced kinetic and potential energy densities be of the form
	$\mathcal{K}=\frac{1}{2}\boldsymbol{\Omega}^{T}\mathbb{I}\boldsymbol{\Omega}$
	and $\mathcal{U}=\frac{1}{2}\boldsymbol{\pi}^{T}\mathbb{N}\boldsymbol{\pi}$
	respectively, where $\mathbb{I},\mathbb{N}\in\mathbb{R}_{+}^{3\times3}$
	are symmetric positive-definite matrices. We write the generalised
	momentum and stress as $\mathbf{L}=\mathbb{I}\boldsymbol{\Omega}$
	and $\mathbf{M}=\mathbb{K}\boldsymbol{\pi}$ respectively, where we
	have used bold-faced notation to signify that these are vectors in
	$T\mathbb{E}^{3}$. In this context we should see $\mathbf{L}(t,u)$
	and $\mathbf{M}(\boldsymbol{\pi}(t,u))$ as the angular momentum of
	the material point $u$ at time $t$, and the moment it is experiencing,
	respectively. These quantities are all in the body frame of the system.
	As $\text{ad}^{*}=\text{ad}$ for $\mathfrak{so}(3)$ (see App.~\ref{app:Vector and Matrix operations})
	we have that $\dot{\mathbf{L}}+\boldsymbol{\Omega}\times\mathbf{L}=\partial_{u}\mathbf{M}+\boldsymbol{\pi}\times\mathbf{M}$,
	which are angular momentum balance equations. Further examples are
	presented in Sec.~\ref{sec:cosserat-as-cartan-media}-\ref{sec:Further-examples-of-cartan-media}.
	
	We now consider the final line of Eq.~\ref{eq:variation of ell 2}.
	The first term in the integrand vanishes, as $\int_{W}\partial_{t}\langle\eta,S\rangle dt\wedge dV=\int_{0}^{T}dt\int_{M}\partial_{t}\langle\eta,S\rangle dV=\int_{M}[\langle\eta,S\rangle]_{0}^{T}dV=0$,
	where we have used that $\eta$ vanishes at the temporal boundaries.
	To simplify the treatment of the second term, we will introduce an
	arbitrary volume form $\omega$ over $M$, which we write in local
	coordinates as $\omega=wdV$, where $w$ is a scalar density. Let
	$A^{\alpha}=w^{-1}\langle\eta,Q^{\alpha}\rangle$, which are components
	of a vector field $A=A^{\alpha}\frac{\partial}{\partial u^{\alpha}}$
	on $M$, then $\int_{M}\partial_{\alpha}\langle\eta,Q^{\alpha}\rangle dV=\int_{M}g^{-1}\partial_{\alpha}(gA^{\alpha})\ \omega=\int_{M}\text{div}A\ \omega$,
	where we have used the definition of the divergence. From the divergence
	theorem, we then have that $\int_{M}\text{div}A\ \omega=\int_{\partial M}(n,A)\ \omega_{S}$,
	where $\omega_{S}$ is the induced volume form on the material surface
	$\partial M$, and where $n:\partial M\to T^{*}M$ is the unique (up
	to scalar multiplication) covector-field normal to $\partial M$,
	and the contraction is $(n,A)=n_{\alpha}A^{\alpha}$ in local coordinates.
	We now impose that this term must vanish for all $\eta$, which gives
	us the boundary condition
	\begin{equation}
		n_{\alpha}Q^{\alpha}=0\label{eq:stress boundary condition}
	\end{equation}
	in local coordinates. This is a coordinate-independent condition,
	as $n_{\alpha}=\frac{\partial u'^{\beta}}{\partial u^{\alpha}}n'_{\beta}$
	under a change of charts, such that $n'_{\alpha}Q'^{\alpha}=|J|n_{\alpha}Q^{\alpha}=0$.
	As the stress is a function of the strain $Q^{\alpha}=Q^{\alpha}(X_{1},X_{2},\dots,X_{d})$,
	then these boundary conditions will in turn induce conditions on $X_{\alpha}$
	at the material boundaries.
	
	\subsection{Generalised body force densities \label{subsec:Generalised-body-force}}
	
	If the assumptions of the previous subsection do not hold, that is,
	if the Lagrangian has explicit configurational dependence, or if the
	generalised strain couples with the spatial frame, \emph{generalised
		body force densities }will appear in the balance equations as a result.
	Furthermore, as the dynamics is no longer purely constitutive, the
	equations of motion will now require explicit knowledge of the structure
	field $\Phi$. Fundamentally, this is due to the explicit involvement
	of the spatial frame in the dynamics, which couples with intrinsic
	geometric formuation of the kinematics. Within the context of the
	geometric mechanics of point-particles \citep{marleHenriPoincareNote2013,marsdenIntroductionMechanicsSymmetry2013,holmGeometricMechanicsRotating2008},
	such symmetry-breaking terms in the Lagrangian are called \emph{advected
		terms}.
	
	Varying the left-trivialised Lagrangian Eq.~\ref{eq:left-trivialised Lagrangian},
	we find
	\begin{equation}
		\begin{aligned} & \delta\int_{W}\check{\mathcal{L}}\ dt\wedge d^{d}u\\
			& =\int_{W}\left\{ \mathcal{T}_{\mu}\delta\Phi^{\mu}+\left\langle \delta N,S\right\rangle -\left\langle \delta X_{\alpha},Q^{\alpha}\right\rangle \right\} \ dt\wedge d^{d}u
		\end{aligned}
		\label{eq:variation of left-trivialised lagrangian 1}
	\end{equation}
	where $S=\frac{\partial\check{\mathcal{L}}}{\partial N}$ and $Q=-\frac{\partial\check{\mathcal{L}}}{\partial X_{\alpha}}$,
	and we have defined $\mathcal{T}_{\mu}=\frac{\partial\check{\mathcal{L}}}{\partial\Phi^{\mu}}$.
	We can identify $\mathcal{T}$ as a covector-field on the Lie group
	$\mathcal{T}\in T^{*}G$, as it contracts with the variational test
	function $\delta\Phi\in TG$. Depending on how $G$ is parameterised,
	$\Phi^{\mu}$ can either be considered coordinates on $G$, or the
	components of a matrix representation $G$. For example, if $G=SO(3)$,
	we may represent Lie group elements in terms of Euler angles $(\alpha,\beta,\gamma)\mapsto R(\alpha,\beta,\gamma)\in SO(3)$,
	or as rotation matrices $R=(\mathbf{e}_{1}\ \mathbf{e}_{2}\ \mathbf{e}_{3})\in\mathbb{R}^{3\times3}$,
	where $\mathbf{e}_{i},\ i=1,2,3$ are an orthonormal set of basis
	vectors in $\mathbb{R}^{3}$. In the latter case, the contraction
	in the first term of Eq.~\ref{eq:variation of left-trivialised lagrangian 1}
	may be written as $\mathcal{T}_{\mu\nu}^{T}\delta\Phi_{\mu\nu}$,
	where $\mathcal{T}_{\mu\nu}=\frac{\partial\check{\mathcal{L}}}{\partial\Phi_{\nu\mu}}$,
	following the numerator-layout matrix derivative convention as defined
	in App.~\ref{app:Vector and Matrix operations}.
	
	Now, recall that the variational test function can be written as $\delta\Phi=\Phi\eta$.
	We thus have that $\mathcal{T}_{\mu\nu}^{T}\delta\Phi_{\mu\nu}=\mathcal{T}_{\mu\nu}^{T}\Phi_{\mu\sigma}\eta_{\sigma\nu}=\text{Tr}\left[\mathcal{T}\Phi\eta\right]=\eta_{i}\text{Tr}\left[\mathcal{T}\Phi b_{i}\right]=\langle\eta,T\rangle$,
	where we have defined the dual Lia algebra-valued \emph{generalised
		body force density}
	\begin{equation}
		\begin{aligned}T & =T_{i}B_{i}\in\mathfrak{g}^{*}\\
			T_{i} & =\mathcal{\text{Tr}}\left[\mathcal{T}\Phi b_{i}\right].
		\end{aligned}
		\label{eq:body force in terms of G}
	\end{equation}
	This equation can be seen as transforming the generalised force in
	the spatial frame $\mathcal{T},$ into a generalised force in the
	body frame $T$, via a pull-back from $T^{*}G$ to $\mathfrak{g}^{*}$.
	That is, $\mathcal{T}\in T^{*}G$ is trivialised to a dual Lie algebra-valued
	vector $T\in\mathfrak{g}^{*}$. The generalised body force density
	transforms as a density under a change of charts, it thus follows
	follows the same transformation law as the Lagrangian Eq.~\ref{eq:L transformation}.
	
	By imposing that the full variation must vanish for all $\eta$, we
	find the momentum balance equations $\mathcal{D}_{t}^{*}S=\mathcal{D}_{\alpha}^{*}Q^{\alpha}+T$.
	Note that in Eq.~\ref{eq:body force in terms of G} we see explicitly
	that the the equations of motion are no longer expressed purely in
	terms of the intrinsic geometry of the system, as $T$ is a function
	of $\Phi$. In numerical simulations, this entails that we need to
	numerically propagate the structure field in time using Eq.~\ref{eq:generalised velocity fields},
	in addition to the generalised strain $X_{\alpha}$ and velocity $N$
	fields.
	
	Let us now illustrate the usage of Eq.~\ref{eq:body force in terms of G}
	by again considering the filament on the sphere. We write a basis
	of the Lie algebra as $\hat{b}_{i}\in\mathfrak{so}(3)$ Consider a
	configurational potential $\mathcal{U}(\mathbf{p})=a\mathbf{p}^{s}\cdot\mathbf{p}$,
	which we can rewrite as $\hat{\mathcal{U}}(R)=a\mathbf{p}^{T}R\mathbf{p}$.
	We write the structure field as an orthonormal frame $R=\left(\mathbf{e}_{1}\ \mathbf{e}_{2}\ \mathbf{e}_{3}\right)$.
	Then, we have that $\mathcal{T}=-g\mathbf{p}\mathbf{p}^{T}$, and
	$T_{i}=\text{Tr}\left[\mathcal{T}R\hat{b}_{i}\right]=-a\text{Tr}\left[\mathbf{p}\mathbf{p}^{T}R\hat{b}_{i}\right]=-a\mathbf{p}^{T}R\hat{b}_{i}\mathbf{p}=-a\mathbf{p}^{T}R(\mathbf{d}_{i}\times\mathbf{p})=a\mathbf{p}^{T}(\mathbf{p}^{s}\times\mathbf{e}_{i})$.
	We can write $\mathbf{T}=aR^{T}(\mathbf{p}\times\mathbf{p}^{s})$.
	Using the results from the previous subsection, the angular momentum
	balance equations are now $\dot{\mathbf{L}}+\boldsymbol{\Omega}\times\mathbf{L}=\partial_{u}\mathbf{M}+\boldsymbol{\pi}\times\mathbf{M}+aR^{T}(\mathbf{p}\times\mathbf{p}^{s})$.
	
	The generalised body force may be derived equivalently in terms of
	the configurational degree of freedom, as $\check{\mathcal{L}}$ may
	be considered a function of $\Phi$ and $x$ interchangeably. We can
	therefore write the variation as $\delta\check{\mathcal{L}}=\mathcal{T}_{\mu}\delta x^{\mu}+\dots$,
	where $\mathcal{T}_{\mu}=\frac{\partial\check{\mathcal{L}}}{\partial x^{\mu}}$
	and $\delta x=\rho(\text{Ad}_{\Phi}\eta)x$ as before. We find an
	alternative equivalent expression for the generalised body force density
	\begin{equation}
		\begin{aligned}T & =T_{i}B_{i}\in\mathfrak{g}^{*}\\
			T_{i} & =\mathcal{T}_{\mu}\rho(\text{Ad}_{\Phi}b_{i})_{\nu}^{\mu}x^{\mu}\\
			& =\mathcal{T}_{\mu}\Pi(\Phi)_{\nu}^{\mu}\rho(b_{i})_{\sigma}^{\nu}x_{0}^{\sigma}.
		\end{aligned}
		\label{eq:body force in terms of x}
	\end{equation}
	Let us now take a closer look at the transformation from $\mathcal{T}$
	to $T$. Recall that we may in general consider the configuration
	space as a subset of some larger space $\mathbb{X}\subseteq\mathbb{R}^{m},$where
	$m>n$. Therefore, we have that $\mathcal{T}=\frac{\partial\mathcal{U}}{\partial x}\in T^{*}\mathbb{R}^{m}$.
	That is, at a given configurational point $q\in\mathbb{X}$, the spatial
	generalised body force $\mathcal{T}\in T_{q}^{*}\mathbb{R}^{m}$ may
	be ``pointing outside'' of the cotangent plane $T_{q}^{*}\mathbb{X}$.
	However, as Eq.~\ref{eq:body force in terms of x} is a pull-back
	from $T^{*}\mathbb{R}^{m}$ to $\mathfrak{g}^{*}$, this leads to
	a generalised force $T$ that is compatible with the geometry of the
	configuration space.
	
	Considering the same example as before, we now find that $\boldsymbol{\mathcal{T}}=-\frac{\partial\mathcal{U}}{\partial\boldsymbol{p}}=-a\mathbf{p}^{s}$.
	From Eq.~\ref{eq:body force in terms of x} we have that $\boldsymbol{\mathcal{T}}^{T}R\hat{b}_{i}\mathbf{p}=\boldsymbol{\mathcal{T}}^{T}(\mathbf{e}_{i}\times\mathbf{p}^{s})=\mathbf{e}_{i}^{T}(\mathbf{p}\times\boldsymbol{\mathcal{T}})$,
	such that $\mathbf{T}=R^{T}(\mathbf{p}^{s}\times\boldsymbol{\mathcal{T}})=aR^{T}(\mathbf{p}\times\mathbf{p}^{s})$,
	which is the same expression as earlier. Comparing the expressions
	for $\boldsymbol{\mathcal{T}}$ and $\mathbf{T}$ we see that the
	latter is of a form such that it is compatible with the geometry of
	the configuration space. That is, where $\boldsymbol{\mathcal{T}}=-a\mathbf{p}$
	is a \emph{force}, $\mathbf{T}=aR^{T}(\mathbf{p}\times\mathbf{p}^{s})$
	is a \emph{moment, }in the traditional sense of classical mechanics.
	The corresponding moment in the spatial frame is then $\mathbf{T}^{s}=R\mathbf{T}=a(\mathbf{p}\times\mathbf{p}^{s})$.
	
	\subsection{Non-conservative dynamics and surface forces \label{subsec:Non-conservative-dynamics}}
	
	Energy dissipation or gain signifies the presence of non-conservative
	dynamics within a continuum system. This implies that, mathematically,
	the forces and stresses that are acting on the system do not arise
	from a scalar principle. In this section, we will generalise the results
	of the previous two subsections to include such non-conservative dynamics.
	
	In classical mechanics, Hamilton's principle serves as a specialized
	instance of the D'Alembert principle, which accommodates non-conservative
	dynamics. By analogy, we will introduce a \emph{generalized integral
		Lagrange-D'Alembert} \emph{principle} tailored to Cartan media. We
	begin by considering classical mechanics. In the framework presented
	in this paper, classical mechanical systems are point-particles, corresponding
	to a $0$-dimensional material base space $\text{dim}(M)=0$. As before,
	the configuration space is $\mathbb{X}$, and the system configuration
	is denoted by $q$ (which in this context are often called \emph{generalised
		coordinates}). The conservative dynamics of such a system is specified
	by a Lagrangian $\mathcal{L}(q,\dot{q})$, and from Hamilton's principle
	$\delta\int_{0}^{T}\mathcal{L}=0$ we find $\int_{0}^{T}\left(\frac{\partial\mathcal{L}}{\partial\dot{q}^{\mu}}\delta\dot{q}^{\mu}+\frac{\partial\mathcal{L}}{\partial q^{\mu}}\delta q^{\mu}\right)dt=0$.
	In most applications the Lagrangian is of the form $\mathcal{L}(q,\dot{q})=\mathcal{K}(\dot{q})-\mathcal{U}(q)$,
	where $\mathcal{K}$ and $\mathcal{U}$ are the kinetic and potential
	energy functions of the system. In which case, we can identify the
	momentum $p_{\mu}=\frac{\partial\mathcal{K}}{\partial\dot{q}^{\mu}}$
	and $f_{\mu}=-\frac{\partial\mathcal{U}}{\partial q^{\mu}}$ as the
	inertial and external forces acting on the system respectively. For
	a point particle, we can model non-conservative dynamics by simply
	relaxing the assumption that $f$ derives from a potential. In which
	case, we arrive at the variational principle $\int_{0}^{T}(p_{\mu}\delta\dot{q}^{\mu}-f_{\mu}\delta q^{\mu})dt=0$,
	which is the integral Lagrange-D'Alembert principle for a point particle
	\citep{marsdenIntroductionMechanicsSymmetry2013,holmEulerPoincareEquations1998}.
	This same conceptual leap can be extended to Cartan media. As we have
	already seen, the dynamics of continua is substantially different
	from that of point-particles, as the former can suffer constitutive
	stresses in addition to body forces. In the non-conservative setting,
	there is a third possible component of the dynamics. The system may
	suffer a surface force density, normal to the material surface.
	
	To construct a Lagrange-D'Alembert principle for Cartan media, we
	suspend the requirement that the dynamical fields derive from the
	gradient of a potential. That is, at the point when the generalised
	stress field $Q^{\alpha}$ and the generalised body force density
	$T$ are introduced in Eq.~\ref{eq:variation of ell 2} and Eq.~\ref{eq:body force in terms of G}
	respectively, we allow for these to be general functions of the variables
	of the kinematic configuration. Lastly, we also introduce a generalised
	surface force density $P$, defined over the material surface of the
	system. The resulting expression is
	\begin{equation}
		\begin{aligned} & \int_{W}\left\{ \langle\delta N,S\rangle-\langle\delta X_{\alpha},Q^{\alpha}\rangle+\langle\eta,T\rangle\right\} dt\wedge dV\\
			& \quad+\int_{[0,T]\times\partial M}PdS=0
		\end{aligned}
	\end{equation}
	which is a generalised Lagrange-D'Alembert principle for Cartan media.
	Here, $dS$ is a surface element induced by the pullback of the volume
	element $dV$ from $M$ to $\partial M$. The resulting equations
	of motion are
	\begin{subequations}
		\label{eq:non-conservative dynamics}
		\begin{align}
			\mathcal{D}_{t}^{*}S & =\mathcal{D}_{\alpha}^{*}Q^{\alpha}+T,\label{eq:non-conservative momentum balance equations}\\
			n_{\alpha}Q^{\alpha} & =P,\quad\text{on }\partial M,\label{eq:non-constitutive boundary conditions}
		\end{align}
	\end{subequations}
	which, together with the kinematics equations of motion Eq.~\ref{eq:kinematic equations of motion},
	defines the non-conservative mechanics of Cartan media.
	
	\subsection{Dynamics with adapted structure fields \label{subsec:Dynamics-under-adapted}}
	
	In Sec.~\ref{subsec:Adapted-frames} we considered the case when
	$r>n$. That is, when the dimension of the symmetry group $G$ is
	larger than that of the configuration space $\mathbb{X}$. The additional
	$n-r$ degrees of freedom in the kinematic description of the system
	can be seen as a gauge freedom, which is eliminated by a constructing
	an adapted structure field. As a result, $X_{\alpha}$ and $N$ will
	each suffer $n-r$ constraints on their components. That is, the kinematic
	fields will now take value in vector subsets $X_{\alpha}\in V^{\alpha}\subseteq\mathfrak{g}$
	and $N\in V^{t}\subseteq\mathfrak{g}$ respectively, where $\text{dim}(V_{\alpha})=\text{dim}(N)=n$.
	
	This in turn has implications on the dynamic conjugate variables $S$
	and $Q^{\alpha}$. Recall that kinematic adaptation can be seen as
	a way to emulate the system by kinematically constraining another
	Cartan system in a larger configuration space, with the same shared
	symmetry group $G$ that accomodates for all the degrees of freedom
	of $G$. This entails that the generalised momentum balance equations
	must be consistent with these kinematic constraints. That is, the
	generalised momentum $S$ must have the same degrees of freedom as
	that of the generalised velocity $N$, and the generalised stresses
	must be such that they enforce the kinematic constraint.
	
	Continuing the discussion from Sec.~\ref{subsec:Adapted-frames},
	let us again consider a spatially adapted structure field. Equation
	\ref{eq:adapted frames constraints} are $r-n$ constraints on $N$.
	Let $N_{i},\ i=1,\dots,n$ designate independent degrees of freedom
	of the velocity, and $N_{j}=N_{j}(X),\ j=n+1,\dots,r$ be components
	of the velocity determined by Eq.~\ref{eq:adapted frames constraints},
	such that $N=N_{i}\tilde{b}_{i}+N_{j}\tilde{b}_{j}=N_{k}\tilde{b}_{k},\ k=1,\dots,r$,
	where $\tilde{b}_{i}$ is a basis for $\mathfrak{g}$ that reflects
	this separation. Let $\tilde{B}_{i}$ be the corresponding dual basis
	for $\mathfrak{g}^{*}$, and we write $S=S_{k}\tilde{B}_{k}$. Then,
	as the independent degrees of freedom in the velocity are $N_{i}$,
	the components of the generalised momentum in this basis are $S_{i}=\frac{\partial\check{\mathcal{L}}}{\partial N_{i}}$
	and $S_{j}=0$. From Eq.~\ref{eq:non-conservative momentum balance equations}
	we then find that $(\mathcal{D}_{\alpha}^{*}Q^{\alpha}+T)_{j}=0$,
	which are $r-n$ constraints on the generalised stress, where the
	subscript denotes the components in the dual basis $\tilde{B}_{j}$.
	
	We now return to the example of the filament on the sphere, discussed
	in Sec.~\ref{subsec:Adapted-frames}. As $\Omega_{2}$ is not a dynamical
	degree of freedom of the filament, we have that $L_{2}=0$ of the
	angular momentum (or, in the general setting, the generalised momentum).
	From Eq.~\ref{eq:non-conservative dynamics} we then find that $M_{1}=(L_{1}\Omega_{3}-L_{3}\Omega_{1}-M_{2}'-m_{2})/\kappa$.
	We thus see that kinematic adaptation introduces a fictitious moment
	$\mathbf{M}^{f}=M_{1}\mathbf{e}_{1}$. We can interpret the fictitious
	moment as acting on the ``Cosserat rod'' on the sphere so as to
	align $\mathbf{e}_{1}$ to be tangent with its center-line, and thus
	effectively reducing its degrees of freedom to that of a filament
	on a sphere.
	
	\section{Geometric integration}
	
	We have derived the geometric mechanics of Cartan media, encapsulated
	in Eq.~\ref{sec:geometric kinematics} and Eq.~\ref{eq:non-conservative dynamics}.
	The temporal evolution of a system under the influence of given generalised
	stresses and body force densities, given initial boundary conditions,
	can be found by integration of these equations of motion. Here we
	derive integrators to this end.
	
	Given an initial boundary condition $q_{0}:M\to\mathbb{X}$, and a
	corresponding structure field $\Phi_{0}:M\to G$ that satisfies $q_{0}=\Phi_{0}\cdot q_{r}$,
	we have that the structure generator at the initial temporal boundary
	is $\xi_{M,0}=\Phi_{0}^{-1}d_{M}\Phi_{0}$. As the equations of motions
	are of second order, we must also specify an initial velocity $\dot{\Phi}_{0}:M\to TG$
	(the dot does not denote a derivative here). Given a chart $(U,\mathbf{u})\in\mathcal{A}_{M}$,
	we have that the initial generalised strain and generalised velocity
	are $X_{\alpha,0}=\Phi_{0}^{-1}\partial_{\alpha}\Phi_{0}$ and $N_{0}=\Phi^{-1}\dot{\Phi}_{0}$.
	Then, over this chart, the structure generator can be solved for by
	integrating
	
	\begin{subequations}
		\label{eq:non-const and phi}
		\begin{align}
			\dot{X}_{\alpha} & =\mathcal{D}_{u}N\label{eq:X eq}\\
			\mathcal{D}_{t}^{*}S & =\mathcal{D}_{\alpha}^{*}Q^{\alpha}+T,\label{eq:S eq}\\
			\dot{\Phi} & =\Phi N,\label{eq:Phi N eq}
		\end{align}
	\end{subequations}
	with initial conditions $X_{\alpha}(0,\mathbf{u})=X_{\alpha,0}(\mathbf{u})$
	and $N(0,\mathbf{u})=N_{0}(\mathbf{u})$, over a patchwork of charts
	that collectively cover $M$. The generalised stress must satisfy
	$Q^{\alpha}n_{\alpha}=P$ at the material boundary $\partial M$,
	where $n$ is a normal covector field on $\partial M$. If the stress
	is a function of the strain $Q^{\alpha}=Q^{\alpha}(X_{1},X_{2},\dots,X_{d})$,
	then these boundary conditions will in turn induce conditions on $X_{\alpha}$
	at the material boundaries. Note that in many relevant applications,
	including $M=S^{2}$ (see the methodological note at the end of Sec.~\ref{subsec:Surfaces}),
	it is possible to simulate systems with a single chart.
	
	The solution of Eq.~\ref{eq:S eq}-\ref{eq:S eq} yields the structure
	generator $\xi:W\to\mathfrak{g}$, and Eq.~\ref{eq:Phi N eq} yields
	the structure field $\Phi:W\to G$, from which the spatio-temporal
	configuration $q:W\to\mathbb{X}$ can be found using Eq.~\ref{eq:x from phi}.
	It is only necessary to integrate Eq.~\ref{eq:Phi N eq} concurrently
	with Eq.~\ref{eq:S eq}-\ref{eq:S eq} if the dynamics couples directly
	with $\Phi$ (or $q$). That is, if the system is purely constitutive
	then Eq.~\ref{eq:S eq}-\ref{eq:S eq} can be solved on its own.
	Equation \ref{eq:Phi N eq} can then be solved separately in order
	to reconstruct the structure field.
	
	Naively, the simplest method by which to numerically solve Eq.~\ref{eq:non-const and phi}
	is using the \emph{Forward-Euler integrator} (FEI) \citep{ascherComputerMethodsOrdinary1998},
	which assumes that the right-hand sides are approximately constant
	over the interval $\Delta t$ between time steps. This effectively
	linearises the equations of motion, such that, for example, the propagator
	of Eq.~\ref{eq:Phi N eq} is $\Phi(t+\Delta t,p)\approx\Phi(t,p)+\Phi(t,p)N(t,p)\Delta t$,
	evaluated point-wise at $p\in M$, where the numerical error grows
	as $O(\Delta t^{2})$. A significant drawback of the FEI, and similar
	but more sophisticated methods like Runge-Kutta schemes \citep{ascherComputerMethodsOrdinary1998},
	is that $\Phi(t+\Delta t,p)\not\in G$ in general, for any finite
	$\Delta t$. That is, numerical errors pushes $\Phi(t+\Delta t,p)$
	out of the Lie group. Fundamentally, this is due to how Lie group
	elements are parameterised as matrices $G\subset\mathbb{R}^{m\times m}$,
	such that $T_{g}G\subset T_{g}\mathbb{R}^{m\times m}$ at any point
	$g\in G$. This means that $\Phi(t,p)N(t,p)\Delta t\in T_{\Phi(t,p)}\mathbb{R}^{m\times m}$,
	such that the sum $\Phi(t,p)+\Phi(t,p)N(t,p)\Delta t$ is only an
	element of $G$ in general if $\Delta t=0$. To circumvent this problem,
	we may instead solve the discretised equation over the time-step.
	The result is known as a \emph{Lie group integrator} (LGI) \citep{iserlesLiegroupMethods2005}.
	In the case of Eq.~\ref{eq:Phi N eq}, the solution is $\Phi(t+\Delta t,p)\approx\Phi(t,p)\exp_{G}(N(t,p)\Delta t)$,
	where $\exp_{G}:\mathfrak{g}\to G$ is the expontential map. The propagator
	ensures that $\Phi$ remains in $G$ regardless of $\Delta t$. That
	is, the error in the numerical scheme, which remains $O(\Delta t^{2})$,
	take value in $G$, and not in $\mathbb{R}^{m\times m}$.
	
	Here we will derive LGIs for the mechanics of Cartan media. We begin
	with Eq.~\ref{eq:Phi N eq}, which outlines the usage of the above
	example in more explicit detail. We then move on to Eq.~\ref{eq:S eq}-\ref{eq:S eq},
	where we derive semi-analytical forms of the LGI in terms of the Lie
	group operations $\exp_{G}$ and $\text{Ad}$. The results we present
	here can be developed further to construct propagators with better
	error scaling; for example, Runge--Kutta--Munthe--Kaas methods
	\citep{munthe-kaasHighOrderRungeKutta1999,buddGeometricIntegrationNumerical1999,engoNumericalIntegrationLie2001}
	can be used, which incorporates Runge-Kutta methods into LGIs.
	
	\subsection{Spatio-temporal reconstruction \label{subsec:Spatio-temporal reconstruction}}
	
	For a given structure generator $\xi$, the frame field is given by
	solving the reconstruction equation $d\Phi=\Phi\xi$, and by specifying
	some point $\Phi(t_{0},u_{0})=\Phi_{0}$. Formally, the frame field
	at $(t,p)\in W$ can be found by integrating the reconstruction equation
	over some curve $\gamma:[0,1]\to W$ that satisfies $\gamma(0)=(t_{0},u_{0})$
	and $\gamma(1)=(t,p)$. That is, $\Phi(\gamma(\alpha))=\Phi(t_{0},p_{0})\mathscr{A}\exp\left\{ \int_{0}^{1}\xi(\gamma(\alpha))\right\} $,
	which reconstructs the frame along $\gamma$, and where $\mathscr{A}$
	signifies a $\alpha$-ordered integral. The single-valuedness of $\Phi(\gamma(1))=\Phi(t,p)$
	under any choice of $\gamma$ is ensured by the integrability conditions
	Eq.~\ref{eq:xi integrability condition}. Numerically, we may approximate
	this integration by discretising the path along the parameter $\alpha$.
	We therefore see that there are in principle an infinite amount of
	ways $\Phi$ can be numerically reconstructed, by the repeated integration
	of $\xi$ along a set of curves. Below we will construct such an integration
	scheme.
	
	Let $\Phi_{0}:M\to SE(3)$ be the structure field of the initial conditions
	of the system. We discretise time uniformly as $t=k\Delta t$, where
	$k=1,\dots,n_{t}$ and $n_{t}=T/\Delta t$. Given a generalised velocity
	$N:W\to\mathfrak{se}(3)$, the numerical integration scheme is
	\begin{equation}
		\Phi^{k+1}(p)=\Phi^{k}(p)\exp\left(N(i\Delta t,p)\Delta t\right),
	\end{equation}
	with initial conditions $\Phi^{0}(p)=\Phi_{0}(p)$, such that $\Phi(i\Delta t,p)\approx\Phi^{k}(p)$,
	and where $k=1,\dots,n_{t}$, and where $\exp:\mathfrak{g}\to G$
	is the expontential map.
	
	\subsection{Geometric integrators for dynamical motion}
	
	The form of the mechanical equations of motion belie a structure that
	is amenable to geometric Lie group integration. To see this, we can
	insert the definition of $\mathcal{D}_{u}$ into Eq.~\ref{eq:X eq},
	and rewrite the kinematic equation of motion as $\dot{X}=V-\text{ad}_{N}X$,
	where $V=N'$ . Similarly, by inserting the definitions of $\mathcal{D}_{t}^{*}$
	and $\mathcal{D}_{u}^{*}$ into Eq.~\ref{eq:S eq}, we can rewrite
	the dynamical equation of motion as $\dot{S}=W-\text{ad}_{N}^{*}S$,
	where $W=\mathcal{D}_{u}^{*}Q+T$. Therefore, we can see that the
	equations of motion are in terms of translations ($V$ and $W$) and
	adjoint actions ($-\text{ad}_{N}$ and $-\text{ad}_{N}^{*}$) on $S$
	and $X$ respectively. Let $\tau_{V}:\mathfrak{se}(3)\to\mathfrak{se}(3)$
	and $\tau_{W}^{*}:\mathfrak{se}(3)^{*}\to\mathfrak{se}(3)^{*}$ be
	operators defined such that $\tau_{V}X=V+X$ and $\tau_{W}^{*}S=W+S$.
	Then we have that $\dot{X}=(\tau_{V}-\text{ad}_{N})X$ and $\dot{S}=(\tau_{W}^{*}-\text{ad}_{N}^{*})S$.
	For short times $\Delta t$, we can approximate as these operators
	as constant over the interval $[t,t+\Delta t]$. Formally, we can
	then construct short-time propagators as
	\begin{subequations}
		\label{eq:mechanical LGIs-1-1}
		\begin{align}
			X(t+\Delta t,p) & \approx e^{\Delta t(\tau_{V}-\text{ad}_{N})}X(t,p),\label{eq:X LGI-1-1}\\
			S(t+\Delta t,p) & \approx e^{\Delta t(\tau_{W}^{*}-\text{ad}_{N}^{*})}S(t,p),\label{eq:S LGI-1-1}
		\end{align}
	\end{subequations}
	which would be evaluated point-wise over $p\in M$. The right-hand
	sides equate to \citep{haleOrdinaryDifferentialEquations2009} \begin{widetext}
		
		\begin{subequations}
			\label{eq:mechanical LGIs-1-1-2}
			\begin{align}
				e^{\Delta t(\tau_{V}-\text{ad}_{N})}X & =e^{-\Delta t\ \text{ad}_{N}}X+\left(\int_{0}^{\Delta t}e^{-(\Delta t-\Delta t')\text{ad}_{N}}d\Delta t'\right)V,\label{eq:X LGI-1-1-1}\\
				e^{\Delta t(\tau_{W}^{*}-\text{ad}_{N}^{*})}S & =e^{-\Delta t\ \text{ad}_{N}^{*}}S+\left(\int_{0}^{\Delta t}e^{-(\Delta t-\Delta t')\text{ad}_{N}^{*}}d\Delta t'\right)W.\label{eq:S LGI-1-1-2}
			\end{align}
		\end{subequations}
	\end{widetext}Using the identities \citep{rossmannLieGroupsIntroduction2006}
	$\text{Ad}_{e^{Y}}=e^{\text{ad}_{Y}}$ and $\text{Ad}_{e^{Y}}^{*}=e^{-\text{ad}_{Y}^{*}}$
	we can succintly write the Lie group integrators of Eq.~\ref{eq:X eq}
	and Eq.~\ref{eq:S eq} as\begin{widetext}
		\begin{subequations}
			\label{eq:mechanical LGIs-1}
			\begin{align}
				X(t+\Delta t,p) & \approx\text{Ad}_{e^{-\Delta tN}}X(t,p)+\mathscr{E}(\Delta t,-N)V,\label{eq:X LGI-1}\\
				S(t+\Delta t,p) & \approx\text{Ad}_{e^{\Delta tN}}^{*}S(t,p)+\mathscr{E}^{*}(\Delta t,N)W,\label{eq:S LGI-1}
			\end{align}
		\end{subequations}
	\end{widetext}where we have defined the maps $\mathscr{E}:\mathbb{R}_{+}\times\mathfrak{g}\times\mathfrak{g}\to\mathfrak{g}$
	and $\mathscr{E}:\mathbb{R}_{+}\times\mathfrak{g}\times\mathfrak{g}^{*}\to\mathfrak{g}^{*}$
	as $\mathscr{E}(\Delta t,Y)=\int_{0}^{\Delta t}\text{Ad}_{e^{(\Delta t-\Delta t')Y}}d\Delta t'$
	and $\mathscr{E}^{*}(\Delta t,A)=\int_{0}^{\Delta t}\text{Ad}_{e^{Y(\Delta t-\Delta t')}}^{*}d\Delta t'$,
	which acts as $\mathscr{E}(\Delta t,Y)C$ and $\mathscr{E}^{*}(\Delta t,Y)Z$
	for any $\Delta t\in\mathbb{R}_{+}$, $Y,C\in\mathfrak{g}$ and $Z\in\mathfrak{g}^{*}$.
	It should be noted that the LGI for the dynamics can be mapped to
	that of the kinematics, meaning that Eq.~\ref{eq:X LGI-1} and Eq.~\ref{eq:S LGI-1}
	do not need to be implemented separately in numerics. The mapping
	can be found by taking the transpose of the latter, leading to $S(t+\Delta t,u)^{T}\approx\text{Ad}_{e^{N\Delta t}}S(t,u)^{T}+\mathscr{E}(\Delta t,N)W^{T}$.
	We thus see that only $\text{Ad}_{e^{Y}}$ and $\mathscr{E}(\Delta t,Y)$
	need to be implemented numerically.
	
	\section{Cosserat media \label{sec:cosserat-as-cartan-media}}
	
	Cosserat media, itself a generalisation of classical continuum mechanics,
	can be seen as a specific instance of Cartan media. In full generality,
	the configuration at a material point $p\in M$ is specified by a
	displacement $\mathbf{r}(p)\in\mathbb{E}^{3}$, the external configuration,
	and a set of directors $\mathbf{v}_{i}(p)\in T\mathbb{E}^{3},\ i=1,\dots,s$,
	the internal configuration, where $s$ is the number of directors.
	The directors are fully independent degrees of freedom, and as such
	the symmetry group of the internal configuration is the general linear
	group $GL(3)$ (or, if $s>3$, products of $GL(3)$). In many applications
	\citep{rendaDynamicModelMultibending2014,boyerMacrocontinuousDynamicsHyperredundant2012,sackBiologicalTissueMechanics2016,graziosoGeometricallyExactModel2019,obrezkovMicropolarBeamlikeStructures2022,aydinNeuromuscularActuationBiohybrid2019,naughtonElasticaCompliantMechanics2021,neffGeometricallyExactCosserat2004,simoStressResultantGeometrically1989,ericksenLiquidCrystalsCosserat1974,krishnaswamyCosserattypeModelRed1996,rangamaniSmallScaleMembrane2014,stefanouThreedimensionalCosseratHomogenization2008,jogHigherorderShellElements2004,dongCosseratInterphaseModels2014,rubinCosseratShellModel2004},
	it suffices to consider inextensible and and orthonormal directors,
	such that the collective symmetry group of the configuration space
	is $SE(3)$. Here we apply the theoretical framework we developed
	in previous sections to Cosserat systems of this kind, in $d=\text{dim}(M)=3$
	(Cosserat solid) $d=2$ (Cosserat surface), and $d=1$ (Cosserat rod)
	material dimensions. In all three examples, the discussion will follow
	the same general sequence: After contextualising the model in the
	broader physics literature, we define the material base space, configuration
	space and symmetry group, and then construct a structure field that
	reconstructs the spatio-temporal configuration. This allows us to
	define the generalised strain and velocity fields of the system, and
	we use Eq.~\ref{eq:dx} to show their relation to the derivatives
	of the spatio-temporal configuration. We also briefly contextualise
	the generalised strain and velocity fields for the system in related
	concepts within differential geometry. The kinematics and dynamics
	of the system follow as a straightforward application of the machinery
	developed in Sec.~\ref{sec:geometric kinematics} and Sec.~\ref{sec:geometric dynamics},
	and we find the physical dimensions of all quantities involved using
	dimensional analysis. As the mechanics of the three systems share
	many similarities, we will abbreviate the exposition where possible
	to avoid repetition.
	
	\subsection{Cosserat solids \label{sec:cosserat 3d bodies}}
	
	A Cosserat solid can be considered a continuum solid, where the point-continua
	possess micropolar degrees of freedom. Here, we consider the case
	where the micropolarity is described by a triad of orthonormal directors.
	Such systems have been used to model, for example, liquid crystals
	\citep{epsteinContinuousDistributionsInhomogeneities2001,leeContinuumTheorySmectic2003},
	electromagnetic and ferromagnetic media \citep{ivanovaNewTheoryCosserat2022,pariaUnifiedTheoryMechanics1978,ivanovaModelingPhysicalFields,koteraCosseratContinuumTheory2000},
	biological materials \citep{onckCosseratModelingCellular2002}, ceramics
	\citep{stefanouThreedimensionalCosseratHomogenization2008}, rocks
	and granular media \citep{besdoInelasticBehaviourPlane1985,besdoNumericalCosseratApproachPredicting1991,ebrahimianNumericalStudyInterface2021,mohanFrictionalCosseratModel1999},
	and other applications in material science \citep{stefanouCosseratApproachLocalization2017,stefanouThreedimensionalCosseratHomogenization2008,iesanDeformationPorousCosserat2011,forestCosseratModellingSize2000}.
	
	The Cosserat solid is an instance of a Cartan system in three material
	dimensions $d=3$, with configuration space $\mathbb{X}=\mathcal{F}(\mathbb{E}^{3})$
	and symmetry group $G=SE(3)$. The configurations of the point continua
	of the system comprises a displacement $\mathbf{r}\in\mathbb{E}^{3}$,
	and a micropolar degree of freedom represented as an orthonormal triad
	of directors $E=(\mathbf{e}_{1}\ \mathbf{e}_{2}\ \mathbf{e}_{3}),\ \mathbf{e}_{i}\in T\mathbb{E}^{3}$.
	The material base space is a connected and bounded subset $M\subset\mathbb{E}^{3}$,
	equipped with coordinates $(u,v,w):M\to\mathbb{R}^{3}$. Configurations
	and group elements will be represented as matrices
	\begin{equation}
		\left(\begin{array}{cc}
			1 & \mathbf{0}^{T}\\
			\mathbf{f} & F
		\end{array}\right)\in\mathcal{F}(\mathbb{E}^{3}),\ \left(\begin{array}{cc}
			1 & \mathbf{0}^{T}\\
			\mathbf{a} & A
		\end{array}\right)\in SE(3),\label{eq:matrix rep of SE(3)}
	\end{equation}
	where $A\in SO(3)$, and in short-hand we write $(\mathbf{f},F)\in\mathcal{F}(\mathbb{E}^{3})$
	and $(\mathbf{a};A)\in SE(3)$. The group action can then be written
	as a matrix multiplication $(\mathbf{a};A)\cdot(\mathbf{f},F)=(\mathbf{a};A)(\mathbf{f},F)=(A\mathbf{f}+\mathbf{a},AF)$.
	The spatio-temporal configuration is $q=(\mathbf{r},E):W\to\mathcal{F}(\mathbb{E}^{3})$.
	Let $q_{r}=(\mathbf{0},D)$, where $D$ is an orthonormal triad of
	fixed basis vectors, and let $R:W\to SO(3)$ satisfy $RD=E$. Then
	$\Phi=(\mathbf{r};R):W\to SE(3)$ is a structure field satisfying
	$q=\Phi q_{r}$. For the sake of simplicity, we will let $D=\mathbbm{1}_{3\times3}$,
	such that $q_{r}=\mathbbm{1}_{4\times4}$, implying that $R=E$, and
	so effectively identifying $\Phi$ and $q$. However, we will still
	distinguish between $E$ and $R$, the former referring to the micropolar
	degree of freedom of the Cosserat solid, and the latter to the rotation
	between the stationary basis $D$ and the moving basis of $E$. Such
	that $\mathbf{v}^{s}=R\mathbf{v}=v_{i}\mathbf{e}_{i}$, relating tangent
	vectors in the spatial and body frames of reference. As we have noted
	in previous examples, $q_{r}=(\mathbf{0},D)$ can be seen as the observed
	position and orientation of a material point $p\in M$ at time $t\in[0,T]$,
	of an observer that is located at $\mathbf{r}(t,p)$ and in a reference
	frame $E(t,p)$.
	
	We write the generalised velocity and strain fields as $N=\Phi^{-1}\dot{\Phi}=\{\mathbf{V};\boldsymbol{\Omega}\}$
	and $X_{\alpha}=\Phi^{-1}\partial_{\alpha}\Phi=\{\boldsymbol{\theta}_{\alpha},\boldsymbol{\pi}_{\alpha}\},\ \alpha=u,v,w$,
	where $\mathbf{V},\boldsymbol{\Omega},\boldsymbol{\theta}_{\alpha},\boldsymbol{\pi}_{\alpha}:W\to T\mathbb{E}^{3}$,
	and where we have introduced the short-hand for the matrix representation
	of $\mathfrak{se}(3)$
	\begin{equation}
		\{\mathbf{a};\mathbf{b}\}:=\left(\begin{array}{cc}
			0 & \mathbf{0}^{T}\\
			\mathbf{a} & \hat{b}
		\end{array}\right)\in\mathfrak{se}(3),
	\end{equation}
	for any $\mathbf{a},\mathbf{b}\in\mathbb{R}^{3}$, where the hat map
	is defined in Eq.~\ref{eq:hat map}. The translational and rotational
	components of $N$ and $X_{\alpha}$ can be contextualised in terms
	of more familiar expressions using Eq.~\ref{eq:dx}. We have that
	$\dot{q}=(\text{Ad}_{\Phi}N)q$, from which we find that $\dot{\mathbf{r}}=\mathbf{V}^{s}=V_{i}\mathbf{e}_{i}$
	and $\dot{\mathbf{e}}_{i}=\boldsymbol{\Omega}^{s}\times\mathbf{e}_{i}=(\Omega_{i}\mathbf{e}_{i})\times\mathbf{e}_{i}=\mathbf{e}_{j}\hat{\Omega}_{ji}$.
	We see that $\mathbf{V}$ and $\boldsymbol{\Omega}$ is a velocity
	and angular velocity on the displacement field $\mathbf{r}$ and the
	director frame $E$ respectively. Along the material coordinates,
	we have $\partial_{\alpha}q=(\text{Ad}_{\Phi}X_{\alpha})q$, from
	which we find $\partial_{\alpha}\mathbf{r}=\boldsymbol{\theta}^{s}=\theta_{\alpha,i}\mathbf{e}_{i}$
	and $\partial_{\alpha}\mathbf{e}_{i}=\boldsymbol{\pi}_{\alpha}^{s}\times\mathbf{e}_{i}=\mathbf{e}_{j}\hat{\pi}_{\alpha,ji}$.
	We see that $\boldsymbol{\theta}_{\alpha}$ and $\boldsymbol{\pi}_{\alpha}$
	encode the strains of $\mathbf{r}$ and $E$ respectively. We should
	emphasise that $\mathbf{V}$, $\boldsymbol{\Omega}$, $\boldsymbol{\theta}_{\alpha}$
	and $\boldsymbol{\pi}_{\alpha}$ are all vector fields in the body
	frame of reference, and comprise therefore the intrinsic spatio-temporal
	geometry of the Cosserat solid.
	
	The strains of the displacement and director fields, $\boldsymbol{\theta}_{\alpha}$
	and $\boldsymbol{\pi}_{\alpha}$, can be related to corresponding
	objects in Riemannian geometry. At a given time $t$, the spatial
	configuration of the Cosserat solid can be seen as a manifold $\mathcal{R}_{t}=\mathbf{r}(t,M)\subset\mathbb{E}^{3}$
	and vector fields $\mathbf{e}_{i}:W\to T\mathcal{R}_{t},\ i=1,2,3$.
	As $d_{M}\mathbf{r}=\boldsymbol{\theta}_{\alpha}^{s}du^{\alpha}$,
	we have that $d_{M}\mathbf{r}\cdot d_{M}\mathbf{r}=g_{\alpha\beta}du^{\alpha}du^{\beta}$,
	where $g_{\alpha\beta}=\boldsymbol{\theta}_{\alpha}^{T}\boldsymbol{\theta}_{\beta}$
	is a time-dependent Riemannian metric on $\mathcal{R}_{t}$. Furthermore,
	the \textit{Christoffel symbols} \citep{misnerGravitation2017} are
	defined as $\partial_{\alpha}\mathbf{e}_{j}=\Gamma_{j\alpha}^{i}\mathbf{e}_{i}$,
	we can thus identify $\hat{\pi}_{\alpha,ij}=\Gamma_{j\alpha}^{i}$,
	and $D_{\alpha}$ as the covariant derivative on $\mathbf{r}(t,M)$.
	If we extend these same arguments to include time, now using the exterior
	derivative $d$ rather than $d_{M}$, we can conclude that $D_{t}$
	is a covariant derivative on the spatio-temporal manifold $\mathbf{r}(W)$
	in the curvi-linear coordinates defined by the orthonormal frame field
	$E$.
	
	As for all Cartan media, the geometrised kinematics of the Cosserat
	solid is given by Eq.~\ref{eq:kinematic equations of motion}. Substituting
	$N=\{\mathbf{V};\boldsymbol{\Omega}\}$ and $X_{\alpha}=\{\boldsymbol{\theta}_{\alpha},\boldsymbol{\pi}_{\alpha}\}$
	into the equations of motion, we find
	\begin{subequations}
		\label{eq:cosserat body kinematics eom}
		\begin{align}
			D_{t}\boldsymbol{\theta}_{\alpha} & =D_{\alpha}\mathbf{V}\\
			\dot{\boldsymbol{\pi}}_{\alpha} & =D_{\alpha}\boldsymbol{\Omega}
		\end{align}
	\end{subequations}
	for $\alpha=u,v,w$, where $D_{t}=\partial_{t}+\hat{\Omega}$ and
	$D_{\alpha}=\partial_{\alpha}+\hat{\pi}_{\alpha}$. The differential
	operators $D_{t}$ and $D_{\alpha}$ should be interpreted as covariant
	derivatives on the manifold $\mathbf{r}(W)$, which we will show below.
	The spatial integrability conditions Eq.~\ref{eq:spatial integrability conditions}
	yields $D_{\alpha}\boldsymbol{\theta}_{\beta}=D_{\beta}\boldsymbol{\theta}_{\alpha}$
	and $\partial_{\alpha}\boldsymbol{\pi}_{\beta}=D_{\beta}\boldsymbol{\pi}_{\alpha}$,
	for $\alpha,\beta=u,v,w$, and must be satisfied at all times $t\in[0,T]$.
	However, as was shown explicitly in Sec.~\ref{subsec:Spatial-integrability},
	it suffices that they hold at $t=0$ as the kinematic equations of
	motion preserves spatial integrability.
	
	Given a kinetic energy density $\mathcal{K}(N)$ density, we write
	the generalised momentum as $S=\frac{\partial\mathcal{K}}{\partial N}=\{\mathbf{P};\mathbf{L}\}^{*}$,
	where the matrix derivative is taken using the numerator-layout convention
	(see Eq.~\ref{eq:numerator-layout convention}), and where we introduced
	a short-hand for the matrix representation of $\mathfrak{se}(3)^{*}$
	as $\{\mathbf{y};\mathbf{z}\}^{*}=\{\mathbf{y};\mathbf{z}\}^{T}\in\mathfrak{se}(3)^{*}$,
	for any $\mathbf{y},\mathbf{z}\in\mathbb{R}^{3}$. We write the generalised
	stress and body force density as $Q^{\alpha}=\{\mathbf{F}^{\alpha};\mathbf{M}^{\alpha}\}^{*}$
	and $T=\{\mathbf{f};\mathbf{m}\}^{*}$, where $\mathbf{F}^{\alpha},\mathbf{M}^{\alpha},\mathbf{f},\mathbf{m}:W\to T\mathbb{E}^{3},\ \alpha=u,v,w$,
	Substituting these expressions into the generalised momentum balance
	equations Eq.~\ref{subsec:Spatial-integrability}, we find
	\begin{subequations}
		\label{eq:cosserat body dynamics}
		\begin{align}
			D_{t}\mathbf{P} & =D_{\alpha}\mathbf{F}^{\alpha}+\mathbf{f},\label{eq:momentum balance cosserat body}\\
			D_{t}\mathbf{L} & =D_{\alpha}\mathbf{M}^{\alpha}+\boldsymbol{\theta}_{\alpha}\times\mathbf{F}^{\alpha}+\mathbf{m}.\label{eq:angular momentum balance cosserat body}
		\end{align}
	\end{subequations}
	and $n_{\alpha}\mathbf{F^{\alpha}}=n_{\alpha}\mathbf{M}^{\alpha}=0$
	on $\partial M$. Given conservative dynamics specified by a Lagrangian
	density $\mathcal{L}(q,\dot{q},\partial_{\alpha}q)$, with corresponding
	left-trivialisation $\check{\mathcal{L}}=\mathcal{K}-\mathcal{U}$,
	the components of the generalised stress $Q^{\alpha}=\{\mathbf{F}^{\alpha};\mathbf{M}^{\alpha}\}^{*}$
	can be found as $\mathbf{F}^{\alpha}=-\frac{\partial\check{\mathcal{L}}}{\partial\boldsymbol{\theta}_{\alpha}}=\frac{\partial\mathcal{U}}{\partial\boldsymbol{\theta}_{\alpha}}$
	and $\mathbf{M}^{\alpha}=-\frac{\partial\check{\mathcal{L}}}{\partial\boldsymbol{\pi}_{\alpha}}=\frac{\partial\mathcal{U}}{\partial\boldsymbol{\pi}_{\alpha}}$.
	An expression for the components of generalised body force density
	$T=\{\mathbf{f};\mathbf{m}\}^{*}$ can be found by evaluating Eq.~\ref{eq:body force in terms of G},
	to find
	\begin{subequations}
		\label{eq:cosserat body,  body force}
		\begin{align}
			\mathbf{f} & =R^{T}\frac{\partial\mathcal{L}}{\partial\mathbf{r}},\\
			\mathbf{m} & =R^{T}\sum_{i=1}^{3}\frac{\partial\mathcal{L}}{\partial\mathbf{e}_{i}}\times\mathbf{e}_{i}.
		\end{align}
	\end{subequations}
	Equation \ref{eq:cosserat body dynamics} and Eq.~\ref{eq:cosserat body dynamics}
	together fully define the mechanics of Cosserat solids, given that
	the map $S=S(N)$ is invertible.
	
	Physical interprations of the components of the generalised strain,
	velocity, momentum, stress and body force fields can be found using
	dimensional analysis. Let $L$, $T$ and $M$ refer to the dimensions
	of material length, time and mass respectively. For all Cartan media,
	the dimensions of all quantities can be inferred from Eq.~\ref{eq:dx},
	given the dimensions of the spatio-temporal configuration $q=(\mathbf{r},E)$
	and its derivatives, as well as the dimensions of the kinetic energy
	density $\mathcal{K}(N)$. We thus let $[\mathbf{r}]=L$, $[E]=1$,
	$[\partial_{\alpha}]=L^{-1}$ and $[\partial_{t}]=T^{-1}$. Now, as
	$\dot{\mathbf{r}}=V_{i}\mathbf{e}_{i}$, for example, we find that
	$[\mathbf{V}]=LT^{-1}$, as is expected of a velocity. Continuing
	in the same vein, we find that $[\boldsymbol{\theta}_{\alpha}]=1$,
	$[\boldsymbol{\Omega}]=\mathrm{T}^{-1}$ and $[\boldsymbol{\pi}_{\alpha}]=\mathrm{L}^{-1}$.
	Now, let the kinetic energy density have units of energy per unit
	material volume $[\mathcal{K}(N)]=ML^{-1}T^{-2}$. From $\mathbf{P}=\frac{\partial\mathcal{K}}{\partial\mathbf{V}}$
	and $\mathbf{L}=\frac{\partial\mathcal{K}}{\partial\boldsymbol{\Omega}}$
	we find that $[\mathbf{P}]=MLT^{-1}$ and $[\vec{L}]=ML^{-1}T^{-1}$.
	From Eq.~\ref{eq:cosserat body dynamics} we find that $[\mathbf{F}^{\alpha}]=ML^{-1}T^{-2}$,
	$[\mathbf{M}^{\alpha}]=MT^{-2}$, $[\mathbf{f}]=ML^{-2}T^{-2}$ and
	$[\mathbf{m}]=ML^{-1}T^{-2}$. We thus see that $\mathbf{P}$ and
	$\mathbf{L}$ have units of momentum and angular momentum per unit
	material volume respectively, and $\mathbf{F}^{\alpha}$ and $\mathbf{M}^{\alpha}$
	force and moment per unit material area respectively. Finally, if
	we assume a form $\mathcal{K}(N)=\frac{1}{2}\rho_{0}^{V}|\mathbf{V}|^{2}+\frac{1}{2}\boldsymbol{\Omega}^{T}\mathbb{I}\boldsymbol{\Omega}$,
	where $\rho_{0}^{V}$ is the mass volume density of the body and $\mathbb{I}\in\mathbb{R}^{3\times3}$
	the moment of inertia of the director frame, then $\mathbf{P}=\frac{\partial\mathcal{K}}{\partial\mathbf{V}}=\rho_{0}^{V}\mathbf{V}$
	and $\mathbf{L}=\mathbb{I}\boldsymbol{\Omega}$. We thus interpret
	$\mathbf{P}$, $\mathbf{F}^{\alpha}$ and $\mathbf{f}$ respectively
	as the linear momentum, stress and body force density on the displacement
	field $\mathbf{r}$. Similarly, $\mathbf{L}$, $\mathbf{M}^{\alpha}$
	and $\mathbf{m}$ are respectively the angular momentum, moment-stress
	and moment density on the director frame $E$.
	
	We now relate the mechanics of Cosserat solids to that of non-micropolar
	continuum mechanics. As $\dot{\mathbf{v}}^{s}=RD_{t}\mathbf{v}$ and
	$\partial_{\alpha}\mathbf{v}^{s}=RD_{\alpha}\mathbf{v}$ for any vector
	field $\mathbf{v}:W\to T\mathbb{E}^{3}$, we find the momentum balance
	equations in the spatial frame of reference $\dot{\mathbf{P}}^{s}=\partial_{\alpha}\mathbf{F}^{s,\alpha}+\mathbf{f}^{s}$
	and $\dot{\mathbf{L}}^{s}=\partial_{\alpha}\mathbf{M}^{s,\alpha}+\partial_{\alpha}\mathbf{r}\times\mathbf{F}^{s,\alpha}+\mathbf{m}^{s}$,
	which are consistent with the literature \citep{rubinCosseratTheoriesShells2000,altenbachCosseratMedia2013}.
	The former of these can be identified as the \textit{Cauchy momentum
		equation}\textit{\emph{ \citep{basarNonlinearContinuumMechanics2013}
			of the displacement field $\mathbf{r}$, where $\Sigma_{\alpha i}=F_{i}^{s,\alpha}$
			is the }}\textit{first Piola-Kirchhoff stress tensor}\textit{\emph{.
	}}The latter equation is the angular momentum balance of the director
	field $E$, which should be seen as a micropolar analogue of the Cauchy
	momentum equation. We can thus see Eq.~\ref{eq:cosserat body dynamics}
	as a generalised and curvilinear Cauchy momentum equation, expressed
	in terms of the moving frame $E$.
	
	\textit{\emph{}}%

	\subsection{Cosserat surfaces \label{sec:cosserat surfaces}}
	
	\begin{figure*}[t]
		\centering \includegraphics[width=0.6\textwidth]{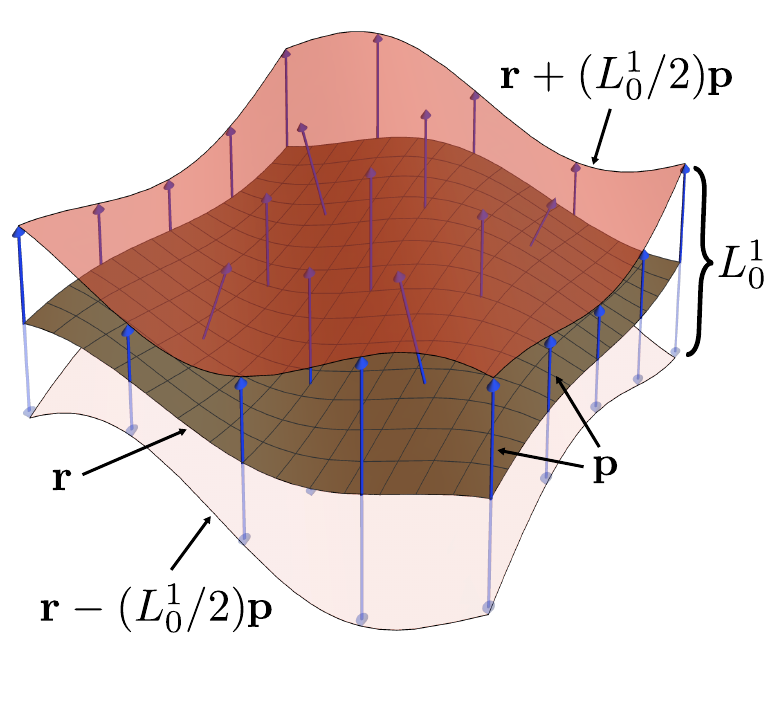}
		\caption{The director field $\mathbf{p}:W\to S^{2}$ (blue arrows) and the
			midsurface $\mathbf{r}:W\to\mathbb{E}^{2}$ (brown surface) of the
			Cosserat surface approximates a thin shell by constraining the material
			fibres along the director to be fixed. The upper and lower boundaries
			of the shell (transparent red surfaces) are given by $\mathbf{r}\pm(L_{0}^{1}/2)\mathbf{p}_{1}$
			respectively.}
		\label{fig:cosserat surface}
	\end{figure*}
	The theory of Cosserat surface was primarily developed for applications
	to thin elastic shells \citep{sansourCosseratSurfaceShell1995,altenbachGeneralizedCosserattypeTheories2010,altenbachCosseratTypeShells2013,naghdiTheoryShellsPlates1972}
	and plates \citep{steinbergCosseratPlateTheory2022,altenbachTheoriesPlatesBased2010,naghdiTheoryShellsPlates1972}.
	The material base space has dimension $d=2$, and the external degree
	of freedom is a mid-surface $\mathbf{r}:W\to\mathbb{E}^{3}$, such
	that $\mathbf{r}(t,M)$ is a two-dimensional surface in $\mathbb{E}^{3}$,
	at each time $t$. As opposed to Cosserat solids, the micropolar degree
	of freedom of Cosserat surfaces is typically a single director $\mathbf{p}^{s}:W\to T\mathbb{E}^{3}$,
	which is often \citep{ericksenLiquidCrystalsCosserat1974,jogHigherorderShellElements2004,dongCosseratInterphaseModels2014,rubinCosseratShellModel2004,greenLinearTheoryCosserat1971}
	inextensible, satisfying $\mathbf{p}^{s}\cdot\mathbf{p}^{s}=1$. In
	the context of shell theory, the director field represents rigid fibres
	from $\mathbf{r}(t,p)-(L_{w}/2)\mathbf{p}^{s}(t,p)$ to $\mathbf{r}(t,p)+(L_{w}/2)\mathbf{p}^{s}(t,p)$
	at each point $u\in M$, where $L_{w}$ is the width of the shell.
	See Fig.~\ref{fig:cosserat surface} for an illustration. The Cosserat
	surface can therefore be be considered the result of a dimensional
	reduction of a three-dimensional continuum body that is thin in one
	material dimension. What remains of the coarse-grained dimension,
	the director field, accomodates for transverse shear deformations
	\citep{sansourCosseratSurfaceShell1995} of the shell boundary relative
	to the mid-surface. The omitted $SO(2)$ rotational degree of freedom
	around the axis of $\mathbf{p}^{s}$, so-called drill rotations \citep{sansourCosseratSurfaceShell1995},
	are in many applications \citep{krishnaswamyCosserattypeModelRed1996,ericksenLiquidCrystalsCosserat1974,jogHigherorderShellElements2004,rangamaniSmallScaleMembrane2014}
	excluded as they are unmotivated physically \citep{antmanGeneralTheoriesShells2005,mohammadisaemInplaneDrillRotations2021,neffGeometricallyExactCosserat2004}.
	In addition to shells, Cosserat surfaces have also been used to model
	cell membranes \citep{rangamaniSmallScaleMembrane2014,krishnaswamyCosserattypeModelRed1996,ericksenTheoryCosseratSurfaces1979},
	liquid crystal and material interphases \citep{ericksenLiquidCrystalsCosserat1974,dongCosseratInterphaseModels2014,rubinCosseratShellModel2004}.
	
	The Cosserat surface is a Cartan system with a material base space
	in $d=2$ dimensions, and configuration space $\mathbb{X}=\mathbb{E}^{3}\times S^{2}$,
	on which $G=SE(3)$ is the symmetry group. If $M$ is homeomorphic
	to a bounded and closed subset in $\mathbb{R}^{2}$, such as $M=[0,1]\times[0,1]$,
	then the Cosserat surface called \emph{open}. If $M$ is homeomorphic
	to $S^{2}$, then the surface is \emph{closed}, as in Fig.~\ref{fig:M to Cosserat surface}.
	There are some subtleties involved in the study of closed surfaces;
	see the end of Sec.~\ref{subsec:Surfaces} for a discussion. If the
	surface is open then $M$ admits global coordinates, but in general
	we have local charts $(u,v):U\to\mathbb{R}^{2}$, where $U\subseteq M$.
	We write the spatio-temporal configuration as
	\begin{equation}
		q=(\mathbf{r},\mathbf{p}^{s}):=\left(\begin{array}{cc}
			1 & 0\\
			\mathbf{r} & \mathbf{p}^{s}
		\end{array}\right)\in\mathbb{R}^{4\times2}.
	\end{equation}
	Let $\Phi:W\to SE(3)$ be $\Phi=(\mathbf{r};R)$, and let $q_{r}=(\mathbf{0},\mathbf{p})$,
	where $R:W\to SO(3)$ satisfies $R\mathbf{p}=\mathbf{p}^{s}$ and
	$\mathbf{p}\in S^{2}$ is a constant unit vector, then $\Phi$ is
	a structure field satisfying $\Phi q_{r}=q$. As before, we will identify
	$R$ with an orthonormal triad in $T\mathbb{E}^{3}$, writing $R=(\mathbf{e}_{1}\ \mathbf{e}_{2}\ \mathbf{e}_{3})=E$.
	As the notation indicates, $\mathbf{p}$ should be considered the
	orientation of any point-continua on the Cosserat surface, relative
	to the moving basis $E$. Correspondingly, $\mathbf{p}^{s}(t,p)$
	is the actual orientation of the point-continua at $p\in M$ at time
	$t$ in the spatial frame of reference. For the sake of simplicity,
	we will therefore set $\mathbf{p}=(0\ 0\ 1)^{T}$, such that $\mathbf{p}^{s}=\mathbf{e}_{3}$.
	
	The generalised velocity and strain are of identical form as those
	of the Cosserat solid, but where the strain fields $X_{\alpha}=\{\boldsymbol{\theta}_{\alpha},\boldsymbol{\pi}_{\alpha}\}$
	are now defined with respect to the two material directions $\alpha=u,v$.
	From Eq.~\ref{eq:dx} we find that $\dot{\mathbf{r}}=V_{i}\mathbf{e}_{i}$,
	$\dot{\mathbf{p}}^{s}=\boldsymbol{\Omega}^{s}\times\mathbf{p}^{s}=\mathbf{e}_{i}\hat{\Omega}_{ij}p_{j}=\mathbf{e}_{i}\hat{\Omega}_{i3}$,
	$\partial_{\alpha}\mathbf{r}=\theta_{\alpha,i}\mathbf{e}_{i}$ and
	$\partial_{\alpha}\mathbf{p}^{s}=\boldsymbol{\pi}_{\alpha}^{s}\times\mathbf{p}^{s}=\mathbf{e}_{i}\hat{\pi}_{\alpha,i3}$.
	
	Briefly, we will now relate the strains of the displacement field,
	$\boldsymbol{\theta}_{u}$ and $\boldsymbol{\theta}_{v}$, to some
	standard concepts from the differential geoemetry of surfaces \citep{chernLecturesDifferentialGeometry1999,clellandFrenetCartanMethod2017}.
	Let $\mathcal{R}_{t}=\mathbf{r}(t,M)$, which is a two-dimensional
	surface in $\mathbb{E}^{3}$, for each time $t$. The \emph{first
		fundamental form} of $\mathcal{R}_{t}$ is $\mathrm{I}=d_{M}\mathbf{r}\cdot d_{M}\mathbf{r}:=g_{\alpha\beta}du^{\alpha}du^{\beta}$,
	where $g_{\alpha\beta}=\boldsymbol{\theta}_{\alpha}^{T}\boldsymbol{\theta}_{\beta}$
	is a time-dependent Riemannian metric on $\mathcal{R}_{t}$ induced
	from the Euclidean metric on $\mathbb{E}^{3}$. The \emph{second fundamental
		form} is $\mathrm{I\!I}=-d_{M}\mathbf{n}\cdot d_{M}\mathbf{r}:=b_{\alpha\beta}du^{\alpha}du^{\beta}$,
	where $b_{\alpha\beta}=\partial_{\alpha}\mathbf{n}\cdot\partial_{\beta}\mathbf{r}$
	and $\mathbf{n}:W\to S^{2}$ is the \emph{Gauss map}, given by $\mathbf{n}=\boldsymbol{\theta}_{u}^{s}\times\boldsymbol{\theta}_{v}^{s}/\left|\boldsymbol{\theta}_{u}^{s}\times\boldsymbol{\theta}_{v}^{s}\right|$,
	which is a vector field that is normal to $\mathcal{R}_{t}$. The
	\emph{Gaussian curvature} of $\mathcal{R}_{t}$ is given by $K=\text{det}(b)/\text{det}(g)$,
	and the \emph{mean curvature }as $H=\text{tr}(g^{-1}b)$.
	
	The kinematic equations of motion of the Cosserat surface in terms
	of the translational and rotational components of $X_{\alpha}$ and
	$N$ are indentical to those of the Cosserat solid, Eq.~\ref{eq:cosserat body kinematics eom},
	but where $\alpha=u,v$. Furthermore the spatial integrability conditions
	reduce to $D_{u}\boldsymbol{\theta}_{v}=D_{v}\boldsymbol{\theta}_{u}$
	and $\partial_{u}\boldsymbol{\pi}_{v}=D_{v}\boldsymbol{\pi}_{u}$.
	Similarly, the generalised momentum balance equation of the Cosserat
	surface is given by Eq.~\ref{eq:cosserat body dynamics}, with $\alpha=u,v$,
	and the explicit expressions for the body force and moment densities
	are given by Eq.~\ref{eq:cosserat body,  body force}. Given $[\mathbf{r}]=L$,
	$[\mathbf{p}^{s}]=1$ and assuming that the kinetic energy density
	has units of energy per unit material area $[\mathcal{K}(N)]=MT^{-2}$,
	we find that $\mathbf{P}$ and $\mathbf{L}$ have units of momentum
	and angular momentum per unit material area respectively, and $\mathbf{F}^{\alpha}$
	and $\mathbf{M}^{\alpha}$ units of force and moment per unit material
	length respectively. The body force and moment densities $\mathbf{f}$
	and $\mathbf{m}$ have units of force and moment per unit material
	area respectively.
	
	The above fully determines the mechanics of Cosserat surfaces. However,
	note that in the process of replacing the configurational variables
	$q$ and $\dot{q}$ with the Lie algebraic strain and velocity fields
	$X_{\alpha}$ and $N$, we have implicitly introduced a superfluous
	degree of freedom. These are the aformentioned drill rotations, which
	is an $SO(2)$ gauge symmetry of $(\mathbf{e}_{1}\ \mathbf{e}_{2})$
	around the axis of the director $\mathbf{p}^{s}=\mathbf{e}_{3}$.
	Fundamentally, this is reflected in the fact that $\text{dim}(G)-\text{dim}(\mathbb{X})=r-n=1$.
	For certain systems, as those of the subsequent section, the superfluous
	degrees of freedom lead to inconveniences for the formulation of the
	mechanics. For example, it may be difficult, or impractical, to construct
	Lagrangians that do not couple with the superfluous degrees of freedom.
	However, for the Cosserat surface, this is generally not an issue.
	To illustrate this, we consider a simple example. Let the Lagrangian
	be separable as $\mathcal{L}(q,\dot{q},\partial_{\alpha}q)=\mathcal{L}_{\mathbf{r}}(\mathbf{r},\dot{\mathbf{r}},\partial_{\alpha}\mathbf{r})+\mathcal{L}_{\mathbf{p}}(\mathbf{p}^{s},\dot{\mathbf{p}}^{s},\partial_{\alpha}\mathbf{p}^{s})$,
	and let $\mathcal{L}_{\mathbf{p}}=\frac{1}{2}r|\dot{\mathbf{p}}^{s}|^{2}+\frac{1}{2}k_{u}|\partial_{u}\mathbf{p}^{s}|+\frac{1}{2}k_{v}|\partial_{v}\mathbf{p}^{s}|$,
	where $r,k_{u},k_{v}\in\mathbb{R}$ are constants. We have shown previously
	that $|\dot{\mathbf{p}}^{s}|^{2}=\boldsymbol{\Omega}^{T}P_{0}\boldsymbol{\Omega}$
	and $|\partial_{\alpha}\mathbf{p}^{s}|^{2}=\boldsymbol{\pi}_{\alpha}^{T}P_{0}\boldsymbol{\pi}_{\alpha}$,
	where $P_{0}=\text{diag}\left\{ 1,1,0\right\} $, such that the reduced
	Lagrangian is $\ell_{\mathbf{p}}(\boldsymbol{\Omega},\boldsymbol{\pi}_{\alpha})=\frac{1}{2}r\boldsymbol{\Omega}^{T}P_{0}\boldsymbol{\Omega}+\frac{1}{2}k_{u}\boldsymbol{\pi}_{u}^{T}P_{0}\boldsymbol{\pi}_{u}+\frac{1}{2}k_{v}\boldsymbol{\pi}_{v}^{T}P_{0}\boldsymbol{\pi}_{v}$.
	We then find that $\mathbf{L}=(r\Omega_{1}\ r\Omega_{2}\ 0)^{T}$,
	$\mathbf{M}_{\alpha}=(k_{\alpha}\pi_{\alpha,1}\ k_{\alpha}\pi_{\alpha,2}\ 0)^{T}$
	and $\mathbf{m}=\mathbf{0}$. Substituting these into Eq.~\ref{eq:angular momentum balance cosserat body}
	we find that the third components of both the left- and right-hand
	sides vanish; that is, $(D_{t}\mathbf{L})_{3}=(D_{\alpha}\mathbf{M}^{\alpha})_{3}=0$.
	The angular rotation around $\mathbf{p}$ may therefore be set to
	any constant $\Omega_{3}(t,p)=\bar{\Omega}_{3}\in\mathbb{R}$ with
	impunity, and $\bar{\Omega}_{3}=0$ is therefore a natural choice.
	This is consistent with the fact that drill rotations do not enter
	into the dynamics. In general, it is not possible for Lagrangians
	$\mathcal{L}(q,\dot{q},\partial_{\alpha}q)$ to, upon left-trivialisation,
	yield dynamics that couple with drill rotations. See \citep{munthe-kaasIntegratorsHomogeneousSpaces2016a}
	for further discussions on the ambiguities of Lie group actions on
	homogeneous spaces when $r>n$, and how to resolve them.
	
	\subsection{Cosserat rods \label{subsec:Cosserat-rods}}
	
	The Cosserat rod can be conceived as a continuum limit of connected
	rigid bodies; that is, the system consists of a center-line curve
	and an orthonormal frame of directors, where the latter represents
	the rigid body cross-sections of rod. As opposed to the pure center-line
	mechanics of filaments (see Sec.~\ref{subsec:Filaments}), Cosserat
	rods can \emph{shear}; that is, the normal of the cross-section need
	not be tangent to the center-line. See Fig.~\ref{fig:M to Cosserat rod}
	for an illustration. Such models are prominently used in soft robotics
	\citep{rendaDiscreteCosseratApproach2018,graziosoGeometricallyExactModel2019,caasenbroodEnergyShapingControllersSoft2022,boyerMacrocontinuousComputedTorque2006,boyerMacrocontinuousDynamicsHyperredundant2012,rendaDynamicModelMultibending2014,naughtonElasticaCompliantMechanics2021,verlSoftRoboticsTransferring2015,paiSTRANDSInteractiveSimulation2002},
	the modelling of muscles and ligaments \citep{sackBiologicalTissueMechanics2016,zhangModelingSimulationComplex2019,kierTonguesTentaclesTrunks1985},
	biological growth models \citep{moultonMultiscaleIntegrationEnvironmental2020,gorielyMathematicsMechanicsBiological2017}
	and active filaments \citep{moultonElasticSecretsChameleon2016,oliveriTheoryDurotacticAxon2021,kaczmarskiActiveFilamentsCurvature2022}.
	In a forthcoming publication we will be treating the geometrisation
	of the Cosserat rod in further detail \footnote{Lukas Kikuchi, Ronojoy Adhikari. In preparation.}.
	
	The exposition of Cosserat solids in Sec.~\ref{sec:cosserat 3d bodies}
	largely subsumes that of Cosserat rods, where the primary difference
	is that the latter is defined in one material dimenion $d=1$. A\emph{
	}rod is \emph{open }if the material base space is an interval $M=[0,L_{0}]$,
	and \emph{closed} if the interval is periodic; that is, if $M\cong\mathbb{T}^{1}$,
	where $\mathbb{T}^{1}$ is the $1$-torus. The frame field, and the
	kinematic and dynamical equations of motion of the Cosserat rod is
	identical to that of the Cosserat solid, but now in a single $\alpha=u$
	dimension. Assuming that the kinetic energy density has units of energy
	per unit material length $[\mathcal{K}(N)]=ML^{-1}T^{-2}$, we can
	find by dimensional analysis that $\mathbf{P}$ and $\mathbf{L}$
	have units of momentum and angular momentum per unit material length
	respectively, and $\mathbf{F}$ and $\mathbf{M}$ units of force and
	moment respectively. The body force and moment densities $\mathbf{f}$
	and $\mathbf{m}$ have units of force and moment per unit material
	length respectively.
	
	\section{Cartan media with adapted structure fields \label{sec:Examples-of-Cartan-media-with-adapted-frames}}
	
	Kinematic adaption, which we introduced in Sec.~\ref{subsec:Adapted-frames},
	is a method by which we reduce superfluous degrees of freedom that
	arise due to the geometrisation process, when the dimension of the
	symmetry group $G$ is larger than that of the configuration space
	$\mathbb{X}$. Furthermore, where the unadapted kinematics does not
	capture the intrisic geometry of the system, an appropiately chosen
	adapted structure field leads to a structure generator that is intrinsic.
	As previously mentioned, the kinematic adaptation discussed draws
	heavily from the theory of moving frames \citep{clellandFrenetCartanMethod2017,cartanTheorieGroupesFinis1951,darbouxLeconsTheorieGenerale1887,frenetCourbesDoubleCourbure1852,felsMovingCoframesPractical1998,felsMovingCoframesII1999,olverModernDevelopmentsTheory,olverSurveyMovingFrames2005},
	which can be consulted for more detailed treatments.
	
	\subsection{Filaments \label{subsec:Filaments}}
	
	Filaments, also known as \emph{Kirchoff rods} \citep{kirchhoffUeberGleichgewichtUnd1859,dillKirchhoffTheoryRods1992},
	can be viewed as thin, slender tubes in the limit of a vanishing cross-sectional
	radius. Such systems appear in many applications \citep{eloyKinematicsMostEfficient2012,tornbergSimulatingDynamicsInteractions2004,goldsteinNonlinearDynamicsStiff1995,sodaDynamicsStiffChains1973,nordgrenComputationMotionElastic1974,hasimotoSolitonVortexFilament1972,laskarFilamentActuationActive2017,laskarBrownianMicrohydrodynamicsActive2015,goldsteinBistableHelices2000,kaczmarskiActiveFilamentsCurvature2022,betchovCurvatureTorsionIsolated1965,liuEffectsIntrinsicCurvature2020,goldsteinDynamicBucklingMorphoelastic2006,lenzMembraneBucklingInduced2009}.
	As opposed to the Cosserat rod, the filament does not posses microstructure;
	and can therefore bend and extend, but not shear. However, in some
	applications \citep{goldsteinNonlinearDynamicsStiff1995,hasimotoSolitonVortexFilament1972,goldsteinBistableHelices2000,parkerDerivationNonlinearRod1984}
	a `fictitious' frame is introduced that is adapted to the curvature
	of the filament. This is akin to a Cosserat rod whose microstructure
	is aligned so as to be tangent to its center-line. The result of this
	is an intrinsic description of the kinematics, where the filament
	is parameterised in terms of the rotations of the frame along its
	material length.
	
	We consider a system with $M=[0,L_{0}]$, $\mathbb{X}=\mathbb{E}^{3}$
	and $G=SE(3)$. We write its the spatio-temporal configuration in
	`homogeneous coordinates' as $q=(1\ \mathbf{r}^{T})^{T}$, where $\mathbf{r}:W\to\mathbb{E}^{3}$
	is a space curve $\mathbf{r}(t,M)\subset\mathbb{E}^{3}$ for every
	time $t\in[0,T].$ We define the structure field $\Phi:W\to SE(3)$
	as $\Phi=\left(\mathbf{r};R\right)$, with matrix representation given
	by Eq.~\ref{eq:matrix rep of SE(3)}, such that $q=\Phi q_{0}$,
	where $q_{0}=(1\ \mathbf{0}^{T})^{T}$. We see that there is an $SO(3)$
	gauge freedom in the choice of $R$. As before we identify the rotation
	with an orthonormal frame field $R=E=(\mathbf{e}_{1}\ \mathbf{e}_{2}\ \mathbf{e}_{3})$.
	We can interpret $(\mathbf{r},E)$ as an element of the configuration
	space of a Cosserat rod $\tilde{\mathbb{X}}=\mathcal{F}(\mathbb{E}^{3})\cong SE(3)$.
	We will now eliminate the $\text{dim}(\tilde{\mathbb{X}})-\text{dim}(\mathbb{X})=3$
	superfluous degrees of freedom by adapting the frame $E$ to $\mathbf{r}$.
	
	At all times $t\in[0,T]$, we let $\mathbf{e}_{1}=\partial_{u}\mathbf{r}/|\partial_{u}\mathbf{r}|$,
	such that $\mathbf{e}_{1}(t,p)$ is tangent to $\mathbf{r}(t,\cdot)$
	at $p\in M$, and $\mathbf{e}_{2}=\partial_{u}\mathbf{e}_{1}/|\partial_{u}\mathbf{e}_{1}|$,
	such that $\mathbf{e}_{2}$ is a vector orthogonal to $\mathbf{e}_{1}$
	in the osculating plane of the curve, and $\mathbf{e}_{3}=\mathbf{e}_{1}\times\mathbf{e}_{2}$.
	These are respectively known as the tangent, normal and binormal vectors.
	We have thus chosen a unique $R=R(\mathbf{r},\partial_{u}\mathbf{r},\partial_{u}^{2}\mathbf{r})$
	for a given space curve $\mathbf{r}(t,\cdot)$. This particular kinematic
	adaption is known as the \emph{Frenet-Serret} frame \citep{frenetCourbesDoubleCourbure1852,clellandFrenetCartanMethod2017}.
	We can interpret this as a kinematically constrained Cosserat rod,
	where its cross-sectional frame is a function of its center-line $\mathbf{r}$.
	We should note that the Frenet-Serret frame is only one particular
	choice of adaptation. In general there is an infinite amount of adapted
	frames, related by an $SO(2)$ rotation around the tangent of the
	center-line. In particular, another common choice of adapted frame
	is the \emph{Bishop frame} \citep{bishopThereMoreOne1975}.
	
	As before, we write the generalised strain and velocity as $X=\{\boldsymbol{\theta};\boldsymbol{\pi}\}$
	and $N=\{\mathbf{V};\boldsymbol{\Omega}\}$ respectively. We can now
	use Eq.~\ref{eq:generalised strain and velocity fields} to find
	what components of $X$ are eliminated by the adaption. From $\partial_{u}\mathbf{r}=\theta_{i}\mathbf{e}_{i}$
	and the adaption of $\mathbf{e}_{1}$ we have that $\theta_{2}=\theta_{3}=0$.
	From $\partial_{u}\mathbf{e}_{1}=\mathbf{e}_{j}\hat{\pi}_{j1}=\mathbf{e}_{2}\pi_{3}-\mathbf{e}_{3}\pi_{2}$
	and the adaption of $\mathbf{e}_{2}$ we find that $\pi_{2}=0$ and
	$\pi_{3}=|\partial_{u}\mathbf{e}_{1}|$. We thus write the remaining
	components of the generalised strain as $\boldsymbol{\theta}=(h\ 0\ 0)^{T}$
	and $\boldsymbol{\pi}=(\tau\ 0\ \kappa)^{T}$, where $\kappa=|\partial_{u}\mathbf{e}_{1}|$
	is the \emph{scalar curvature}, $\tau:W\to\mathbb{R}$ is the \emph{torsion}
	and $h=|\partial_{u}\mathbf{r}|$ is the square-root of the metric
	on the filament induced from the Euclidean metric. To see the latter,
	note that the length of the filament is $L=\int_{0}^{L_{0}}|\partial_{u}\mathbf{r}|du\int_{0}^{L_{0}}|\boldsymbol{\theta}|du=\int_{0}^{L_{0}}hdu$.
	Within the context of the theory of moving frames, $\kappa$ and $\tau$
	(and $h$, although it is often set to unity) are known as \emph{differential
		invariants}, which have the desired property of being invariant under
	rigid transformations. Furthermore, we note that $\kappa$ and $\tau$
	are extrinsic, and $h$ is intrinsic, to the geometry of the filament.
	
	Let us assume that we have initial boundary conditions on the structure
	field that are kinematically adapted using the Frenet-Serret frame;
	that is, $\mathbf{e}_{1}(0,u)=(\partial_{u}\mathbf{r}/h)\big|_{t=0}$
	and $\mathbf{e}_{2}(0,u)=(\partial_{u}\mathbf{e}_{1}/\kappa)\big|_{t=0}$.
	Then $X(0,u)=(\Phi^{-1}\partial_{u}\Phi)\big|_{t=0}=\{(h(0,u)\ 0\ 0)^{T};(\tau(0,u)\ 0\ \kappa(0,u))^{T}\}$.
	We will now derive conditions on the generalised velocity such that
	the system remains kinematically adapted in time. From Eq.~\ref{eq:kinematic equations of motion and integrability conditions}
	we have that $D_{t}\boldsymbol{\theta}=D_{u}\mathbf{V}$ and $\dot{\boldsymbol{\pi}}=D_{u}\boldsymbol{\Omega}$,
	from which we find that $\Omega_{1}=\kappa^{-1}(\Omega_{3}\tau-\partial_{u}\Omega_{2})$,
	$\Omega_{2}=-h^{-1}(V_{t}\tau+\partial_{u}V_{3})$ and $\Omega_{3}=h^{-1}(V_{1}\kappa-V_{3}\tau+\partial_{u}V_{2})$.
	The three components of the angular velocity $\boldsymbol{\Omega}$
	can no longer be specified independently, but are functions of the
	strain and translational velocity $\boldsymbol{\Omega}=\boldsymbol{\Omega}(h,\kappa,\tau,\mathbf{V})$.
	
	As the angular velocity of the frame $\boldsymbol{\Omega}$ is not
	a dynamical degree of freedom of the filament, we must have that the
	generalised momentum is $S=\left\{ \mathbf{P};\mathbf{0}\right\} ^{*}$,
	this will in turn lead to constraints on the generalised stress $Q=\{\mathbf{F};\mathbf{M}\}^{*}$.
	From Eq.~\ref{eq:non-conservative dynamics} we arrive at the constraints
	$M_{2}=\kappa^{-1}(m_{1}+\partial_{u}M_{1})$, $F_{3}h+M_{3}\tau-M_{1}\kappa=m_{2}+\partial_{u}M_{2}$
	and $F_{2}h+M_{2}\tau=-m_{3}+\partial_{u}M_{3}$. As expected, only
	three components of $Q$ can be specified independently, from which
	the remaining components are determined from the constraints.
	
	For example, the constitutive law of an Bernoulli-Euler beam \citep{nordgrenComputationMotionElastic1974}
	is $\mathbf{M}=B\kappa\mathbf{e}_{3}$, where $B$ is the bending
	stiffness of the beam. Having specified two of the components of $\mathbf{M}$
	(note that $M_{2}$ is always determined from the constraint), we
	find that $F_{2}=-h^{-1}(B\partial_{u}\kappa+m_{3})$ and $F_{3}=h^{-1}((1-B)\kappa\tau+m_{2})$.
	This determines the mechanics of a Bernoulli-Euler beam with bending
	stiffness $B$, under the influence of an external moment $\mathbf{m}$,
	as modelled using filament theory.
	
	In general, the force on the filament and the moment on the fictitious
	frame decomposes as $\mathbf{F}=\mathbf{F}^{r}+\mathbf{F}^{f}$ and
	$\mathbf{M}=\mathbf{M}^{r}+\mathbf{M}^{f}$, where $\mathbf{F}^{f}$
	and $\mathbf{M}^{f}$ are respectively a fictitious force and moment.
	We should interpret $\mathbf{F}^{f}$ and $\mathbf{M}^{f}$ as the
	force and moment that arise by necessity due to the kinematic adaption.
	In other words, they act so as to ensure that the fictitious frame
	$E$ remains adapted to the filament in time.
	
	\subsection{Surfaces \label{subsec:Surfaces}}
	
	Kirchhoff-Love theory \citep{loveSmallFreeVibrations1888,bassetDeformationThinElastic1894,altenbachTheorySimpleElastic2004}
	assumes thin, flat structures with the hypothesis that transverse
	shear deformation is negligible. Such systems have a wide range of
	applications \citep{niordsonShellTheory2012,koiterStabiliteitVanHet1945,altenbachTheorySimpleElastic2004,pauloseBucklingPathwaysSpherical2013,budianskyBucklingCircularCylindrical1972,hutchinsonBucklingSphericalShells2016}.
	Kirchhoff-Love surfaces can be understood as kinematically constrained
	Cosserat surfaces \citep{steigmannRelationshipCosseratKirchhoffLove1999},
	where the director field of the latter is fixed to be normal to its
	mid-surface. Here, by kinematic adaption of the Cosserat surface,
	we will derive a geometric theory of Kirchoff-Love surfaces. Using
	the concept a prinicpal adapted frame from the theory of moving frames
	\citep{darbouxLeconsTheorieGenerale1887,clellandFrenetCartanMethod2017},
	we eliminate all superfluous degrees of freedom of the Cosserat surface,
	and avoid so-called `drill rotation formulations' \citep{foxDrillRotationFormulation1992,hughesDrillingDegreesFreedom1989,simoStressResultantGeometrically1992}.
	
	We consider a system in $d=2$ material dimensions, with configuration
	space $\mathbb{X}=\mathbb{E}^{3}$ and symmetry group $G=SE(3)$.
	As for the Cosserat surface, a closed surface has a material base
	space homeomorphic to the sphere $M\cong S^{2}$, and is otherwise
	referred to as open. If $M$ is homeomorphic to $S^{2}$, then the
	surface is \emph{closed}, as in Fig.~\ref{fig:M to Cosserat surface}.
	We will work in local coordinates $\mathbf{u}:U\to\mathbb{R}^{2},\ U\subseteq M$,
	which we write as $\mathbf{u}=(u,v)$. Analogously to the filament,
	we write the spatio-temporal configuration as $q=(1\ \mathbf{r}^{T})^{T}$,
	where $\mathbf{r}:W\to\mathbb{E}^{3}$ is the mid-surface. We define
	the structure field $\Phi:W\to SE(3)$ as $\Phi=\left(\mathbf{r};R\right)$,
	with matrix representation Eq.~\ref{eq:matrix rep of SE(3)}, such
	that $q=\Phi q_{0}$. As for the filament, we must eliminate the $SO(3)$
	gauge freedom in $R$ to construct the kinematic adaptation. We identify
	the rotation with an orthonormal frame field $R=E=(\mathbf{e}_{1}\ \mathbf{e}_{2}\ \mathbf{e}_{3})$.
	Now, note that though Cosserat surfaces are most often conceptualised
	as having a single director $\mathbf{p=}\mathbf{e}_{3}$, we may also
	consider an orthonormal frame of directors $E$, where the orientation
	of $\mathbf{e}_{1}$ and $\mathbf{e}_{2}$ represent drill rotations.
	The process of adaptation will thus be a matter of kinematically constraining
	the director frame field of the Cosserat surface.
	
	As for the Cosserat surface, we write the generalised strain and velocity
	as $X_{\alpha}=\{\boldsymbol{\theta}_{\alpha};\boldsymbol{\pi}_{\alpha}\}$
	and $N=\{\mathbf{V};\boldsymbol{\Omega}\}$ respectively. We will
	first constrain the frame such that $\mathbf{e}_{3}(t,p)$ is normal
	to $\mathbf{r}(t,\cdot)$ at any $p\in M$ and all times $t\in[0,T]$,
	by letting $\mathbf{e}_{3}=\boldsymbol{\theta}_{u}^{s}\times\boldsymbol{\theta}_{v}^{s}/\left|\boldsymbol{\theta}_{u}^{s}\times\boldsymbol{\theta}_{v}^{s}\right|$.
	That is, $\mathbf{e}_{3}$ is now the \emph{Gauss map} of the surface.
	This implies that $\theta_{u,3}=\theta_{v,3}=0$. As for the Cosserat
	surface, the spatial integrability conditions are $D_{u}\boldsymbol{\theta}_{v}=D_{v}\boldsymbol{\theta}_{u}$
	and $\partial_{u}\boldsymbol{\pi}_{v}=D_{v}\boldsymbol{\pi}_{u}$.
	The latter contains what are known as the \emph{Gauss} and \emph{Codazzi-Mainardi
		equations} for surfaces. From the former, we find that $\pi_{\alpha,3}=\sum_{r=1}^{2}\theta_{\alpha,r}(\partial_{u}\theta_{v,r}-\partial_{v}\theta_{u,r})/\psi$
	where we have defined $\psi=(\boldsymbol{\theta}_{u}\times\boldsymbol{\theta}_{v})_{3}$.
	As an aside, we note that in the theory of moving frames the $1$-form
	$\pi_{u,3}du+\pi_{v,3}dv$ is called the \emph{Levi-Civita connection
		form}, which dictates the parallel transport of tangent vectors on
	the surface.
	
	There remains an $SO(2)$ gauge freedom in the adaptation, corresponding
	to rotations around the normal $\mathbf{e}_{3}$. One particular guage
	choice leads to a \emph{principal adapted frame} \citep{clellandFrenetCartanMethod2017},
	in which $\mathbf{e}_{1}(t,p)$ and $\mathbf{e}_{2}(t,p)$ are aligned
	with the geodesic lines of the principal curvatures respectively.
	Such frame fields can be found by diagonalising the \emph{shape operator}
	$\mathcal{S}^{t}:TM\to T\mathcal{R}_{t}$, defined as $\mathcal{S}^{t}(v)=(d_{M}\mathbf{e}_{3}(t,\cdot))(v)$,
	where $v:M\to TM$ is a vector field on $M$, $\mathcal{R}_{t}=\mathbf{\mathbf{r}}(t,M)$
	and $p\in M$. Note that the action of a vector-valued $1$-form $\phi=\mathbf{a}_{\alpha}du^{\alpha}$
	on a vector field $v=v^{\alpha}\frac{\partial}{\partial u^{\alpha}}$
	is given by $\phi(v)=\mathbf{a}_{\alpha}du^{\alpha}(v^{\beta}\partial_{\beta})=\mathbf{a}_{\alpha}\delta_{\beta}^{\alpha}v^{\beta}=v^{\alpha}\mathbf{a}_{\alpha}$.
	The shape operator yields a measure of the extrinsic curvature of
	the surface; that is, it shows how the normal $\mathbf{e}_{3}$ varies
	along tangent vectors in $M$. Now, let $v_{r}=v_{r}^{\alpha}\frac{\partial}{\partial u^{\alpha}}\in TM$
	be defined such that $d_{M}\mathbf{r}(v_{r})=\mathbf{e}_{r},\ r=1,2$.
	Using $d_{M}\mathbf{r}=\boldsymbol{\theta}_{\alpha}^{s}du^{\alpha}=\theta_{\alpha,i}\mathbf{e}_{i}du^{\alpha}$,
	we find that $v_{1}^{u}=\theta_{v,2}/\psi$, $v_{1}^{v}=-\theta_{u,2}/\psi$,
	$v_{2}^{u}=-\theta_{v,1}/\psi$ and $v_{2}^{v}=\theta_{u,1}/\psi$.
	A principal adapted frame diagonalises the shape operator, such that
	it satisfies $\mathcal{S}^{t}(v_{r})\cdot\mathbf{e}_{r}=\kappa_{r}\mathbf{e}_{r}$,
	where $\kappa_{u}$ and $\kappa_{v}$ are the principal curvatures
	of the surface. Let $\mathcal{S}_{rs}^{t}=\mathcal{S}^{t}(v_{r})\cdot\mathbf{e}_{s},\ r,s=1,2$,
	which is a symmetric matrix if the spatial integrability conditions
	is satisfied. If the frame is principal, we must have that $\mathcal{S}_{12}^{t}=\mathcal{S}_{21}^{t}=0$.
	If this condition is satisfied, then the principal curvatures are
	given by $\kappa_{r}=(\pi_{u,r}\theta_{v,r}-\pi_{v,r}\theta_{u,r})/\psi$.
	The Gauss and mean curvatures of the surface is then given by $K=\kappa_{1}\kappa_{2}$
	and $H=(\kappa_{1}+\kappa_{2})/2$ respectively, which are intrinsic
	and extrinsic measures respectively.
	
	Having determined the conditions for a principal adapted frame at
	fixed times, we now move on to kinematics. The kinematic equations
	of motion are $D_{t}\boldsymbol{\theta}_{\alpha}=D_{\alpha}\mathbf{V}$
	and $\dot{\boldsymbol{\pi}}_{\alpha}=D_{\alpha}\boldsymbol{\Omega}$.
	From the former, we find that $\Omega_{r}=(V_{1}(\pi_{u,2}\theta_{v,r}-\pi_{v,2}\theta_{u,r})-V_{2}(\pi_{u,1}\theta_{v,r}+\pi_{v,1}\theta_{u,r})+\frac{d}{dv}V_{3}(\theta_{u,r}-\theta_{v,r}))/\psi$
	for $r=1,2$. Presuming that the frame is principal at $t=0$, such
	that $S_{12}^{0}=S_{21}^{0}=0$, we can find a constraint on $\Omega_{3}$
	by demanding that $\dot{\mathcal{S}}_{12}^{0}=0$. We leave this as
	an exercise for the reader. It should be noted that the resulting
	expression diverges at the umbilical points of the surface; these
	are points $p_{u}\in M$ where $\kappa_{1}(t,p_{u})=\kappa_{2}(t,p_{u})$.
	However, one can show that $\lim_{p\to p_{u}}\Omega_{3}=0$ near such
	points, and we can therefore set $\Omega_{3}=0$ at the umbilical
	points. As for dynamics, the adapted frame induces fictitious forces
	and moments that maintains the adaption of the frame in time. These
	can be found in a manner analogous to the filament, using $\mathbf{0}=D_{\alpha}\mathbf{M}^{\alpha}+\boldsymbol{\theta}_{\alpha}\times\mathbf{F}^{\alpha}+\mathbf{m}=0$.
	
	We conclude with a remark on closed surfaces. There are two main considerations
	when applying the geometrisation procedure, as described above, to
	closed surfaces. Firstly, $S^{2}$ does not admit a global chart.
	This is not a major obstacle, as one can find two complementary charts
	that cover $S^{2}$, and then ensure that the generalised strain,
	velocity, momentum and stress transform accordingly between the charts
	(See Sec.~\ref{subsec:Local-charts}). Secondly, we know from the
	\emph{hairy ball theorem} \citep{eisenbergProofHairyBall1979} that
	vector fields on any manifold diffeomorphic to $S^{2}$, such as $\mathcal{R}_{t}$,
	must vanish at, at least, one point. This implies that the adapted
	frame field can \emph{not} be globally defined over a closed surface.
	We will briefly mention a method by which both of the aformentioned
	issues can be dealt with simultaneously. Let $\mathbf{r}_{0}:M\to\mathbb{E}^{3}$
	be the spatial configuration of the surface at an initial time slice,
	where $M=S^{2}$. We will use spherical coordinates, with polar angle
	$\psi:S^{2}\to[0,\pi]$ and azimuthal angle $\phi:S^{2}\to[0,2\pi]$,
	where the latter is periodic. This is a local chart on $S^{2}$, as
	we have coordinate singularities at $\psi=0,\pi$. However, if we
	extend the domain of the coordinates to also be defined at the singularities,
	the material base space is effectively a cylinder $M=S^{1}\times[0,\pi]$.
	There is in principle nothing that would prevent a closed surface
	to be mathematically parameterised with a cylindrical material base
	space. This would however entail that $\mathbf{r}_{0}(\phi,0)$ and
	$\mathbf{r}_{0}(\phi,\pi)$ must be constant along $\phi$. We can
	then construct a `principally adapted' structure field $\Phi^{i}:S^{1}\times[0,\pi]\to SE(3)$,
	$\Phi_{0}=(\mathbf{r}_{0},E_{0})$, at the initial time slice, using
	the procedure that we have outlined above. We should note that $\Phi_{0}$
	will in general not be constant along $\phi$ at $\psi=0,\pi$. However,
	importantly, the normal $\mathbf{e}_{3}^{i}$ will be constant along
	$\phi$ at $\psi=0,\pi$ by construction. By introducing some redundancy
	into our description of the system, we have thus evaded the issues
	of coordinate singularities and the hairy ball theorem. Mathematically,
	collapsed the circular ends of the cylinder onto the poles of the
	sphere. The geometrisation of the system proceeds as usual with no
	alteration. The resulting mechanics will be consistent with that of
	a closed surface. For instance, the generalised stress will satisfy
	$Q^{\phi}=0$ at $\psi=0,\pi$ by necessity, as there can be no strain
	along $\phi$ at the poles. This method can be refined further, see
	for instance \citep[ch.~11]{trefethenSpectralMethodsMATLAB2000}.
	
	\section{Further examples \label{sec:Further-examples-of-cartan-media}}
	
	The systems of the previous two sections all had configuration spaces
	with $SE(3)$ as their symmery groups. Here we provide some additional
	examples of more exotic systems.
	
	\subsection{Cosserat rods on 2-spheres \label{sec:Cosserat rods on 2-spheres}}
	
	We consider the constitutive dynamics of a microstructured filament
	on a sphere, which will be thought of as a kind of generalised Cosserat
	rod. The microstructure takes the form of rigid-body cross-sections,
	that can orient themselves in the tangent planes of the sphere. We
	briefly considered the kinematics and dynamics of filament on spheres
	in the examples of Sec.~\ref{subsec:Adapted-frames} and Sec.~\ref{subsec:Dynamics-under-adapted}.
	Some recent applications of such systems can be found in \citep{mannaEmergentTopologicalPhenomena2019,hsuActivityinducedPolarPatterns2022}.
	
	\begin{figure*}
		\begin{minipage}[t]{0.48\linewidth}%
			\includegraphics[width=0.9\linewidth]{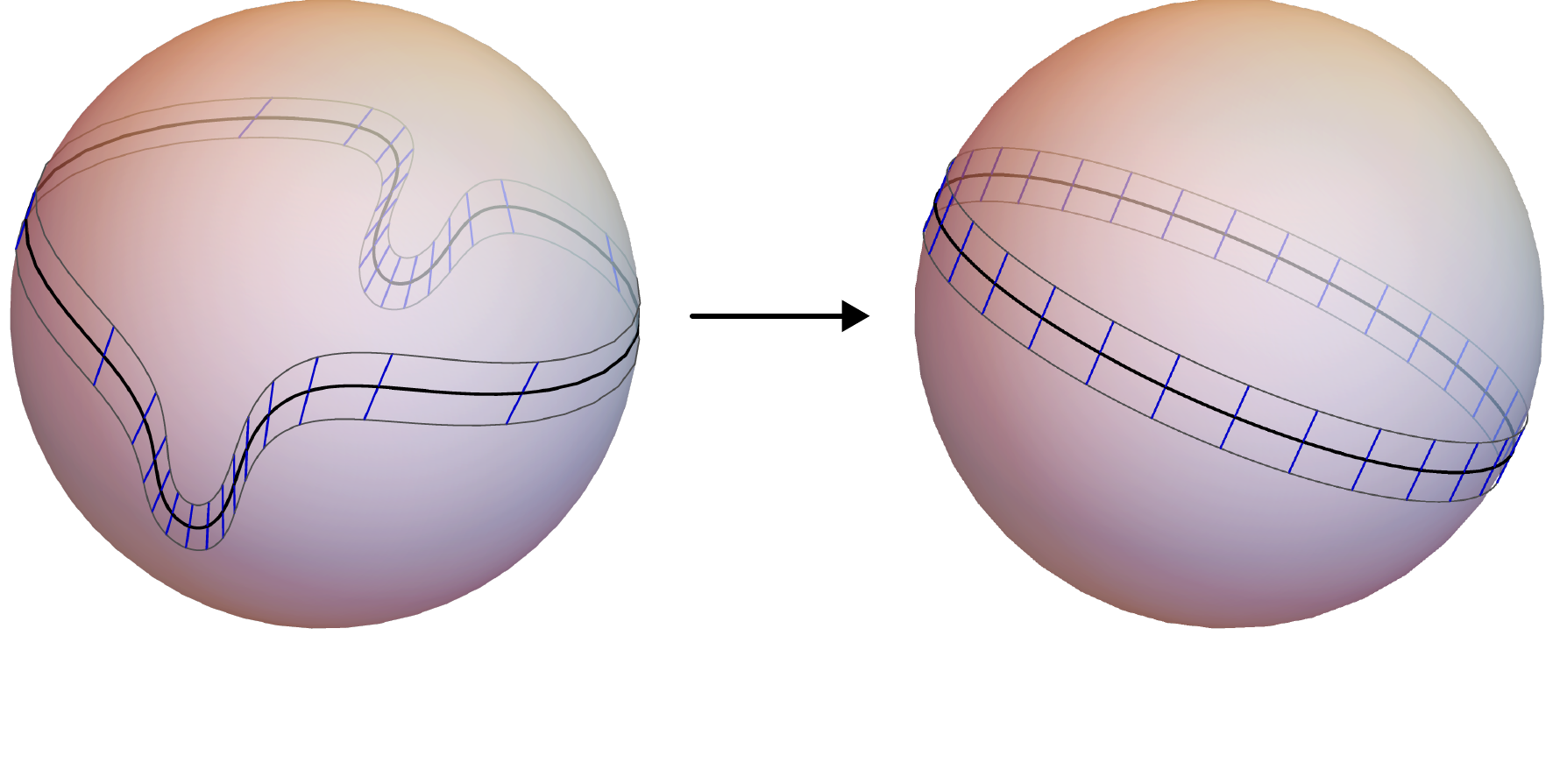}
			\caption{{\small{}Overdamped Cosserat rod on a sphere relaxing from a deformed
					initial configuration (left) to a ground-state (right). Solid black
					lines are the rod center-lines, and the blue lines are the directors
					$\mathbf{e}_{2}$.}}
			\label{fig:rod on sphere simulation}%
		\end{minipage}\hfill{}%
		\begin{minipage}[t]{0.48\linewidth}%
			\includegraphics[width=0.9\linewidth]{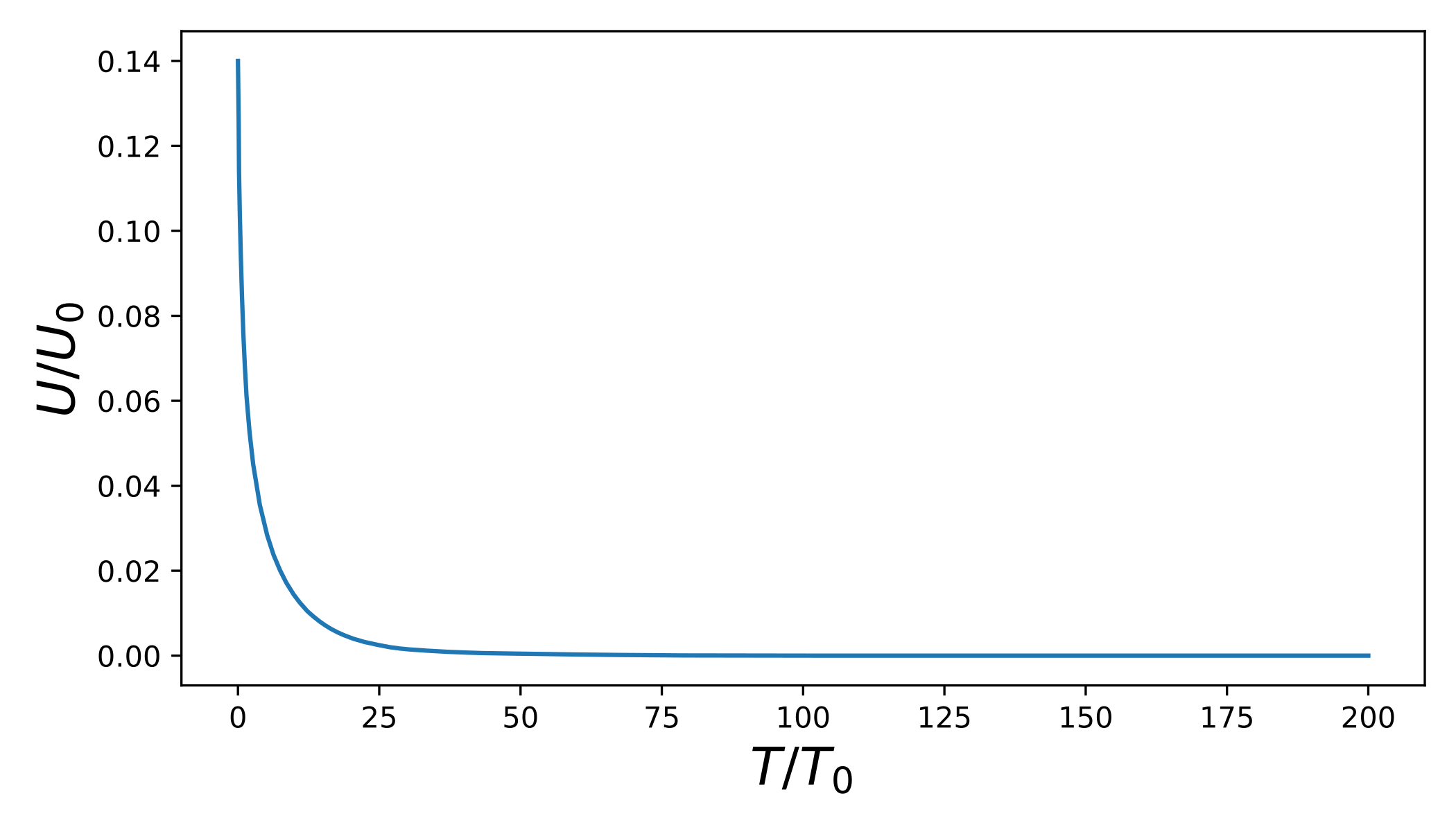}
			\caption{{\small{}Potential energy of the Cosserat rod as a function of simulation
					time $T$.}}
			\label{fig:rod on sphere U}%
		\end{minipage}
	\end{figure*}
	
	We consider a system with material base space $M=[0,L_{0}]$ and configuration
	space $\mathbb{X}=\mathcal{F}(S_{r}^{2})$, where $S_{r}^{2}\subset\mathbb{E}^{3}$
	is the sphere of radius $r$ and $\mathcal{F}(S_{r}^{2})$ its frame
	bundle. We write the spatio-temporal configuration $q:W\to\mathcal{F}(S_{r}^{2})$
	as $q(t,u)=(\mathbf{e}_{1}(t,u)\ \mathbf{e}_{2}(t,u)\ \mathbf{r}(t,u))\in\mathbb{R}^{3\times3}$,
	where $\mathbf{r}(t,u)\in S_{r}^{2}$ is the rod center-line and $\mathbf{e}_{1}(t,u),\mathbf{e}_{2}(t,u)\in T_{\mathbf{r}(t,u)}S_{r}^{2}$
	are the directors of the rod. Let $\mathbf{d}_{i}=(\delta_{i1}\ \delta_{i2}\ \delta_{i3})^{T}$
	and $q_{r}=(\mathbf{d}_{1}\ \mathbf{d}_{2}\ r\mathbf{d}_{3})$, and
	let $\Phi:W\to SO(3)$ satisfy $Rq_{r}=q$, which defines a structure
	field. This further implies that $R=(\mathbf{e}_{1}\ \mathbf{e}_{2}\ \mathbf{e}_{3})$
	and $\mathbf{r}=r\mathbf{e}_{3}$.
	
	We write the generalised velocity and strain fields respectively as
	vectors $\mathbf{X}=(V_{2}/r\ -V_{1}/r\ \Omega_{n})^{T}$ and $\mathbf{N}=(\theta_{2}/r\ -\theta_{1}/r\ \pi_{n})^{T}$,
	where $V_{1},V_{2},\Omega_{n},\theta_{1},\theta_{2},\pi_{n}:W\to\mathbb{R}$,
	such that $\hat{X}(t,u),\hat{N}(t,u)\in\mathfrak{so}(3)$. From $\dot{\mathbf{e}}=\mathbf{e}_{j}\hat{\Omega}_{ji}$,
	we have that $\dot{\mathbf{r}}=V_{1}\mathbf{e}_{1}+V_{2}\mathbf{e}_{2}$,
	$\dot{\mathbf{e}}_{1}=-\Omega_{n}\mathbf{e}_{2}-(V_{1}/r)\mathbf{e}_{3}$
	and $\dot{\mathbf{e}}_{2}=\Omega_{n}\mathbf{e}_{1}-(V_{2}/r)\mathbf{e}_{3}$.
	We can interpret $V_{1}(t,u)\mathbf{e}_{1}(t,u)+V_{1}(t,u)\mathbf{e}_{2}(t,u)\in T_{\mathbf{r}(t,u)}S_{r}^{2}$
	as the velocity of the material point $p\in M$ at time $t\in[0,T]$,
	and $\Omega_{n}(t,u)$ the angular velocity of the frame $(\mathbf{e}_{1}(t,u)\ \mathbf{e}_{2}(t,u))$
	around the axis $\mathbf{e}_{3}(t,u)$. The second terms in the expressions
	for $\dot{\mathbf{e}}_{1}$ and $\dot{\mathbf{e}}_{2}$ arise as a
	result of the parallel transport of the frame on the sphere. The corresponding
	expressions along the material derivative can be replicated from $\partial_{u}\mathbf{e}_{i}=\mathbf{e}_{j}\hat{\pi}_{ji}$.
	
	As before, we use Eq.~\ref{eq:kinematic equations of motion} to
	derive the equations of motion of the geometrised kinematics, from
	which we find that $\dot{\mathbf{X}}=\partial_{u}\mathbf{N}+\mathbf{X}\times\mathbf{N}$.
	From Eq.~\ref{eq:non-conservative dynamics} we find the generalised
	momentum balance condition $\dot{\mathbf{S}}+\boldsymbol{\Omega}\times\mathbf{S}=\partial_{u}\mathbf{Q}+\boldsymbol{\pi}\times\mathbf{Q}+\mathbf{T}$,
	where $\hat{S},\hat{Q},\hat{T}:W\to\mathfrak{so}(3)^{*}$ are the
	generalised momentum, stress and body force densities, and where we
	have used that $\mathfrak{so}(3)=\mathfrak{so}(3)^{*}$. The generalised
	stress must satisfy $\mathbf{Q}(t,0)=\mathbf{Q}(t,L_{0})=0$ for all
	$t\in[0,T]$. We write the components of the dynamical fields as $\mathbf{S}=(rP_{2}\ -rP_{1}\ L_{n})^{T}$,
	$\mathbf{Q}=(rF_{2}\ -rF_{1}\ M_{n})^{T}$ and $\mathbf{T}=(rf_{2}\ -rf_{1}\ m_{n})^{T}$,
	where the first two components of the fields relate to the momenta
	and forces on the center-line, whilst the third component relates
	to the angular momentum and moment on the director frame. Now, let
	$\mathbf{V}=(V_{1}\ V_{2}\ 0)^{T}$, $\boldsymbol{\theta}=(\theta_{1}\ \theta_{2}\ 0)^{T}$,
	$\boldsymbol{\Omega}=(0\ 0\ \Omega_{n})^{T}$, $\boldsymbol{\pi}=(0\ 0\ \pi_{n})^{T}$,
	$\mathbf{P}=(P_{1}\ P_{2}\ 0)^{T}$, $\mathbf{F}=(F_{1}\ F_{2}\ 0)^{T}$,
	$\mathbf{L}=(0\ 0\ L_{n})^{T}$, $\mathbf{M}=(0\ 0\ M_{n})^{T}$,
	$\mathbf{f}=(f_{1}\ f_{2}\ 0)^{T}$ and $\mathbf{m}=(0\ 0\ m_{n})^{T}$.
	We also assume a kinetic energy density of the form $\mathcal{K}(N)=\frac{1}{2}\rho_{0}|\mathbf{V}|^{2}+\frac{1}{2}\mathbb{I}_{n}\Omega_{n}^{2}$,
	such that $\mathbf{P}=\rho_{0}\mathbf{V}$ and $\mathbf{L}=\mathbb{I}_{n}\boldsymbol{\Omega}$,
	where $\rho_{0}\in\mathbb{R}_{+}$ is a mass density per unit material
	length, and $\mathbb{I}_{n}\in\mathbb{R}_{+}$ is the moment of inertia
	of the director frame. Then, the generalised momentum balance conditions
	can be put into the form
	
	\begin{subequations}
		\label{eq:cosserat rod on sphere dynamics}
		\begin{align}
			D_{t}\mathbf{P} & =D_{u}\mathbf{F}+\frac{1}{r^{2}}\mathbf{L}\times\mathbf{V}+\frac{1}{r^{2}}\boldsymbol{\theta}\times\mathbf{M}+\mathbf{f},\\
			\dot{\mathbf{L}} & =\partial_{u}\mathbf{M}+\boldsymbol{\theta}\times\mathbf{F}+\mathbf{m}.\label{eq:rod on sphere angular momentum balance}
		\end{align}
	\end{subequations}
	Equation \ref{eq:cosserat rod on sphere dynamics} are the linear
	and angular momentum balance equations of the Cosserat rod on the
	sphere. The two terms with the factor of $1/r^{2}$ arise due to the
	effects of the curvature on the sphere. To see this, note that as
	$r\to\infty$ we have that $S_{r}^{2}\to\mathbb{E}^{2}$, and in that
	limit Eq.~\ref{eq:cosserat rod on sphere dynamics} becomes the expected
	balance equations of Cosserat rods constricted to the plane. Compare
	Eq.~\ref{eq:cosserat rod on sphere dynamics} to Eq.~\ref{eq:momentum balance cosserat body}
	and the results of Sec.~\ref{subsec:Cosserat-rods}.
	
	To illustrate the mechanics of Cosserat rods on spheres, we now consider
	a simple example of consitutive dynamics. Consider a potential energy
	density of the form $\mathcal{U}(X)=\frac{1}{2}(\theta_{1}-1\ \theta_{2})^{T}\mathbb{K}(\theta_{1}-1\ \theta_{2})+\frac{1}{2}\mathbb{N}_{n}\pi_{n}^{2}$,
	where $\mathbb{K}\in\mathbb{R}_{+}^{2\times2}$ and $\mathbb{N}_{n}\in\mathbb{R}_{+}$
	are stiffness coefficients for the center-line and director frame
	respectively. The form of the potential is such that the rest state
	is $\boldsymbol{\theta}=(1\ 0\ 0)^{T}$ and $\pi_{n}=0$, corresponding
	to a greater circle centerline aligned with the director $\mathbf{e}_{1}$.
	We also include a dissipative force $\mathbf{f}=-\gamma_{T}\mathbf{V}$
	and moment $\mathbf{m}=-\gamma_{R}\boldsymbol{\Omega}$, where $\gamma_{T},\gamma_{R}\in\mathbb{R}_{+}$,
	so that the any initial configuration at $t=0$ reaches the rest state
	at $t\to\infty$. See Fig.~\ref{fig:rod on sphere simulation} and
	Fig.~\ref{fig:rod on sphere U} for an illustration of the results
	of a simulation in an overdamped regime. To simulate the system we
	used the set of codes in \citep{lukaskikuchiPyCoss2022}.
	
	As a concluding remark, we note that it would be a straightforward
	exercise to extend the treatment discussed here to consider Cosserat
	rods on general two-dimensional radial manifolds. A radial manifold
	refers to a surface wherein every point can be linked to the origin
	using a straight line segment without crossing the surface. To do
	this, one should first promote the spherical radius $r$ to a radial
	map $r:S^{2}\to\mathbb{R}^{+}$, and then derive what will correspond
	to Eq.~\ref{eq:cosserat rod on sphere dynamics}, from the generalised
	momentum balance equation.
	
	\subsection{$O(n)$-NLSM field theory \label{sec:The O(3) non-linear sigma model}}
	
	Our exposition thus far has been dedicated to study of continuum mechanics
	in homogeneous spaces. Accordingly, we have viewed Cartan media as
	sub-manifolds of their homogeneous configuration spaces. However,
	an equally ppint of view is to see our work as a framework for geometrising
	general field theories, with base space $[0,T]\times M$ and target
	space $\mathbb{X}$. Here we give an example of such an application,
	where we apply the geometrisation procedure on an \emph{$O(n)$ non-linear
		$\sigma$ model} (NLSM) \citep{ketovQuantumNonlinearSigmaModels2013}.
	
	We will construct a field theory for configurations $\mathbf{n}:W\to S^{n}\subset\mathbb{E}^{n+1}$,
	where $W=[0,T]\times\mathbb{R}^{d}$ is the base space of the field
	theory, and $S^{n}$ the target space. Let $\Phi:W\to SO(n)$ satisfy
	$\mathbf{n}=\Phi\mathbf{n}_{0}$. We will restrict the symmetry group
	to be $SO(n)$, rather than $O(n)$, assuming that $\mathbf{n}$ suffers
	no discontinuities. Let $t,u^{1},\dots,u^{d}:W\to\mathbb{R}^{d+1}$
	be coordinates on $W$, and we write partial derivatives as $\partial_{\gamma},\ \gamma=0,1,\dots,d$,
	where $\partial_{0}=\frac{\partial}{\partial t}$ and $\partial_{\alpha}=\frac{\partial}{\partial u^{\alpha}},\ \alpha=1,\dots,d$.
	
	The \textit{$O(n)$ non-linear $\sigma$ model} (NLSM) is defined
	by the Lagrangian density 
	\begin{equation}
		\mathcal{L}=\frac{1}{2}g^{\gamma\kappa}(\partial_{\gamma}\mathbf{n})\cdot(\partial_{\kappa}\mathbf{n})\label{eq:O(n) model}
	\end{equation}
	where $g^{\gamma\kappa}$ is a metric on $W$. In general the metric
	has a signature $(v,p,r)$, corresponding to the number of positive,
	negative and zero eigenvalues, we will however assume that $v=d+1$
	here for simplicity. We will also assume we work in coordinates such
	that $g=\mathbbm{1}_{(d+1)\times(d+1)}$.
	
	Let $Z_{\gamma}=\Phi^{-1}\partial_{\gamma}\Phi$, such that $Z_{0}=N$
	and $Z_{\alpha}=X_{\alpha}$. We assume that $Z_{\gamma}:W\to\mathfrak{so}(n)$
	are in their fundamental matrix representation, and that they act
	on $S^{n}$ accordingly. Now, we have that $(\partial^{\gamma}\mathbf{n})\cdot(\partial_{\gamma}\mathbf{n})=(Z_{\gamma}\mathbf{n}_{0})\cdot(Z_{\gamma}\mathbf{n}_{0})$.
	Let $\mathrm{A}:\mathfrak{so}(3)\to\mathfrak{so}(3)$ be the linear
	operator defined as $\langle Z_{\gamma},\mathrm{A}Z_{\gamma}\rangle=\mathbf{n}_{0}^{T}Z_{\gamma}^{T}Z_{\gamma}\mathbf{n}_{0}$.
	We can then write the Lagrangian in its reduced form as $\ell(N,X)=\frac{1}{2}\langle Z_{\gamma},\mathrm{A}Z_{\gamma}\rangle$.
	The corresponding generalised momentum and stress fields are then
	$S=\mathrm{A}N^{T}$ and $Q^{\alpha}=\mathrm{A}X_{\alpha}^{T}$. The
	geometrised kinematics and dynamics of the field theory are then a
	straightforward application of Eq.~\ref{sec:geometric kinematics}
	and Eq.~\ref{eq:constitutive momentum balance equations}.
	
	\subsection{Relativistic Cosserat rods \label{sec:Relativistic Cosserat rods}}
	
	Here we consider Cosserat rods in relativistic space-times. Such systems
	have previously been developed in \citep{delphenichMechanicsCosseratMedia2015},
	wherein applications for modelling free Dirac electrons and the Weyssenhoff
	fluid were provided using relativistic Cosserat media.
	
	We work in units where the speed-of-light constant is set to unity
	$c=1$. The Minkowski space $\mathbb{M}^{1,3}$ is the vector space
	$\mathbb{R}^{4}$ equipped with the \textit{Minkowski inner product}
	$\langle\cdot,\cdot\rangle_{M}:\mathbb{M}^{1,3}\times\mathbb{M}^{1,3}\to\mathbb{R}$
	with signature $(1,3,0)$. In other words, given some basis $D=(\mathbf{d}_{0},\mathbf{d}_{1},\mathbf{d}_{2},\mathbf{d}_{3})$
	for $\mathbb{R}^{4}$, the \textit{Minkowski metric} \textit{$\eta_{ij}=\langle\mathbf{d}_{i},\mathbf{d}_{j}\rangle_{M}$
	}has $1$ negative eigenvalue and $3$ positive eigenvalues. Henceforth
	we will assume that the basis is defined such that $\eta=\text{diag}(-1,1,1,1)$.
	Any basis that diagonalises $\eta$ in this way will be called an
	\textit{orthonormal} basis. A vector $\mathbf{v}\in\mathbb{M}^{1,3}$
	is known as \textit{time-like} if $\langle\mathbf{v},\mathbf{v}\rangle_{M}<0$,
	\textit{space-like} if $\langle\mathbf{v},\mathbf{v}\rangle_{M}>0$
	and \textit{light-like} if $\langle\mathbf{v},\mathbf{v}\rangle_{M}=0$.
	We can thus identify $\mathbf{d}_{0}$ as the time-like direction
	in this basis, and $(\mathbf{d}_{1},\mathbf{d}_{2},\mathbf{d}_{3})$
	as the space-like directions.
	
	The \textit{space-time coordinates} $\mathbf{x}(\tau)$ of an observer
	is a function $\mathbf{x}:[0,T]\to\mathbb{M}^{1,3}$ where $\tau$
	is the time measured by clocks co-moving with the observer, known
	as the \textit{proper time}. The \textit{$4$-velocity} of the observer
	is given by the time-like vector $\mathbf{U}^{s}=\partial_{\tau}\mathbf{x}$,
	and the proper time is defined such that $\langle\mathbf{U}^{s},\mathbf{U}^{s}\rangle=|\mathbf{U}^{s}|^{2}=-1$.
	The \textit{inertial frame} of the observer at proper time $\tau$
	is an orthonormal basis $E(\tau)=(\mathbf{e}_{0}(\tau)\ \mathbf{e}_{1}(\tau)\ \mathbf{e}(\tau)\ \mathbf{e}_{3}(\tau))$
	such that, if we write $\mathbf{U}^{s}=U_{\gamma}\mathbf{e}_{\gamma},\ \gamma=0,1,2,3$,
	then $\mathbf{U}=(U_{0}\ U_{1}\ U_{2}\ U_{3})=(1\ 0\ 0\ 0)^{T}$.
	Intuitively, this corresponds to the fact that an observer is always
	stationary in its own co-moving inertial reference frame. We can thus
	construct such a basis by setting $\mathbf{e}_{0}=\mathbf{U}^{s}$,
	and the remaining three basis elements $(\mathbf{e}_{1},\mathbf{e}_{2},\mathbf{e}_{3})$
	will specify the spatial orientation of the observer. Hencerforth,
	we will expand vectors $\mathbf{v}^{s}\in T\mathbb{M}^{1,3}$ as $\mathbf{v}^{s}=v_{\gamma}\mathbf{e}_{\gamma}$
	and $\mathbf{v}=(v_{0}\ \vec{v})^{T}$, where $\vec{v}=(v_{1}\ v_{2}\ v_{3})\in\mathbb{R}^{3}$
	are the spatial components of $\mathbf{v}$.
	
	Any two inertial frames $E_{1}$ and $E_{2}$ can be related by a
	\textit{Lorentz transformation} \textit{$E_{2}=\Lambda E_{1}$ }where
	$\Lambda\in SO(1,3)$, and where $SO(1,3)$ is the \textit{Lorentz
		group} on $\mathbb{M}^{1,3}$, defined as $SO(1,3)=\{\Lambda\in\mathbb{R}^{4\times4}\ |\ \langle\Lambda\mathbf{v},\Lambda\mathbf{v}\rangle_{M}=\langle\mathbf{v},\mathbf{v}\rangle_{M}\ \forall\mathbf{v}\in\mathbb{M}^{3,1}\}$,
	which is the group of rotations in space and \textit{Lorentz boosts}.
	The Lorentz group is thus the set of linear transformations that preserves
	the Minkowski inner product. Combined with the group of translations
	on $\mathbb{M}^{1,3}$, we have the \textit{Poincaré} group
	\begin{equation}
		\begin{aligned}M(1,3)= & \bigg\{\begin{pmatrix}1 & \mathbf{0}^{T}\\
				\mathbf{t} & \Lambda
			\end{pmatrix}\in\mathbb{R}^{5\times5}\ |\ \\
			& \mathbf{t}\in\mathbb{M}^{1,3},\ \Lambda\in SO(3,1)\bigg\}
		\end{aligned}
	\end{equation}
	of space-time translations and rotations, which is a semi-direct product
	$M(1,3)=T(4)\rtimes SO(1,3)$. We will write elements of \textit{\emph{Poincaré
			group using the short-hand $(\mathbf{t};\Lambda)\in M(1,3)$.}}
	
	Now consider a one-dimensional continuum of inertial observers, parameterised
	by a material coordinate $u$. At proper time $\tau$, relative to
	the observer at material coordinate $u$, we write their space-time
	coordinates as $\mathbf{r}(\tau,u)$ and their inertial frame as $E(\tau,u)$.
	We may assume that at proper time $\tau=-\infty$, the continuum of
	observers were co-moving in the same inertial reference frame, at
	which point their clocks were synchronised. We can then consider this
	system a \textit{relativistic} Cosserat rod. The configuration space
	of the system is thus the frame bundle of Minkowski space $\mathbb{X}=\mathcal{F}(\mathbb{M}^{1,3})$,
	and we write elements as $(\mathbf{a},A)\in\mathbb{X}$, where $\mathbf{a}\in\mathbb{M}^{1,3}$
	and $A\in SO(1,3)$, to be understood as a shorthand for their matrix
	representations as $\mathbb{R}^{5\times5}$-matrices. The kinematic
	base space is $W=[0,\mathcal{T}]\times[0,L_{0}]$ where $[0,\mathcal{T}]$
	is the proper time domain in consideration. The rod thus has spatio-temporal
	configuration $q=(\mathbf{r},E)$, where $(\mathbf{e}_{1},\mathbf{e}_{2},\mathbf{e}_{3})$
	is the cross-sectional frame of the rod. Let $q_{0}=(\mathbf{0},D)$,
	we can then define a structure field $\Phi:W\to M(1,3)$ as $\Phi=(\mathbf{r};\Lambda)$,
	where $\Lambda D=E$, satisfying $q=\Phi q_{0}$. Analogously to what
	we have done in previous examples, we let $D=\mathbbm{1}_{4\times4}$,
	thereby identifying $\Lambda=E$ as the inertial frame field of the
	rod.
	
	We introduce a short-hand for the matrix representation of $\mathfrak{m}(1,3)$
	\begin{equation}
		\{\mathbf{y};Z\}:=\left(\begin{array}{cc}
			0 & \mathbf{0}^{T}\\
			\mathbf{y} & Z
		\end{array}\right)\in\mathfrak{m}(1,3),
	\end{equation}
	for any $\mathbf{y}\in\mathbb{M}^{1,3}$ and $Z\in\mathfrak{so}(1,3)$.
	The fundamental matrix representation of Lie algebra elements $Z\in\mathfrak{so}(1,3)$
	is
	
	\begin{equation}
		Z=\begin{pmatrix}0 & \vec{c}^{T}\\
			\vec{c} & \hat{d}
		\end{pmatrix}
	\end{equation}
	for any $\vec{c},\vec{d}\in\mathbb{R}^{3}$, and we introduce the
	short-hand $Z=\left[\vec{c};\vec{d}\right]$. We write the generalised
	velocity and strain field as $N=\{\mathbf{U};O\}$ and $X=\{\boldsymbol{\phi};\chi\}$,
	where $\mathbf{U},\boldsymbol{\phi}:W\to T\mathbb{M}^{1,3}$ and $O,\chi:W\to\mathfrak{so}(1,3)$,
	and we write $O=\left[\vec{a};\vec{\Omega}\right]$ and $\chi=\left[\vec{w};\vec{\pi}\right]$,
	where $\vec{a},\vec{\Omega},\vec{w},\vec{\pi}:W\to\mathbb{R}^{3}$.
	
	We now proceed to interpret the components of the generalised velocity
	and strain fields. using Eq.~\ref{eq:dx}, we have that $\partial_{\tau}\mathbf{r}=\mathbf{U}^{s}=U_{\gamma}\mathbf{e}_{\gamma}$,
	$\partial_{\tau}\mathbf{e}_{\gamma}=\mathbf{e}_{\kappa}O_{\kappa\gamma}$,
	$\partial_{u}\mathbf{r}=\boldsymbol{\phi}^{s}=\phi_{\gamma}\mathbf{e}_{\gamma}$
	and $\partial_{u}\mathbf{e}_{\gamma}=\mathbf{e}_{\kappa}\chi_{\kappa\gamma}$,
	which we will further decompose into time-like and space-like equations.
	We saw previously that $\partial_{\tau}\mathbf{r}=\mathbf{U}^{s}=\mathbf{e}_{0}$,
	and the \emph{$4$-acceleration} is thus given by $\mathbf{a}\equiv\partial_{\tau}^{2}\mathbf{r}=\partial_{\tau}\mathbf{e}_{0}=a_{i}\mathbf{e}_{i}$,
	we can thus write $\mathbf{a}=(0\ \vec{a})^{T}$. This is the correct
	expression for the co-moving 4-acceleration in special relativistic
	kinematics \citep[p.~99]{rindlerRelativitySpecialGeneral2001}. We
	have that $\partial_{\tau}\mathbf{e}_{i}=\mathbf{e}_{j}\hat{\Omega}_{ji}$
	and $\partial_{u}\mathbf{e}_{i}=\mathbf{e}_{j}\hat{\pi}_{ji}$, and
	can therefore identify $\vec{\Omega}$ and $\vec{\pi}$ as the angular
	velocity and angular rate-of-change along $u$ of the frame.
	
	Having constrained $E(\tau,u)$ to be an inertial frame, we have in
	effect kinematically adapted the system. However, notably, the constraint
	is not with respect to the spatial derivatives of $\mathbf{r}$, as
	in previous examples. As a consequence, as $\mathbf{U}=(1\ 0\ 0\ 0)^{T}$,
	the entirety of the generalised velocity is encoded in $O=\left[\vec{a};\vec{\Omega}\right]$.
	In other words, the kinematics of the relativistic Cosserat rod is
	specified by the spatial \emph{acceleration} $\vec{a}$ of the center-line,
	as well as the angular velocity $\vec{\Omega}$ of the cross-section.
	This stands in contrast to the non-relativistic Cosserat rod, where
	we instead specify the \emph{velocity} of the center-line. We can
	understand this difference by noting that velocity is \emph{itself}
	a kinematic degree of freedom in special relativity, in addition to
	position and orientation. Only the latter two are kinematic degrees
	of freedom in non-relativistic systems. This is therefore the reason
	why we must specify the \emph{acceleration} of the frame, as opposed
	to its \emph{velocity}, in the kinematics.
	
	Finally, the equations of motion of the generalised strain can be
	found as usual from Eq.~\ref{eq:kinematic equations of motion},
	which decomposes as $\partial_{\tau}\boldsymbol{\phi}=\chi\mathbf{U}-O\boldsymbol{\phi}$
	and $\partial_{\tau}\chi=(\partial_{u}+\text{ad}_{\chi})O$.
	
	\section{Summary and conclusion}
	
	We now summarise the key steps in expressing the mechanics of Cartan
	media in geometric form. The first step, of course, is to identify
	a material base space $M$ and a configuration space $\mathbb{X}$.
	The former encodes the topology of the continuum while the latter
	specifies the space of configurations of a constitituent at, say,
	$p\in M$. We note that $M$ serves as continuous index set and hence
	its topology, rather than its differential structure, is what is relevant
	for the mechanics of the system. The second step is to identify the
	symmetry group $G$ of $\mathbb{X}$ and to identify the lift \textbf{$\Phi:W\to G$
	}into the symmetry group of the map $q:W\to\mathbb{X}$ from the space-time
	manifold to the homogeneous space. In the third step, the local structure
	of $G$ can then be used to express the kinematics in a geometrised
	form given by Eq.~\ref{eq:kinematic equations of motion}. The spatial
	integrability conditions, which must be satisfied at all times $t\in[0,T]$,
	are given by Eq.~\ref{eq:spatial integrability conditions}. This
	completes the kinematic aspects of geometrisation. The dynamics is
	constructed using a kinetic energy density $\mathcal{K}(N)$ in terms
	of the generalised velocity field $N$. The corresponding generalised
	momentum is then $S=\frac{\partial\mathcal{K}}{\partial N}$. The
	geometrised dynamics of a system under the influence of generalised
	stresses $Q^{\alpha}$, body $T$ and surface force $P$ is then given
	by Eq.~\ref{eq:spatial integrability conditions}. In the conservative
	case, the stress and body force are derived from a potential. In certain
	cases, when $\text{dim}(\mathbb{X})<\text{dim}(G)$, the superfluous
	degrees of freedom may be eliminated by adapting the structure field
	accordingly. The procedure was described in Sec.~\ref{subsec:Adapted-frames}
	and Sec.~\ref{subsec:Dynamics-under-adapted} for the kinematics
	and dynamics respectively, and we provided examples in Sec.~\ref{sec:Examples-of-Cartan-media-with-adapted-frames}.
	
	The principal advantages of lifting the mechanics from the homogeneous
	space into the symmetry group $G$ and then into the Lie algebra are
	as follows. \emph{The geometrisation process: }In the geometrisation
	of the kinematics and dynamics, we use the trivialisation $TG\cong G\times\mathfrak{g}$
	to trivialise the equations of motion. This results in equations in
	terms of vector-space valued (i.e. linear) variables. This can be
	contrasted to equations of motions expressed on the $\mathbb{X}$-
	or $G$-level, which are inherently non-linear. The conservative,
	and non-conservative, geometrised kinematic and dynamic equations
	of motion of any Cartan media follow systematically given choices
	of $M$, $\mathbb{X}$and $G$. In particular using the results of
	Sec.~\ref{subsec:Generalised-body-force}, the body forces on any
	Cartan media that results from a potential (i.e. a conservative body
	force) can be found explicitly in a very systematic and straighforward
	way. For example, we do not have to worry about what the torque is
	on the micropolarity $E$, if there is a term that couples $E$ and
	gravity in the potential. As the geometry of the configuration is
	encoded in $\mathbb{X}$, notions such as moments or torque appear
	as an immediate consequence of the geometrisation. This has been recognised
	in the geometric mechanics literature for Lie groups, but not in the
	systematic way that we present here for the continuum setting and
	arbitrary $\mathbb{X}$. The geometrisation procedure has particular
	advantages for numerical simulations as errors accrue in the Lie algebra,
	as opposed to in $\mathbb{X}$ (or $G$), meaning that the geometric
	character of the configuration space is respected at all times (e.g.
	orthonormal frames remain orthogonal and normalised in time). Much
	work remains to be done in exploring the structure-preserving aspect
	of geometric integration for Cartan media. The geometrisation leads
	very naturally to a distinction between intrinsic and extrinsic descriptions
	of strain when constructing adapted structure fields (in the case
	that $\text{dim}\mathbb{X}<\text{dim}G$). This framework is in particular
	a natural choice for the description of the \emph{constitutive} mechanics
	of Cartan media. Via the geometrisation process, the system is parameterised
	\emph{in terms} of its intrinsic (and extrinsic, when $\text{dim}\mathbb{X}<\text{dim}G$)
	geometry. External body forces are pulled-back into this geometrised
	description as well. We conclude by mentioning that an invariant theory
	of topological defects \footnote{We anticipate that the failure of spatial integrability, Eq.~\ref{eq:spatial integrability conditions},
		can be considered a topological defect.} and the inclusion of stochasticity in the generalised stresses are
	important avenues for further research.
	\begin{acknowledgments}
		This work was funded in part by the European Research Council under
		the Horizon 2020 Programme, ERC grant agreement number 740269.
	\end{acknowledgments}

	\appendix
	
	\section{Vector and matrix operations \label{app:Vector and Matrix operations}}
	
	Here we present a list of definitions in vector and matrix algebra
	and calculus. In three-dimensional space, there is a natural isomorphism
	between $3$-vectors and anti-symmetric $3\times3$-matrices. This
	isomorphism is known as the hat map. For a column vector $\mathbf{v}\in\mathbb{R}^{3}$,
	the corresponding anti-symmetric matrix $\hat{v}\in\mathbb{R}^{3\times3}$
	is defined as 
	\begin{equation}
		\hat{v}=\begin{pmatrix}0 & -v_{3} & v_{2}\\
			v_{3} & 0 & -v_{1}\\
			-v_{2} & v_{1} & 0
		\end{pmatrix},\label{eq:hat map}
	\end{equation}
	which satisfies $\mathbf{v}\times\mathbf{w}=\hat{v}\mathbf{w}$ for
	any $\mathbf{w}\in\mathbb{R}^{3}$. Conversely, for a given anti-symmetric
	matrix $\hat{b}\in\mathbb{R}^{3\times3}$, the corresponding vector
	is
	\begin{equation}
		\mathbf{b}=(b_{32}\ b_{13}\ b_{21})^{T}\in\mathbb{R}^{3}.
	\end{equation}
	Further, we identify anti-symmetric matrices $\hat{b}$ as elements
	of the Lie algebra $\mathfrak{so}(3)$ of the orthogonal group $SO(3)$.
	
	The fundamental matrix representation of an element $A\in\mathfrak{se}(3)$
	of the special Euclidean transformations in $3$-dimensions as 
	\begin{equation}
		A=\begin{pmatrix}0 & \mathbf{0}^{T}\\
			\mathbf{a}_{1} & \hat{a}_{2}
		\end{pmatrix}\in\mathfrak{se}(3)\label{eq:(app) se(3) fundamental matrix rep}
	\end{equation}
	where $\mathbf{a}_{1},\mathbf{a}_{2}\in\mathbb{R}^{3}$ and $\mathbf{0}\in\mathbb{R}^{3}$,
	and we write this in a short-hand notation as $A=\{\mathbf{a}_{1};\mathbf{a}_{2}\}$.
	Similarly, for dual Lie algebra elements $Y\in\mathfrak{se}(3)^{*}$,
	we write 
	\begin{equation}
		Y=\left\{ \mathbf{y}_{1};\mathbf{y}_{2}\right\} =\begin{pmatrix}0 & \mathbf{y}_{1}^{T}\\
			\vec{0} & \hat{y}_{2}^{T}
		\end{pmatrix}
	\end{equation}
	where $\mathbf{y}_{1},\mathbf{y}_{2}\in\mathbb{R}^{3}$. We can define
	a basis for $\mathfrak{se}(3)$ as
	\begin{equation}
		\begin{aligned}b_{i} & =\left\{ (\delta_{i1}\ \delta_{i2}\ \delta_{i3})^{T};(\delta_{i4}\ \delta_{i5}\ \delta_{i6})^{T}\right\} \\
			& =\left(\begin{array}{cccc}
				0 & 0 & 0 & 0\\
				\delta_{i1} & 0 & -\delta_{i6} & \delta_{i5}\\
				\delta_{i2} & \delta_{i6} & 0 & -\delta_{i4}\\
				\delta_{i3} & -\delta_{i5} & \delta_{i4} & 0
			\end{array}\right)\in\mathfrak{se}(3)
		\end{aligned}
	\end{equation}
	and the corresponding dual basis is then $B_{i}=(b_{i})^{T}=\left\{ (\delta_{i1}\ \delta_{i2}\ \delta_{i3})^{T};(\delta_{i4}\ \delta_{i5}\ \delta_{i6})^{T}\right\} ^{*}\in\mathfrak{se}(3)^{*}$.
	Then, any Lie algebra and dual Lie algebra element can be expanded
	as $C=C_{i}b_{i}$ and $Y=Y_{i}B_{i}$, where $C_{i},Y_{i}\in\mathbb{R},\ i=1,\dots,6$.
	We can then define an inner product $\langle C,Y\rangle=C_{i}Y_{i}$.
	
	The adjoint representation of $\mathfrak{se}(3)$ can be written in
	matrix-form as
	\[
	[\text{ad}_{A}]=\left(\begin{array}{cc}
		\hat{a}_{2} & \hat{a}_{1}\\
		0_{3\times3} & \hat{a}_{2}
	\end{array}\right)\in\mathbb{R}^{6\times6}
	\]
	for any $Y=\{\mathbf{a}_{1};\mathbf{a}_{2}\}$ such that, if $B=\{\mathbf{b}_{1};\mathbf{b}_{2}\}$,
	$C=\{\mathbf{c}_{1};\mathbf{c}_{2}\}$ and $\text{ad}_{A}B=C$, then
	$[\text{ad}_{A}](\mathbf{b}_{1}^{T}\ \mathbf{b}_{2}^{T})^{T}=(\mathbf{c}_{1}^{T}\ \mathbf{c}_{2}^{T})^{T}$.
	The corresponding dual adjoint matrix representation is $[\text{ad}_{A}^{*}]=-[\text{ad}_{A}]^{T}$.
	
	Throughout the text we will often differentiate scalars with respect
	to vectors and matrices. We carry out matrix derivatives using the
	numerator-layout convention. For a matrix $X\in\mathbb{R}^{p\times q}$
	and $f:\mathbb{R}^{p\times q}\to\mathbb{R}$ a scalar function $f(X)$,
	then 
	\begin{equation}
		\frac{\partial y}{\partial X}=\begin{pmatrix}\frac{\partial f}{\partial X_{11}} & \frac{\partial f}{\partial X_{21}} & \dots & \frac{\partial f}{\partial X_{p1}}\\
			\frac{\partial f}{\partial X_{12}} & \frac{\partial f}{\partial X_{22}} & \dots & \frac{\partial f}{\partial X_{p2}}\\
			\vdots & \ddots & \vdots & \vdots\\
			\frac{\partial f}{\partial X_{1q}} & \frac{\partial f}{\partial X_{2q}} & \dots & \frac{\partial f}{\partial X_{pq}}
		\end{pmatrix}.\label{eq:numerator-layout convention}
	\end{equation}
	For any $\mathbf{x}\in\mathbb{R}^{d}$ and a function $y:\mathbb{R}^{d}\to\mathbb{R}$
	we write
	\begin{equation}
		\frac{\partial y}{\partial\mathbf{x}}=\begin{pmatrix}\frac{\partial f}{\partial x_{1}}\\
			\frac{\partial f}{\partial x_{2}}\\
			\vdots\\
			\frac{\partial f}{\partial x_{d}}
		\end{pmatrix}.
	\end{equation}
	If $A\in\mathfrak{se}(3)$ is a Lie algebra element in the fundamental
	matrix representation Eq.~\ref{eq:(app) se(3) fundamental matrix rep},
	then not all of the $16$ elements of the matrix are independent degrees
	of freedom. Strictly speaking, this entails that matrix derivatives
	with respect to matrix functions on $\mathfrak{se}(3)$ are not well-defined.
	However we will introduce, for any $A=\left\{ \mathbf{a}_{1};\mathbf{a}_{2}\right\} \in\mathfrak{se}(3)$
	and any function $m:\mathfrak{se}(3)\to\mathbb{R}$, the short-hand
	\[
	\frac{\partial m}{\partial Y}=\left(\begin{array}{cc}
		0 & \mathbf{y}_{1}^{T}\\
		\vec{0} & \hat{y}_{2}^{T}
	\end{array}\right)\in\mathfrak{se}^{*}(3)
	\]
	where $\mathbf{y}_{1}=\frac{\partial m}{\partial\mathbf{a}_{1}}$
	and $\mathbf{y}_{2}=\frac{\partial m}{\partial\mathbf{a}_{2}}$, and
	where the matrix derivative has been taken with respect to the non-zero
	elements of $A$, and the remaining elements of $\frac{\partial g}{\partial Y}$
	are set to zero.
\end{document}